\documentclass[acmsmall, screen]{acmart}

\usepackage{multirow}

\usepackage{subfigure} 
\usepackage{pifont}

\usepackage{bm}
\usepackage{url}
\usepackage{hyperref}
\hypersetup{
    colorlinks=true,
    linkcolor=blue,
    filecolor=magenta,      
    urlcolor=cyan,
}

\graphicspath{{pictures/}}

\usepackage{algorithm}
\usepackage{algorithmic}
\usepackage{setspace}

\usepackage[normalem]{ulem}
\newcommand{\revise}[1]{{\color{black}{#1}}}
\newcommand{\delete}[1]{}

\newcommand{\revision}[1]{{\color{black}{#1}}}
\newcommand{\deletion}[1]{}

\newcommand{\fcr}[1]{{\color{green}{\textbf{[FCR: #1]}}}}

\AtBeginDocument{%
  \providecommand\BibTeX{{%
    \normalfont B\kern-0.5em{\scshape i\kern-0.25em b}\kern-0.8em\TeX}}}

\setcopyright{acmcopyright}
\copyrightyear{2022}
\acmYear{2022}

\acmJournal{TOSEM}
\acmVolume{0}
\acmNumber{0}
\acmArticle{1}
\acmMonth{0}

\newcommand{\toolname}{EACS}

\title{An Extractive-and-Abstractive Framework for Source Code Summarization}

\begin{document}

\author{Weisong Sun} \email{weisongsun@smail.nju.edu.cn}
\orcid{0000-0001-9236-8264}
\affiliation{
  \institution{State Key Laboratory for Novel Software Technology, Nanjing University}
  \city{Nanjing}
  \state{Jiangsu}
  \country{China}
  \postcode{210093}
}
\author{Chunrong Fang} \email{fangchunrong@nju.edu.cn}
\orcid{0000-0002-9930-7111}
\authornote{\textbf{Chunrong Fang is the corresponding author.}}
\affiliation{
  \institution{State Key Laboratory for Novel Software Technology, Nanjing University}
  \city{Nanjing}
  \state{Jiangsu}
  \country{China}
  \postcode{210093}
}
\author{Yuchen Chen}     \email{yuc.chen@outlook.com}
\orcid{0000-0002-3380-5564}
\affiliation{
  \institution{State Key Laboratory for Novel Software Technology, Nanjing University}
  \city{Nanjing}
  \state{Jiangsu}
  \country{China}
  \postcode{210093}
}
\author{Quanjun Zhang}    \email{quanjun.zhang@smail.nju.edu.cn}
\orcid{0000-0002-2495-3805}
\affiliation{
  \institution{State Key Laboratory for Novel Software Technology, Nanjing University}
  \city{Nanjing}
  \state{Jiangsu}
  \country{China}
  \postcode{210093}
}
\author{Guanhong Tao}    \email{taog@purdue.edu}
\orcid{0000-0002-4701-1327}
\affiliation{
  \institution{Purdue University}
  \city{West Lafayette}
  \state{Indiana}
  \country{USA}
  \postcode{47907}
}

\author{Yudu You}
\email{nju_yyd@163.com}
\orcid{0009-0006-2193-3696}
\affiliation{
  \institution{State Key Laboratory for Novel Software Technology, Nanjing University}
  \city{Nanjing}
  \state{Jiangsu}
  \country{China}
  \postcode{210093}
}

\author{Tingxu Han}    \email{txhan@smail.nju.edu.cn}
\orcid{0000-0003-1821-611X}
\affiliation{
  \institution{State Key Laboratory for Novel Software Technology, Nanjing University}
  \city{Nanjing}
  \state{Jiangsu}
  \country{China}
  \postcode{210093}
}

\author{Yifei Ge}
\email{gyf991213@126.com}
\orcid{0009-0009-0957-854X}
\affiliation{
  \institution{State Key Laboratory for Novel Software Technology, Nanjing University}
  \city{Nanjing}
  \state{Jiangsu}
  \country{China}
  \postcode{210093}
}

\author{Yuling Hu}
\email{yulinghu@smail.nju.edu.cn}
\orcid{0009-0001-0168-8842}
\affiliation{
  \institution{State Key Laboratory for Novel Software Technology, Nanjing University}
  \city{Nanjing}
  \state{Jiangsu}
  \country{China}
  \postcode{210093}
}

\author{Bin Luo}
\email{luobin@nju.edu.cn}
\orcid{0000-0002-9036-0063}
\affiliation{
  \institution{State Key Laboratory for Novel Software Technology, Nanjing University}
  \city{Nanjing}
  \state{Jiangsu}
  \country{China}
  \postcode{210093}
}

\author{Zhenyu Chen}
\email{zychen@nju.edu.cn}
\orcid{0000-0002-9592-7022}
\affiliation{
  \institution{State Key Laboratory for Novel Software Technology, Nanjing University}
  \city{Nanjing}
  \state{Jiangsu}
  \country{China}
  \postcode{210093}
}

\renewcommand{\shortauthors}{W. Sun, C. Fang, Y. Chen, Q. Zhang, G. Tao, Y. You, T. Han, Y. Ge, Y. Hu, B. Luo, and Z. Chen.}

\begin{abstract}
(Source) Code summarization aims to automatically generate summaries/comments for given code snippets in the form of natural language. Such summaries play a key role in helping developers understand and maintain source code. Existing code summarization techniques can be categorized into \textit{extractive methods} and \textit{abstractive methods}. 
The \textit{extractive methods} extract a subset of important statements and keywords from the code snippet using retrieval techniques and generate a summary that preserves factual details in important statements and keywords. However, such a subset may miss identifier or entity naming, and consequently, the naturalness of the generated summary is usually poor. The \textit{abstractive methods} can generate human-written-like summaries leveraging encoder-decoder models. However, the generated summaries often miss important factual details.

To generate human-written-like summaries with preserved factual details, we propose a novel extractive-and-abstractive framework. The extractive module in the framework performs the task of extractive code summarization, which takes in the code snippet and predicts important statements containing key factual details. The abstractive module in the framework performs the task of abstractive code summarization, which takes in the code snippet and important statements in parallel and generates a succinct and human-written-like natural language summary. We evaluate the effectiveness of our technique, called {\toolname}, by conducting extensive experiments on three datasets involving six programming languages. Experimental results show that {\toolname} significantly outperforms state-of-the-art techniques for all three widely used metrics, including \deletion{BLEU-4}\revision{BLEU}, METEOR, and ROUGH-L. In addition, the human evaluation demonstrates that the summaries generated by {\toolname} have higher naturalness and informativeness and are more relevant to given code snippets.
\end{abstract}



\maketitle

\section{Introduction}
\label{sec:introduction}
\begin{figure}[htbp]
  \centering
  \includegraphics[width=0.7\linewidth]{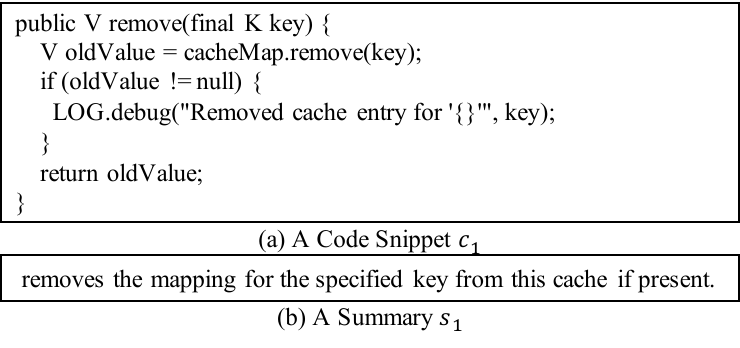}
  \caption{Example of Code Snippet and Summary}
  \label{fig:example_code_summary}
\end{figure}

Code comments play a significant role in facilitating code comprehension~\cite{1981-Comments-on-Program-Comprehension, 1988-Program-Readability, 2018-Measuring-Program-Comprehension, 2020-Code-to-Comment-Translation} and software maintenance~\cite{1993-Maintenance-Productivity, 2005-Documentation-Essential-Software-Maintenance, 2020-CPC, 2021-Why-My-Code-Summarization-Not-Work}. Writing high-quality code comments has been recognized as a good programming practice~\cite{2005-Documentation-Essential-Software-Maintenance, 2020-CPC}. The code comment is one of the most common summaries used during software development. However, writing code comments is a labor-intensive and
time-consuming task~\cite{2005-Documentation-Essential-Software-Maintenance, 2005-Survey-of-Documentation-Practice}. As a result, good comments are often absent, unmatched, and outdated during code evolution~\cite{2018-TL-CodeSum}. 
The prior work~\cite{2018-TL-CodeSum} shows that the lack of high-quality code comments is a common problem in the software industry. 
(Source) code summarization is a hot research topic~\cite{2010-Automated-Text-Summarization-Summarizing-Code, 2013-Evaluating-Source-Code-Summarization, 2018-Study-of-StaQC-Code-Summarization, 2019-Automatic-Code-Summarization, 2019-Datasets-for-Code-Summarization, 2019-Eye-Movements-in-Code-Summarization, 2020-Human-Study-Code-Summarization, 2020-Code-to-Comment-Translation, 2021-Action-Word-Prediction-forCode-Summarization, 2021-Neural-Code-Summarization-How-Far, 2021-Why-My-Code-Summarization-Not-Work, 2021-Reassessing-Metrics-for-Code-Summarization, 2021-Adversarial-Robustness-Deep-Comment-Generation}, which aims to design advanced techniques to support automatic generation of code summaries (i.e., comments).  For a code snippet (a method or function) given by the developer, code summarization techniques can automatically generate natural language summaries related to it. Figure~\ref{fig:example_code_summary} shows an example. The code snippet in Figure~\ref{fig:example_code_summary}(a) is provided by the developer. The summary ``removes the mapping for the specified key from this cache if present.'' in Figure~\ref{fig:example_code_summary}(b) is a summary that satisfies the developer's requirement. The summary is usually a succinct description in natural language summarizing the intention/functionality of the desired code snippet~\cite{2021-Why-My-Code-Summarization-Not-Work}. 

Code summarization can be viewed as a special kind of text summarization task, where the text is not written by a traditional natural language but a programming language. Indeed, as early as ten years ago, automatic text summarization techniques were introduced to automatic code summarization~\cite{2010-Automated-Text-Summarization-Summarizing-Code}, detailed in Section~\ref{subsec:automatic_text_summarization}. Therefore, similar to text summarization~\cite{2021-Automatic-Text-Summarization, 2017-Recent-Automatic-Text-Summarization}, code summarization can be subdivided into extractive code summarization (\textit{extractive methods}) and abstractive code summarization (\textit{abstractive methods}). Most of the early code summarization techniques are \textit{extractive methods}, which widely use an indexing-retrieval framework to generate summaries~\cite{2010-Program-Comprehension-with-Code-Summarization, 2010-Automated-Text-Summarization-Summarizing-Code, 2013-Evaluating-Source-Code-Summarization, 2014-Code-Summarization-Eye-tracking-Study}. They first index terms in code snippets and then retrieve top-$n$ key terms as summaries. The terms in summaries are extracted from the current code snippet~\cite{2010-Program-Comprehension-with-Code-Summarization, 2010-Towards-Generating-Summary-Java-Methods}, context code snippets~\cite{2016-Code-Summarization-of-Context, 2013-Automatic-Generation-Summaries-for-Java-Classes}, or similar code snippets~\cite{2013-Evaluating-Source-Code-Summarization, 2015-CloCom}. Therefore, \textit{extractive methods} can produce a summary that preserves the conceptual integrity and factual information of the input code snippet. 
\revise{In addition, \textit{extractive methods} do not require ontology~\cite{2010-Summarizing-Software-Concerns} or training data~\cite{2010-Automated-Text-Summarization-Summarizing-Code}. However, the terms extracted from code snippets may be meaningless words or abbreviations when the identifiers and methods are poorly named. The summaries consisting of these terms are not informative.
Besides, the naturalness of the summaries generated by \textit{extractive methods} is usually poor and not as human-written~\cite{2016-CODE-NN, 2021-EditSum}, detailed in Section~\ref{sec:motivating_example}.
}

Recently, with the success of deep learning (DL) in abstractive text summarization, the DL-based code summarization techniques have been proposed one after another~\cite{2016-CODE-NN, 2018-Structured-Neural-Summarization, 2018-TL-CodeSum, 2018-Improving-Code-Summarization-via-DRL, 2019-Ast-attendgru, 2019-Convolutional-Neural-Network-Code-Summarization, 2020-Hybrid-DeepCom, 2020-RL-Guided-Code-Summarization, 2020-Rencos, 2020-R2Com, 2020-Improved-Code-Summarization-via-GNN, 2020-Transformer-based-Approach-for-Code-Summarization, 2021-BASTS, 2021-API2Com, 2021-CoTexT}. Compared with \textit{extractive methods}, similar to abstractive text summarization~\cite{2021-Neural-Abstractive-Summarization}, the DL-based code summarization techniques have stronger abstract expression capabilities and can generate human-written-like summaries. 
Therefore, we call the DL-based code summarization techniques as \textit{abstractive methods}. \textit{Abstractive methods} widely adopt the neural network model based on the encoder-decoder architecture and then train the model on a large code-comment corpus. The encoder first transforms code snippets into embedding representations (also called context vectors), and then the decoder decodes the context vectors into short natural language summaries. As in many research areas and as chronicled by Allamanis et al.~\cite{2018-Survey-ML-for-Big-Code}, traditional methods have largely given way to deep learning methods based on big data input. Although the \textit{abstractive methods} have the ability to generate novel words and phrases not featured in the code snippet – as a human-written abstract usually does, the generated summaries often miss important factual details in the code snippet, detailed in Section~\ref{sec:motivating_example}.

In this paper, we propose an extractive-and-abstractive framework for code summarization, which inherits the advantages of extractive and abstractive methods and shields their respective disadvantages. Specifically, we utilize pairs of code snippets and comments to train an extractor (an extractive method) and an abstracter (an abstractive method). The well-trained extractor can be used to predict important statements in code snippets. These important statements and the entire code snippet are input to the abstracter to generate a short natural language summary. The well-trained abstracter first utilizes two separate encoders to transform important statements and the entire code snippet into two context vectors. Then, the two context vectors are fused to produce a fusion vector, which will be passed to a decoder to generate a natural language summary. Compared with existing \textit{abstractive methods}, our extractive-and-abstractive framework is equipped with an extractor, substantially balancing attention on important information and global contextual information, reducing the risk of missing important factual details and improving the overall performance.

In summary, we make the following contributions.
\begin{itemize}
    \item To the best of our knowledge, we are the first to propose an extractive-and-abstractive framework for code summarization, which inherits the advantages of the extractive and abstractive methods and shields their respective disadvantages. It is a general framework and can be combined with multiple advanced models~(see experimental results in Section~\ref{subsubsec:results_of_combine_with_diff_models}).

	\item We implement a code summarization prototype called {\toolname} based on the extractive-and-abstractive framework. {\toolname} is able to generate a succinct natural language summary that is not only human-written-like, but also preserves important factual details.
	\item We conduct extensive quantitative experiments on three widely used datasets to evaluate {\toolname}. Experimental results show that {\toolname} significantly outperforms state-of-the-art baselines in terms of all three widely used automatic metrics, \deletion{BLEU-4}\revision{BLEU}, METEOR, and ROUGE-L (detailed in Section~\ref{subsubsec:result_of_RQ1}). The source code of {\toolname} and all the data used in this paper are released and can be downloaded from the website~\cite{2022-EACS}.
	\item We conduct a qualitative human evaluation to evaluate the summaries generated by {\toolname} and baselines in terms of four aspects: similarity, naturalness, informativeness, and relevance. The statistical results of human scores show that the summaries generated by {\toolname} are more informative and relevant to code snippets (detailed in Section~\ref{subsubsec:human_evaluation}).
\end{itemize}

The remainder of this paper is organized as follows. Section~\ref{sec:background} provides the background of automatic text summarization and neural machine translation. Section~\ref{sec:motivating_example} shows the motivating example. Section~\ref{sec:methodology} introduces our methodology, i.e., the design of {\toolname}. Section~\ref{sec:evaluation} presents automatic evaluation, human evaluation, and case studies in detail. Section \ref{sec:threats_to_validity} introduces some threats to validity. Section~\ref{sec:related_work} presents the related work. We conclude the paper in Section~\ref{sec:conclusion}.

\section{Background}
\label{sec:background}

\subsection{Automatic Text Summarization}
\label{subsec:automatic_text_summarization}
When designing {\toolname}, we drew on the advanced ideas and techniques of automatic text summarization. Therefore, we first introduce the background of automatic text summarization.

Automatic text summarization is the task of automatically condensing a piece of text to a shorter version summary while maintaining the important points~\cite{ 2017-Recent-Automatic-Text-Summarization, 2021-Automatic-Text-Summarization}. 
According to technical characteristics, text summarization is subdivided into extractive text summarization (\textit{extractive methods}) and abstractive text summarization (\textit{abstractive methods})~\cite{1999-Advances-Automatic-Text-Summarization, 2021-Neural-Abstractive-Summarization}. \textit{Extractive methods} assemble summaries by directly selecting words, phrases, and sentences from the source text \revision{that capture its most salient content}. The generated summaries usually persist salient information of source text~\cite{2017-Get-To-Point, 2018-Extract-Coherent-Summary, 2018-Neural-Document-Summarization}. \revision{In the early days, researchers adopted various similarity scores based on specific sentence features (keywords, position, length, frequency, linguistic) and metrics (structure-based, vector-based, and graph-based) to estimate salience between a sentence in a text and its reference summary~\cite{2004-LSA-in-Text-Summarization, 2004-LexRank}. Recently, with advances in distributed representations of words, phrases, and sentences, researchers have proposed to use these distributed representations to compute similarity scores~\cite{2020-Extractive-Abstractive-Neural-Document-Summarization}. Such techniques are further refined by~\cite{2016-Classify-or-Select, 2016-Neural-Summarization-by-Extracting, 2018-Fast-Abstractive-Summarization} where the representations learned by the encoder are used to choose the most salient sentences.} 
In contrast, \textit{abstractive methods} can generate novel words and phrases not featured in the source text -- as a human-written abstract usually does~\cite{2017-Get-To-Point}. \revision{The abstractive methods can be categorized into three categories~\cite{2019-An-Overview-Abstractive-Summarization, 2021-Automatic-Text-Summarization}: 1) structure-based: using pre-defined structures (e.g. graphs~\cite{2010-Opinosis}, trees~\cite{2017-Abstractive-Summarization-by-Partial-Tree}, rules~\cite{2012-Fully-Abstractive-Approach}, and templates~\cite{2014-Template-based-Abstractive-Summarization}), 2) semantic-based: using the text semantic representation and the natural language generation systems (e.g. based on information items, predicate arguments, and semantic graphs)~\cite{2015-Multi-document-Abstractive-Summarization}, and 3) deep-learning-based methods~\cite{2016-Abstractive-Sentence-Summarization, 2016-Summarization-with-Read-Again}. Recently, research in abstractive text summarization has
made significant progress with the help of large
pre-trained models~\cite{2020-BART, 2020-PEGASUS}.} \deletion{Thus, they}\revision{These works show that abstractive methods} have a strong potential to produce high-quality summaries that are verbally innovative~\cite{2021-Neural-Abstractive-Summarization}. 

Automatic text summarization techniques were introduced to automatic code summarization as early as ten years ago~\cite{2010-Automated-Text-Summarization-Summarizing-Code, 2013-Evaluating-Source-Code-Summarization}. 
For example, in 2010, Sonia Haiduc et al.~\cite{2010-Program-Comprehension-with-Code-Summarization} proposed an extractive code summarization technique for the automatic generation of extractive summaries for source code entities. Extractive summaries are generated by selecting the most important terms in code snippets. In addition, they present a study in~\cite{2010-Automated-Text-Summarization-Summarizing-Code} to investigate the suitability of various text summarization techniques for generating code summaries. The study results indicate that a combination of text summarization techniques is most appropriate for code summarization and that developers generally agree with the produced summaries. \revision{Existing extractive methods usually use text retrieval (TR) techniques to determine the most important $n$ terms for each code snippet. The common TR techniques include Vector Space Model~\cite{1975-Vector-Space-Model}, Latent Semantic Indexing~\cite{1990-Latent-Semantic-Analysis}, and Hierarchical PAM~\cite{2007-Mixtures-of-Hierarchical-Topics}. Considering that the quality of the summaries generated by extractive methods depends heavily on the process of extracting the subset, Paige Rodeghero et al.~\cite{2014-Code-Summarization-Eye-tracking-Study} present an eye-tracking study of programmers and propose a tool for selecting keywords based on the findings of the eye-tracking study.} Recently, DL-based code summarization techniques have been proposed one after another~\cite{2018-DeepCom, 2019-Code-Generation-Summarization, 2020-Code-to-Comment-Translation, 2021-SiT}. \revision{For example, Srinivasan Iyer et al.~\cite{2016-CODE-NN} present the first DL-based method for generating code comments. To produce semantic-preserving code embedding representations, multiple aspects of the code snippet have been explored, including tokens~\cite{2018-TL-CodeSum, 2019-Convolutional-Neural-Network-Code-Summarization, 2020-Hybrid-DeepCom, 2020-Transformer-based-Approach-for-Code-Summarization, 2021-CoTexT}, abstract syntactic trees (ASTs)~\cite{2018-Code2seq, 2018-DeepCom, 2019-Code-Summarization-with-Extended-Tree-LSTM, 2019-Ast-attendgru, 2020-Hybrid-DeepCom, 2020-Rencos, 2021-BASTS}, control flows~\cite{2020-RL-Guided-Code-Summarization} and code property graphs~\cite{2020-FusionGNN}. In addition, these DL-based methods have tried various neural network architectures, such as LSTM~\cite{2016-CODE-NN, 2018-DeepCom, 2020-RL-Guided-Code-Summarization}, Bidirectional-LSTM~\cite{2021-EditSum, 2020-Rencos, 2020-R2Com}, GRU~\cite{2020-Hybrid-DeepCom, 2019-Ast-attendgru}, Transformer~\cite{2020-Transformer-based-Approach-for-Code-Summarization, 2021-BASTS} and GNN~\cite{2020-Improved-Code-Summarization-via-GNN, 2020-FusionGNN}.
}
Similar to abstractive text summarization, these techniques also widely adopt the encoder-decoder models borrowed from neural machine translation (detailed in Section~\ref{subsec:neural_machine_translation}) to generate natural language summaries. 
Therefore, we can consider DL-based code summarization techniques as abstractive methods. 
\revision{More extractive and abstractive code summarization techniques are introduced in Section~\ref{sec:related_work}.}

In this paper, we utilize extractive and abstractive methods together. The former is responsible for extracting important factual details while the latter is responsible for generating human-written-like natural language summary, detailed in Section~\ref{sec:methodology}.

\revise{Existing works combining extractive and abstractive summarization methods are mainly proposed in NLP, such as~\cite{2017-Integrating-Extractive-and-Abstractive, 2018-Unified-Model-for-Extractive-Abstractive-Summarization, 2020-Extractive-Abstractive-Neural-Document-Summarization}. Different from these works whose extractors select important sentences from the source text, our extractor extracts important statements from the source code. Additionally, our abstracter takes in important statements and the source code in parallel to generate summaries instead of considering only important sentences as in~\cite{2017-Integrating-Extractive-and-Abstractive, 2020-Extractive-Abstractive-Neural-Document-Summarization} or only source text with sentence-level attention as in~\cite{2018-Unified-Model-for-Extractive-Abstractive-Summarization}.}

\subsection{Neural Machine Translation}
\label{subsec:neural_machine_translation}
\begin{figure}[htbp]
  \centering
  \includegraphics[width=0.9\linewidth]{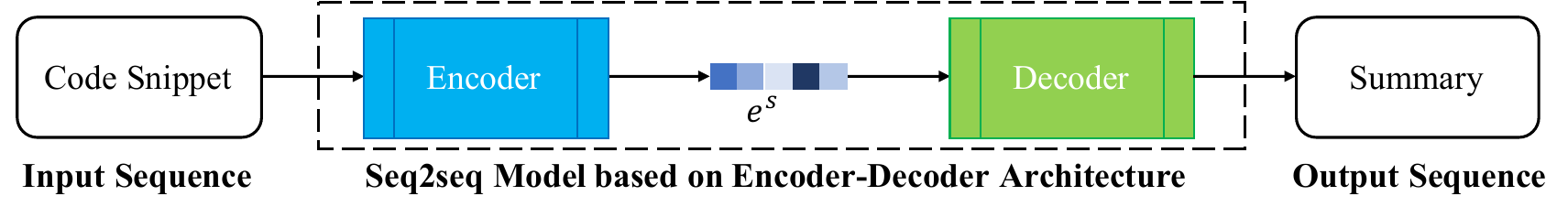}
  \caption{Framework of DL-based Code Summarization}
  \label{fig:DL-based_CS}
\end{figure}

Neural machine translation (NMT) aims to automatically translate one language (e.g., French) into another language (e.g., English) while preserving semantics~\cite{2014-GRU, 2015-NMT-Jointly-Learning-to-Align--Translate, 2018-DeepCom}. NMT has been shown to achieve great success for natural language corpora~\cite{2015-NMT-Jointly-Learning-to-Align--Translate, 2018-DeepCom}. It is typically thought of in terms of sequence to sequence (seq2seq) learning, in which a natural language sentence (e.g., French sentence) is one sequence and is converted into a semantically equivalent target sequence (e.g., English sentence)~\cite{2019-Ast-attendgru}. Code summarization can also be regarded as a kind of translation task, which translates programming language into natural language~\cite{2020-Code-to-Comment-Translation}. Several recent papers~\cite{2018-DeepCom,2016-CODE-NN, 2019-Ast-attendgru, 2018-Code2seq} have explored the idea of applying the seq2seq model to translate code snippets into natural language comments. Similar to NMT, code summarization methods based on seq2seq models also widely adopt the DL-based encoder-decoder architectures. Figure~\ref{fig:DL-based_CS} shows the general framework of the code summarization technique based on the encoder-decoder architecture. From the figure, we can observe that the DL-based code summarization technique usually consists of two key components, an encoder and a decoder. Both encoder and decoder are neural networks, so the DL-based code summarization is also known as neural code summarization~\cite{2020-Rencos, 2021-Neural-Code-Summarization-How-Far}. The encoder is an embedding network that can encode the code snippet $c$ given by the developer into a $d$-dimensional embedding representation $\bm{e}^c \in \mathbb{R}^d$. 
To train such an encoder, existing DL-based code summarization techniques have tried various neural network architectures, such as LSTM~\cite{2016-CODE-NN, 2018-DeepCom, 2020-RL-Guided-Code-Summarization}, Bidirectional-LSTM~\cite{2021-EditSum, 2020-Rencos, 2020-R2Com}, GRU~\cite{2020-Hybrid-DeepCom, 2021-CoCoSum, 2019-Ast-attendgru}, Transformer~\cite{2020-Transformer-based-Approach-for-Code-Summarization, 2021-BASTS, 2021-API2Com} and GNN~\cite{2020-Improved-Code-Summarization-via-GNN, 2020-FusionGNN}. The decoder is also a neural network that can decode the embedding representation $\bm{e}^c$ into a natural language summary. To train such a decoder, existing DL-based code summarization techniques usually adopt the same neural network architecture as the encoder. In DL-based code summarization studies, it is a common practice to use code comments as summaries during the training process~\cite{2018-DeepCom, 2021-EditSum}. Code comments are natural language descriptions used to explain what the code snippets want to do~\cite{2018-DeepCom}. For example, Figure~\ref{fig:example_code_summary}(b) is a comment for the code snippet $c_1$. Therefore, we do not strictly distinguish the meaning of the two terms \textit{comment} and \textit{summary}, and use the term \textit{comment} during the training process, and \textit{summary} at other time.

\section{Motivating Example}
\label{sec:motivating_example}
\begin{figure}[htbp]
  \centering
  \includegraphics[width=0.9\linewidth]{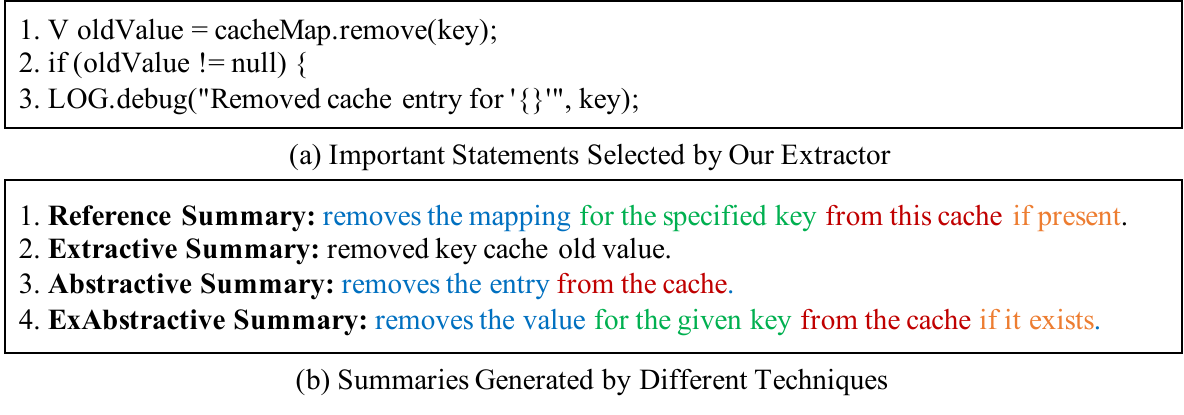}
  \caption{Motivating Example}
  \label{fig:motivation_example}
\end{figure}

In this section, we take the code snippet $c_1$ in Figure~\ref{fig:example_code_summary}(a) as an example and apply different tools to generate summaries for comparison. It is a real-world example from the CodeXGLUE dataset~\cite{2021-CodeXGLUE} (see details in Section~\ref{subsubsec:dataset}). Figure~\ref{fig:example_code_summary}(b) shows the comment ($s_1$) written by the developer for $c_1$. We consider this as a reference summary (the ground truth), as shown in the first line of Figure~\ref{fig:motivation_example}(b). According to the grammar rules in natural language, we can simply divide the reference summary into four parts: ``removes the mapping'' (Blue font), ``for the specified key'' (Green font), ``from this cache'' (Red font), and ``if present''(Orange font).

We study two existing techniques, an extractor method~\cite{2010-Program-Comprehension-with-Code-Summarization} and an abstracter method~\cite{2020-CodeBERT}, in generating summaries for the given example. The extractor method~\cite{2010-Program-Comprehension-with-Code-Summarization} adopts the Latent Semantic Analysis (LSA) techniques~\cite{2009-Update-Summarization-based-LSA} to determine the informativity of every term in the code snippet and then select the top $k$ important terms to compose the summary. The second summary in Figure~\ref{fig:motivation_example}(b) (Extractive Summary) is generated by~\cite{2010-Program-Comprehension-with-Code-Summarization}. Observe that although the extractive summary has a poor naturalness and is far from the reference summary, it contains important factual details that should be included in the summary, e.g., the important terms ``key'' and ``cache''. The abstractive method~\cite{2020-CodeBERT} first trains a model called CodeBERT for obtaining code representations, and then fine-tunes it on the code summarization task. The third summary in Figure~\ref{fig:motivation_example}(b) (Abstractive Summary) shows the result by~\cite{2020-CodeBERT}. Observe that 1) intuitively, the abstractive summary has a good naturalness and is like written by a human; 2) the abstractive summary can cover the first and the third parts (Blue and Red fonts) of the reference summary; 3) the abstractive summary can not cover the second and the fourth parts (Green and Orange fonts), i.e., missing some factual details. 
\revision{In summary, the main drawbacks of current extractive and abstractive methods are the poor naturalness of the generated code summaries and the loss of important factual details, respectively.}

\revision{To overcome the drawbacks of existing extractive and abstractive methods, we propose an extractive-and-abstractive framework, i.e., {\toolname}. {\toolname} simultaneously inherits the advantages of both extractive and abstractive methods. For example, the last summary (i.e., ExAbstractive Summary) in Figure~\ref{fig:motivation_example}(b) is generated by {\toolname} for $c_1$. We can observe that 1) compared with the extractive summary generated by~\cite{2010-Program-Comprehension-with-Code-Summarization}, the exabstractive summary has a good naturalness and is like written by a human; 2) compared with the abstractive summary generated by ~\cite{2020-CodeBERT}, the exabstractive summary can cover all of the four parts and is closer to the reference summary. 
In the next section, we will introduce the design details of {\toolname}.
}

\section{Design}
\label{sec:methodology}
\begin{figure}[htbp]
  \centering
  \includegraphics[width=0.9\linewidth]{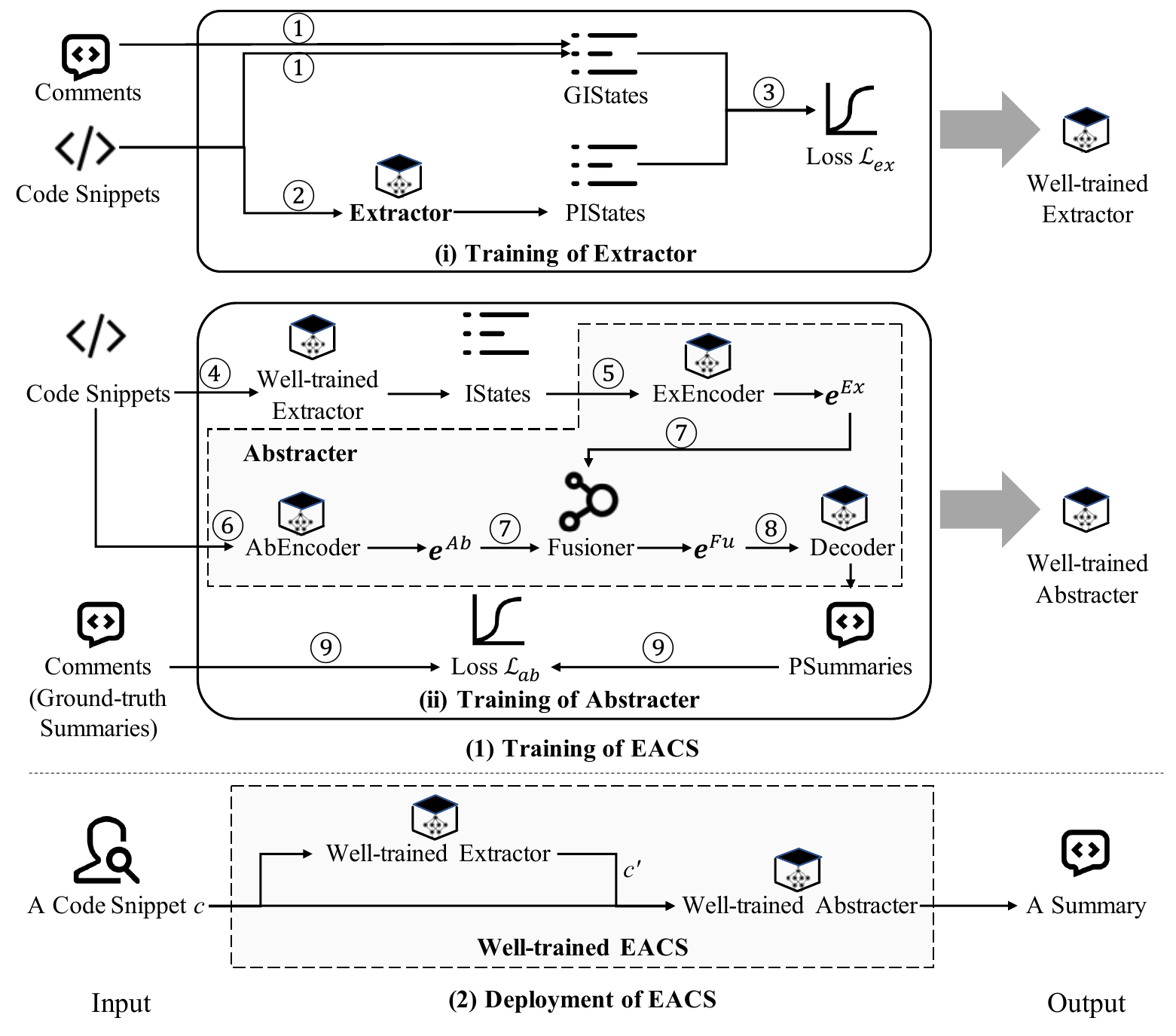}
  \caption{The Overview of {\toolname}}
  \label{fig:overview_of_our_approach}
  \vspace{-0.5cm}
\end{figure}

\subsection{Overview}
\label{subsec:overview}
Figure \ref{fig:overview_of_our_approach} illustrates the overview of our approach {\toolname}. 
\deletion{The top part shows the training process of {\toolname}, and the bottom part shows the deployment (usage) of {\toolname} for a given code snippet.}\revision{The top and bottom parts show the training and deployment of {\toolname}, respectively.}
{\toolname} decomposes the training process into two phases: (i) training of extractor and (ii) training of abstracter. \revision{Both phases leverage the same two types of input data: comments and code snippets.}

\deletion{The training of the extractor aims to produce a well-trained extractor capable of extracting important statements from a given code snippet.}\revision{The goal of phase (i) is to train an extractor capable of extracting important statements from a given code snippet.} 
\deletion{The training of the abstracter aims to produce a well-trained abstracter capable of generating a succinct natural language summary for a given code snippet. Both phases leverage the same two types of input data: comments and code snippets.} 
To train \deletion{the}\revision{such an} extractor, {\toolname} first produces ground-truth important statements (GIStates) based on the informativity of each statement in the code snippet\deletion{, detailed in Section~\ref{subsubsec:training_of_extractor}-\textbf{Step~\ding{192}}}. Then, {\toolname} uses the extractor to extract predicted important \deletion{sentences}\revision{statements} (PIStates)\deletion{, detailed in Section~\ref{subsubsec:training_of_extractor}-\textbf{Step~\ding{193}}}. During this \deletion{procedure}\revision{phase}, \deletion{the model parameters of the extractor are randomly initialized. Based on the loss ($\mathcal{L}_{Ex}$) computed based on PIStates and GIStates, {\toolname} can iteratively update the model parameters of the extractor}\revision{the model parameters of the extractor are initialized randomly and updated iteratively according to the loss $\mathcal{L}_{Ex}$. $\mathcal{L}_{Ex}$ is computed based on PIStates and GIStates}\deletion{, detailed in Section~\ref{subsubsec:training_of_extractor}-\textbf{Step~\ding{194}}}. \revision{Details of training of the extractor are described in Section~\ref{subsubsec:training_of_extractor}.}

\revision{The goal of phase (ii) is to train an abstracter capable of generating a succinct natural language summary for a given code snippet.} To train \deletion{the}\revision{such an} abstracter, given a code snippet, {\toolname} first uses the well-trained extractor to extract the important statements (IStates)\deletion{, detailed in Section~\ref{subsubsec:training_of_Abstracter}-\textbf{Step~\ding{195}}}. \deletion{These important statements}\revision{IStates} will be further transformed into the embedding representation $\bm{e}^{Ex}$ \deletion{by}\revision{through} an encoder named ExEncoder\deletion{, detailed in Section~\ref{subsubsec:training_of_Abstracter}-\textbf{Step~\ding{196}}}.
\deletion{Then}\revision{Meanwhile}, {\toolname} uses another encoder named AbEncoder to transform the entire code snippet into the embedding representation $\bm{e}^{Ab}$\deletion{, detailed in Section~\ref{subsubsec:training_of_Abstracter}-\textbf{Step~\ding{197}}}. 
\deletion{Further}\revision{Then}, {\toolname} produces the fused embedding representation $\bm{e}^{Fu}$ by fusing $\bm{e}^{Ex}$ and $\bm{e}^{Ab}$\deletion{, detailed in Section~\ref{subsubsec:training_of_Abstracter}-\textbf{Step~\ding{198}}}. 
\deletion{The fused embedding representation}\revision{Further,} $\bm{e}^{Fu}$ will be passed to a decoder (Decoder) to generate predicted summaries (PSummaries). During this \deletion{procedure}\revision{phase}, the model parameters of the abstracter (including ExEncoder, AbEncoder, and Decoder) are also randomly initialized. \deletion{Finally, based on the loss ($\mathcal{L}_{Ab}$) between the predicted summaries (PSummaries) and ground-truth summaries (i.e., comments), {\toolname} can iteratively update the model parameters of the abstracter}\revision{{\toolname} iteratively updates the abstracter's parameters according to the loss $\mathcal{L}_{Ab}$. $\mathcal{L}_{Ab}$ is computed based on PSummaries and ground-truth summaries (i.e., comments)}\deletion{, detailed in Section\ref{subsubsec:training_of_Abstracter}-\textbf{Step~\ding{200}}}. \revision{Details of training of the abstracter are described in Section~\ref{subsubsec:training_of_Abstracter}.} 

The well-trained extractor and abstracter are the two core components of {\toolname} to support the code summarization service. When {\toolname} is deployed for usage, it takes in a code snippet from the developer and produces a succinct natural language summary for the code snippet, detailed in Section~\ref{subsec:deployment_of_EACS}.

\subsection{Training of {\toolname}}

\subsubsection{Part (i): Training of Extractor}
\label{subsubsec:training_of_extractor}
\
\newline
Let $c$ denote a code snippet containing a set of statements $stat = [stat_1, stat_2, \cdots, stat_n]$\deletion{,} where $stat_i$ is the $i$-th statement in $c$. Similar to extractive text summarization~\cite{2019-Fine-tune-BERT-for-Extractive-Summarization}, extractive code summarization can be defined as the task of assigning a label $l_i \in \{0, 1\}$ to each $stat_i$\deletion{, indicating}\revision{. The label indicates} whether the factual details contained in the statement $stat_i$ should be included in the summary. It is assumed that the summary represents the important content of the code snippet. Therefore, we can consider the extractor as a classifier that takes in all statements of a code snippet and predicts \deletion{which statements should be selected as}\revision{the} important ones.

As shown in part (i) of Figure~\ref{fig:overview_of_our_approach}, the training of the extractor is completed through three steps: \ding{192} producing ground-truth important Statements (GIStates)\deletion{ based on pairs of comments and code snippets}, \ding{193} producing predicted important statements (PIStates)\deletion{ using the extractor}, and \ding{194} calculating the loss ($\mathcal{L}_{Ex}$) based on GIStates and PIStates and updating the model parameters\deletion{ of the extractor}. We discuss the three steps in detail in the following sections.

\textbf{Step~\ding{192}: Producing Ground-truth Important Statements.}
As mentioned earlier, \deletion{we aim to train an extractor with the ability to take in a sequence of statements of the code snippet $c$ and output the predicted labels of the statements in $c$.}\revision{we aim to train an extractor that can predict the important statements in the code snippet $c$.} We denote the labels of the statements in $c$ as $l = [l_1, l_2, \cdots, l_n]$\deletion{, where}\revision{.} $l_i \in \{0, 1\}$ is the label of the statement $stat_i$. $l_i = 1$ means the $stat_i$ is an informative (important) statement; otherwise, it is not.  
To train such an extractor, we \deletion{need to }build a training dataset\deletion{,} where each sample is a pair of $stat$ and the corresponding labels $\hat{l}$. We consider $\hat{l}$ as the ground-truth label of $stat$.

To obtain the ground-truth labels, like~\cite{2018-Unified-Model-for-Extractive-Abstractive-Summarization}, we first measure the informativity of each statement $stat_i \in stat$\deletion{ by computing the ROUGE-L score~\cite{2004-ROUGE} between the statement $stat_i$ and the reference statements}. \revision{Specifically, the informativity is measured by the ROUGE-L score~\cite{2004-ROUGE} between the statement $stat_i$ and the reference statements.} We consider comments of code snippets as reference statements. \deletion{Second}\revision{Then}, we sort the statements by their informativity and select them in \deletion{the} order of high to low informativity. \deletion{We add one statement each time when the new statement can increase the informativity of all the selected statements.}\revision{We select one statement each time. If a newly selected statement can increase the informativity of all the selected statements, it will be regarded as a ground-truth important statement.} Finally, we obtain the ground-truth labels, i.e., all ground-truth important statements (GIStates).

\revise{
In the above process, the ROUGE-L score is computed based on the longest common subsequence. The previous work~\cite{2020-OCoR} shows that there are often overlapped strings between identifiers (e.g., ``QuickSort'') in code snippets and words (e.g., ``sort'') in summaries. After code preprocessing, all identifiers will be split into multiple words (also called tokens). For example, ``QuickSort'' would be split into ``quick'' and ``sort''. Hence, both token sequences of code snippets and word sequences of summaries can be treated as string sequences (i.e., the same modality) and used to compute the ROUGE-L score. 
}

\revise{
Existing extractors~\cite{2018-Fast-Abstractive-Summarization, 2017-SummaRuNNer, 2019-Text-Summarization-Pretrained-Encoders, 2018-Unified-Model-for-Extractive-Abstractive-Summarization} in NLP widely adopt the ROUGE-L score to select important sentences from the source text. The work~\cite{2018-Fast-Abstractive-Summarization} gives a reasonable explanation of why the ROUGE score is used, that is, ``If the extractor chooses a good sentence, after the abstractor rewrites it the ROUGE match would be high, and thus the action is encouraged. If a bad sentence is chosen, though the abstractor still produces a compressed version of it, the summary would not match the ground truth, and the low ROUGE score discourages this action''.  
In other words, \deletion{since the ROUGE Score is widely used to evaluate the performance of text summarization models, }if the ROUGE \deletion{Score}\revision{score} is used to guide the selection of important sentences, the summaries generated based on such important sentences can get higher ROUGE \deletion{Scores}\revision{scores}.
We follow the production process of ground-truth labels widely adopted in existing extractors~\cite{2018-Fast-Abstractive-Summarization, 2017-SummaRuNNer, 2019-Text-Summarization-Pretrained-Encoders, 2018-Unified-Model-for-Extractive-Abstractive-Summarization}. This process utilizes a greedy strategy to select important sentences~\cite{2019-Text-Summarization-Pretrained-Encoders}. In this way, the ROUGE score between the selected important statements and the reference summary is the highest\deletion{, which}\revision{. And it also} indicates that the selected important statements contain the most factual detail. 
}

\textbf{Step~\ding{193}: Producing Predicted Important Statements.} As shown in part (i) Figure~\ref{fig:overview_of_our_approach}, we use an extractor (Extractor) to predict important statements in the code snippet. \deletion{The extractor is a neural network model that consists of an encoder that transforms statements into embedding representations and a classification layer for predicting the labels of statements.}\revision{The extractor is a neural network model consisting of an encoder and a classification layer. The encoder transforms statements into embedding representations. The classification layer predicts labels of statements based on their embedding representations.} 

The encoder essentially performs the code representation task of transforming complex source code into a numerical (embedding) representation\revision{ while preserving semantics. Such embedding representations are} convenient for program computation\deletion{ while preserving semantics}. Therefore, a large number of existing deep learning-based code representation techniques can be adopted by our {\toolname} to design the encoder. Specifically, the embedding representations of the statements $\bm{e}^{stat}$ can be formalized as $\bm{e}^{stat} = encoder(stat)$\deletion{, where the encoder}\revision{ . $encoder(\cdot)$} is a \deletion{deep learning-based }neural network architecture (e.g., LSTM~\cite{1997-LSTM}) or a pre-trained model (e.g., CodeBERT~\cite{2020-CodeBERT}) that can numericalize sequences of code statements with preserving semantics (i.e., embedding representations). In practice, we tried multiple neural network architectures (e.g., LSTM~\cite{1997-LSTM} and Transformer~\cite{2017-Transformer}) and pre-trained models (e.g., CodeBERT~\cite{2020-CodeBERT} and CodeT5~\cite{2021-CodeT5})\deletion{, and}\revision{. We experimentally} found that, in the code summarization task, the encoder obtained by fine-tuning a pre-trained model performed better than that trained from scratch based on the neural network architecture, detailed in Section~\ref{subsubsec:results_of_combine_with_diff_models}. We do not repeatedly present the design of the neural network architectures or pre-training models involved in the paper. Please read the corresponding paper for more details. The classification layer is connected to the embedding representation $\bm{e}^{stat}$. In practice, we adopt $softmax$ as the classification layer, i.e., $l = softmax(\bm{e}^{stat})$.

\textbf{Step~\ding{194}: Model Training}.
During training, we update the model parameters $\Theta$ of the extractor regarding the sigmoid cross-entropy loss, i.e., $\mathcal{L}_{Ex}(\Theta)$ computed as:
\begin{equation}
	\mathcal{L}_{Ex}(\Theta) = -\frac{1}{N}\sum_{n=1}^{N}{\hat{l}_{n}log{l_n} + (1-\hat{l}_{n})log(1 - l_n)}
	\label{equ:loss_ex}
\end{equation}
where $\hat{l}_n \in {0, 1}$ is the ground-truth label for the
$n$th statement, and $N$ is the number of statements. $\hat{l}_n = 1$ indicates that the model should pay attention to the factual details contained in the $n$th statement to facilitate final summary generation. 

\revision{
Once the well-trained extractor is produced, we can use it to extract important statements from the given code snippet. Taking the code snippet $c_1$ shown in Figure~\ref{fig:example_code_summary}(a) as an example, our well-trained extractor extracts three important statements for it, as shown in Figure~\ref{fig:motivation_example}(a). These important statements play a key role in generating the final summary. For example, the ExAbstractive summary shown in the last line of Figure~\ref{fig:motivation_example}(b) is generated based on these important statements. Intuitively, the second part (Green font) is a translation or summary of the factual details (key terms ``key'') contained in the important statements \textit{V oldValue = cacheMap.remove(key);} and \textit{Log.debug(``Removed cache entry for`\{\}' '', key);}. The fourth part (Orange font) is a translation or summary of the factual details contained in the important conditional statement \textit{if(oldValue) != null)}. We also find that the last two important statements may be related to the logging behavior for debugging purposes and may not be part of the essential code of the function. Considering that the human developers would expect exactly the same summary even without those lines related to logging, we further study the summaries generated by {\toolname} by deleting each important statement in the motivating example and obtain the following results:

(1) deleting the first line yields a summary ``removes a value from the cache if it exists'';

(2) deleting the second line yields ``removes an entry from the cache''; 

(3) deleting the last line yields ``removes the value associated with the key from the cache if present''; 

Observe that deleting either the first or the second line causes the missing factual detail ``for the specified key'', while without the last line, the summary is the same as the ground truth. It indicates that the first two lines are critical for having a completely accurate summary. Additionally, the factual detail ``if it exists'' can only be obtained when the second line presents. This is because the second line checks the return value of the remove function, which validates the existence of the key. Our well-trained extractor precisely captures these details, and {\toolname} can produce the same summary as the reference without the logging line. This example demonstrates the effectiveness of our well-trained extractor. 
}

\subsubsection{Part(ii): Training of Abstracter}
\label{subsubsec:training_of_Abstracter}
\
\newline
As shown in part (ii) of Figure~\ref{fig:overview_of_our_approach}, the training of the abstracter is completed through \deletion{five}\revision{six} steps: \ding{195} extracting important statements (IStates)\deletion{ from the given code snippet using the well-trained extractor}, \ding{196} and \ding{197} producing embedding representations ($\bm{e}^{Ex}$ and $\bm{e}^{Ab}$) of the important statements and the entire code snippet, \ding{198} producing the fused representation $\bm{e}^{Fu}$ based on $\bm{e}^{Ex}$ and $\bm{e}^{Ab}$, \ding{199} generating predicted summary\deletion{ with the help of the decoder}, and \ding{200} computing the loss $\mathcal{L}_{Ab}$ based on the predicted summaries (PSummaries) and the ground-truth summaries (comments) to update the model parameters\deletion{ of the abstracter}. We discuss the six steps in detail as follows.

\textbf{Step~\ding{195}: Extracting Important Statements.} To generate summaries without missing factual details, our {\toolname} pays more attention to the important statements of code snippets\deletion{ where the factual details are contained in}\revision{. These important statements contain factual details that should be included in final generated summaries}. Therefore, different from abstracters in existing abstract code summarization techniques, the abstracter of {\toolname} treats important \deletion{sentences}\revision{statements} as part of the input. In this step, we first use the well-trained extractor \deletion{produced at the phase of training of the extractor }to predict the labels of statements of the given code snippet. Then, the statements with the label 1 will be selected as important statements (IStates).

\textbf{Step~\ding{196} and Step~\ding{197}: Producing Embedding Representations.} Step~\ding{196} and Step~\ding{197} do a similar thing, i.e. leveraging an encoder to transform the source code into an embedding representation. The difference is that Step~\ding{196} deals with important statements selected by the extractor\deletion{ and}\revision{, while} Step~\ding{197} deals with the entire code snippet. Therefore, in {\toolname}, we can use the same neural network architecture or pre-trained model to design ExEncoder and AbEncoder. Given a code snippet $c = [stat_1, stat_2, \cdots, stat_n]$, let $c' \subseteq c$ denote a set of the important statements selected by the extractor from $c$, the tasks performed by ExEncoder and AbEncoder can be formalized as follows:
\begin{equation}
\bm{e}^{Ex} = encoder(c'), \;\;\;\; \bm{e}^{Ab} = encoder(c)
  \label{equ:ExEncoder_AbEncoder}
\end{equation}
where $\bm{e}^{Ex}$ and $\bm{e}^{Ab}$ represent the embedding representations of $c'$ and $c$; \deletion{the encoder}\revision{$encoder(\cdot)$} is a neural network architecture (e.g., LSTM~\cite{1997-LSTM}, Transformer~\cite{2017-Transformer}) or pre-trained model (e.g., CodeBERT~\cite{2020-CodeBERT}) that can process sequential input. ExEncoder and AbEncoder perform the similar task --- producing embedding representations of the code, so we design them with the same code representation techniques as the extractor's encoder.

\textbf{Step~\ding{198}: Producing Fused Representation.} In this step, {\toolname} fuses $\bm{e}^{Ex}$ and $\bm{e}^{Ab}$ to produce a fused embedding representation $e^{Fu}$ through \deletion{the component Fusioner}\revision{the Fusioner component}. Considering that $\bm{e}^{Ex}$ and $\bm{e}^{Ab}$ are not aligned, we fuse them in a concatenated fashion. We try two concatenated ways as follows:
\begin{equation}
   \bm{e}^{Fu} = [\bm{e}^{Ex};\bm{e}^{Ab}] ~\text{or}~ [\bm{e}^{Ab};\bm{e}^{Ex}]
  \label{equ:e_Fu}
\end{equation}
where $[\cdot;\cdot]$ denotes the concatenation of two vectors. The effects of both ways on the performance of {\toolname} are discussed in Section~\ref{subsubsec:effect_of_fusion_fashion}.

\textbf{Step~\ding{199}: Generating Predicted Summaries.} In this section, we utilize the decoder to generate natural language \deletion{summary, which}\revision{summaries. The decoder} takes in the fused embedding representation $\bm{e}^{Fu}$ and predicts word\revision{s} one by one. Specifically, the decoder based on a neural network (e.g., LSTM) is to unfold the context vector $\bm{e}^{Fu}$ into the target sequence (i.e., the word sequence of the summary), through the following dynamic model,
\begin{equation}
  \begin{aligned}
      \bm{h}_t = f(y_{t-1}, \bm{h}_{t-1}, \bm{e}^{Fu})\\ p(y_t|Y_{<t}, X) = g(y_{t-1}, \bm{h}_t, \bm{e}^{Fu})
  \end{aligned}
  \label{equ:sequence_prediction}
\end{equation}
where $f(\cdot)$ and $g(\cdot)$ are activation functions\deletion{,}\revision{;} $\bm{h}_t$ is the hidden state of the neural network at time $t$\deletion{,}\revision{;} $y_t$ is the predicted target word at $t$ through $g(\cdot)$ with $Y_{<t}$ denoting the history $\{y_1, y_2, \cdots , y_{t-1}\}$. The prediction process is typically a classifier over the vocabulary. It can be seen from Equation~(\ref{equ:sequence_prediction}) \deletion{that}\revision{where} the probability of generating \revision{a} target word is related to the current hidden state, the history of the target sequence and the context $\bm{e}^{Fu}$. The essence of the decoder is to classify the vocabularies by optimizing the loss function in order to generate the vector representing the feature of the target word $y_t$. After the vector passes through a \textit{softmax} function, the word corresponding to the highest probability is the result to be output.

\textbf{Step~\ding{200}: Model Training.} During \revision{the} training of the abstracter, the three components (ExEncoder, AbEncoder, and Decoder) are jointly trained to minimize the negative conditional log-likelihood, i.e, $\mathcal{L}_{Ab}(\Theta)$ computed as:
\begin{equation}
\small
	\mathcal{L}_{Ab}(\Theta) = -\frac{1}{N}\sum_{n=1}^{N}logp(\bm{y}_n|\bm{x}_n; \Theta)
	\label{equ:loss_ab}
\end{equation}
where $\Theta$ is the model parameters of the abstracter and each $(\bm{x}_n,\bm{y}_n)$ is \deletion{an}\revision{a} (code snippet, comment) pair from the training set.

\subsection{Deployment of {\toolname}}
\label{subsec:deployment_of_EACS}
After {\toolname} is trained, we can deploy it online for code summarization service. Part (2) of Figure~\ref{fig:overview_of_our_approach} shows the deployment of {\toolname}. For a code snippet $c$ given by the developer, {\toolname} first uses the well-trained extractor to extract important statements from $c$, represented $c'$. Then, {\toolname} uses the well-trained abstracter to generate the summary. In practice, we can consider the well-trained {\toolname} as a black-box  tool that takes in a code snippet given by the developer and generates a succinct natural language summary.

\section{Evaluation}
\label{sec:evaluation}
To evaluate our approach, in this section, we aim to answer the following \deletion{four}\revision{six} research questions:
\begin{description}
    \item[\textbf{RQ1:}] How does {\toolname} perform compared to the state-of-the-art baselines?
    \item[\textbf{RQ2:}] \deletion{How does {\toolname} perform when combined with different neural network architectures and pre-trained models?}\revision{How does {\toolname} perform in terms of generality?}
    \item[\textbf{RQ3:}] How does the fusion way of the extractor and abstracter affect the performance of {\toolname}?
    
    \item[\textbf{RQ4:}] How does the robustness of {\toolname} perform when varying the code length and comment length?
    \revise{\item[\textbf{RQ5:}] How does {\toolname} perform in human evaluation?}

    \revision{\item[\textbf{RQ6:}] How does the similarity metric (e.g., BLEU, METEOR, and ROUGE-L) used in the extractor affect the performance of {\toolname}?}
\end{description}

\subsection{Experimental Setup}
\label{subsubsec:experimental_setup}

\subsubsection{Dataset}
\label{subsubsec:dataset}
We conduct experiments on three datasets, including a Java dataset (JCSD)~\cite{2018-TL-CodeSum}, a Python dataset (PCSD)~\cite{2017-Corpus-Python-Functions}, and a CodeSearchNet corpus~\cite{2019-CodeSearchNet-Challenge}, which have been widely used by existing code summarization studies~\cite{2019-Code-Generation-Summarization, 2020-CodeBERT, 2020-Transformer-based-Approach-for-Code-Summarization, 2020-Rencos, 2021-SiT, 2022-SCRIPT}.
\revise{JCSD is provided by Hu et al.~\cite{2018-TL-CodeSum}, which contains 69,708 pairs of Java methods and their comments collected from GitHub~\cite{2008-GitHub}. 
They have split training/validation/test sets with 69,708/8,714/8,714.
}
PCSD is provided by Barone et al.~\cite{2017-Corpus-Python-Functions}, which contains 108,726 pairs of Python functions and their comments collected from GitHub. It uses docstrings (document strings) as comments. 
\delete{This dataset includes 108,726 code-comment pairs and has been used for training and evaluation in~\cite{2018-Improving-Code-Summarization-via-DRL}, which has split training/validation/test sets.}
\revise{We follow SiT~\cite{2021-SiT} and evaluate our {\toolname} on the PCSD dataset provided by Wan et al.~\cite{2018-Improving-Code-Summarization-via-DRL}, which has split training/validation/test sets with 65,236/21,745/21,745.}
The CodeSearchNet corpus provided by Husain et al.~\cite{2019-CodeSearchNet-Challenge} contains a large number of pairs of code snippets and comments across six programming languages, including Go, Java, JavaScript, PHP, Python, and Ruby. 
\revise{
Lu et al.~\cite{2021-CodeXGLUE} showed that some comments contain content unrelated to the code snippets and performed data cleaning on the CodeSearchNet corpus. Therefore, in this paper, we follow~\cite{2020-CodeBERT, 2021-CodeT5} and use the clean version of the CodeSearchNet corpus called the CodeXGLUE dataset provided by Lu et al.~\cite{2021-CodeXGLUE}. The statistics of the CodeXGLUE dataset are listed in Table~\ref{tab:statistics_of_datasets}.
}
\delete{The statistics of the three datasets are shown in Table~\ref{tab:statistics_of_datasets}.}\revise{Existing works~\cite{2022-Evaluation-Neural-Code-Summarization, 2021-Neural-Code-Summarization-How-Far} show that code pre-processing choices can have a large impact on the summarization performance and should not be neglected. For code pre-processing, we follow the work~\cite{2022-Evaluation-Neural-Code-Summarization} and use the same pre-processing result in {\toolname} and baselines.
}

\delete{\textbf{Note that} existing studies~\cite{2020-CodeBERT, 2021-CodeT5} evaluate the effectiveness of pre-trained models for code representation (e.g., CodeBERT~\cite{2020-CodeBERT} and CodeT5~\cite{2021-CodeT5}) through the following two steps: 1) fine-tuning them on the training set of downstream tasks (e.g. code summarization task) to produce task-specific models; 2) evaluating the effectiveness of the task-specific models on the corresponding test set, which indirectly indicates the effectiveness of the pre-trained models. Therefore, to verify the effectiveness of the pre-trained model, code representation techniques (e.g., CodeBERT) divide the CodeSearchNet corpus into two parts before training the model. One part is used for training the pre-trained model, and the other part is used as downstream task data to evaluate the effectiveness of the pre-trained model. As mentioned earlier, we can use pre-trained models as the encoders of {\toolname}. Therefore, the encoders have seen (learned) the part of the CodeSearchNet corpus used for training the pre-trained models. For a fair comparison, in this paper, we only use the unseen data used as downstream code summarization task data. In practice, we use the cleaned dataset for code summarization task published by CodeBERT~\cite{2020-CodeBERT}.}

\begin{table}[htbp]
\small
  \renewcommand{\arraystretch}{1.1}
  \caption{Statistics of three datasets}
  \label{tab:statistics_of_datasets}
  \centering
  \begin{tabular}{|c|l|c|c|c|}
    \hline
    \multicolumn{2}{|c|}{Dataset} & Training Set Size & Validation Set Size & Test Set Size \\
    \hline
    \multicolumn{2}{|c|}{JCSD} & 69,708 & 8,714 & 8,714 \\
    \multicolumn{2}{|c|}{PCSD} & 65,236 & 21,745 & 21,745\\
    \hline
    \hline
    \multirow{6}{*}{CodeXGLUE} & Go & 167,288 & 7,325 & 8,122 \\
    & Java & 164,923 & 5,183 & 10,955 \\
    & JavaScript & 58,025 & 3,885 & 3,291 \\
    & PHP & 241,241 & 12,982 & 14,014 \\
    & Python & 251,820 & 13,914 & 14,918 \\
    & Ruby & 24,927 & 1,400 & 1,262 \\
    \hline
 \end{tabular}
\end{table}

\subsubsection{Evaluation Metrics}
\label{subsubsec:evaluation_metrics}
We use three metrics BLEU~\cite{2002-BLEU}, METEOR~\cite{2005-METEOR}, and ROUGE~\cite{2004-ROUGE}, to evaluate the model, which are widely used in code summarization~\cite{2022-SCRIPT, 2021-SiT, 2018-Improving-Code-Summarization-via-DRL, 2018-TL-CodeSum, 2017-Transformer, 2016-CODE-NN}. 

\textbf{BLEU}, the abbreviation for BiLingual Evaluation Understudy~\cite{2002-BLEU}, is widely used for evaluating the quality of generated code summaries~\cite{2018-Improving-Code-Summarization-via-DRL, 2018-TL-CodeSum, 2016-CODE-NN}. It is a variant of precision metric, which calculates the similarity by computing the n-gram precision of a generated summary to the reference summary, with a penalty for the overly short length~\cite{2002-BLEU}. It is computed as:
\begin{equation}
	BLEU = BP * exp(\sum_{n=1}^N{w_nlogp_n})
	\label{equ:bleu}
\end{equation}

\begin{equation}
	BP = 
	\begin{cases}
	1, &\text{if $|g| > |r|$} \\
	e^{(1 - \frac{|r|}{|g|})}, &\text{if $|g| \leq |r|$}
	\end{cases}
	\label{equ:bp}
\end{equation}
where $N = 1, 2, 3, 4$ and $w_n = \frac{1}{N}$. $p_n$ is the n-gram precision~\cite{2022-Evaluation-Neural-Code-Summarization}. $BP$ represents the brevity penalty. $g$ and $r$ denote a generated (predicted) summary and a reference summary, respectively. $|g|$ and $|r|$ denote the lengths of $g$ and $r$, respectively.\deletion{BLEU typically uses BLEU-1, BLEU-2, BLEU-3, and BLEU-4 (calculated by 1-gram, 2-gram, 3-gram, and 4-gram precision, respectively) to measure the precision.} In this paper, we follow~\cite{2021-SiT, 2022-SCRIPT} and show \deletion{BLEU-4 score, which is often of interest as it can reflect the weighted results from 1 through 4}\revision{the standard BLEU score which provides a cumulative score of 1-, 2-, 3-, and 4-grams}~\cite{2021-Why-My-Code-Summarization-Not-Work}.

\textbf{METEOR}, the abbreviation for Metric for Evaluation of Translation with Explicit ORdering~\cite{2005-METEOR}, is also widely used to evaluate the quality of generated code summaries~\cite{2021-Code-Summarization-for-Smart-Contracts, 2020-Rencos, 2020-RL-Guided-Code-Summarization}. For a pair of summaries, METEOR creates a word alignment between them and calculates the similarity scores. Suppose $m$ is the number of mapped unigrams between the reference summary $r$ and the generated summary $g$, respectively. Then, precision ($P_{unig}$), recall ($P_{unig}$), and METEOR are computed as:

\begin{equation}
	P_{unig} = \frac{m}{|g|},\;\; R_{unig} = \frac{m}{|r|}
	\label{equ:P_unigram}
\end{equation}

\begin{equation}
	METEOR = (1 - \gamma * frag^\beta) * \frac{P_{unig} * R_{unig}}{\alpha * P_{unig} + (1 - \alpha) * R_{unig}}
	\label{equ:meteor}
\end{equation}
where $frag$ is the fragmentation fraction. As in~\cite{2020-Rencos}, $\alpha$, $\beta$, and $\gamma$ are three penalty parameters whose default values are 0.9, 3.0 and 0.5, respectively.

\textbf{ROUGE-L.} ROUGE is the abbreviation for Recall-oriented Understudy for Gisting Evaluation~\cite{2004-ROUGE}. ROUGE-L, a variant of ROUGE, is computed based on the longest common subsequence
(LCS). ROUGE-L is also widely used to evaluate the quality of generated code summaries~\cite{2021-Project-Level-Encoding-Code-Summarization, 2021-BASTS, 2021-API2Com}. Specifically, the LCS-based F-measure ($F_{lcs}$) is called ROUGE-L~\cite{2004-ROUGE}, and $F_{lcs}$ is computed as:
\begin{equation}
	R_{lcs} = \frac{LCS(r,g)}{|r|}, \;\; P_{lcs} = \frac{LCS(r,g)}{|g|}
	\label{equ:p_r_lcs}
\end{equation}

\begin{equation}
	F_{lcs} = \frac{(1+\beta^2)R_{lcs}P_{lcs}}{R_{lcs}+\beta^2P_{lcs}}
	\label{equ_f-lcs}
\end{equation}
where $r$ and $g$ also denote the reference summary and the generated summary, respectively. Notice that ROUGE-L is 1 when $g = r$; while ROUGE-L is 0 when $LCS(r,g) = 0$, i.e., which means $r$ and $g$ are completely different. $\beta$ is set to 1.2 as in~\cite{2021-CoCoSum, 2018-Improving-Code-Summarization-via-DRL, 2020-Rencos}.

The scores of BLEU, ROUGE-L and METEOR are in the range of [0,1] and usually reported in percentages. The higher the scores, the closer the generated summary is to the reference summary, and the better the code summarization performance. 
\revise{All scores are computed by the same implementation provided by~\cite{2020-Rencos}.
}

\begin{table*}[htbp]
    \small
    \centering
    \renewcommand{\arraystretch}{1.2}
    \caption{Parameter settings}
    \label{tab:parameter_setting}
    
    \begin{tabular}{|c|c|c|c|c|}
    \hline
    Model & mini-batch size & word embedding size & learning rate & dropout \\
    \hline
    LSTM & 256 & 256 & 0.2 & 0.2 \\
    \hline
    
    Transformer & 32 & 512 & 5e-5 & 0.1 \\
    \hline
    
    CodeBERT & 32 & 512 & 5e-5 & 0.1\\
    \hline
    
    CodeT5 & 32 & 512 & 5e-5 & 0.1\\
    \hline
 \end{tabular}
\end{table*}

\subsubsection{Experimental Settings}
\label{subsubsec:experimental_settings} \deletion{To train models, we first shuffle the training data and set the mini-batch size to 32. 
For each batch, the code snippets are padded with a special token $\langle PAD \rangle$ to the maximum length. We set the word embedding size to 512. For LSTM unit, we set the hidden size to 512. We update the parameters via AdamW optimizer~\cite{2015-Adam} with the learning rate 0.0003. To prevent over-fitting, we use dropout with 0.1.}
\revision{
Our {\toolname} is a general framework. In this paper, we try to combine {\toolname} with different neural networks (e.g., LSTM and Transformer) and pre-trained models of code (CodeBERT and CodeT5). Therefore, during the experiments, when {\toolname} is combined with different network architectures/pre-trained models, our experimental settings are different. As we all know, neural network hyperparameter setting is a complex task, not only to consider the model size and data size, but also very dependent on the developer's experience. Therefore, the common practice of hyperparameter setting is to follow the successful experience of existing work. Of course, this paper is no exception. For example, when combining {\toolname} with CodeT5, we follow CodeT5~\cite{2021-CodeT5} and set the mini-batch size to 32, the word embedding size to 512, the learning rate to 5e-5, and the dropout to 0.1, and update the parameters via AdamW optimizer~\cite{2015-Adam}. The code snippets are padded with a special token $\langle PAD \rangle$ to the maximum length. Table~\ref{tab:parameter_setting} presents the settings of several crucial parameters when {\toolname} is combined with LSTM, Transformer, CodeBERT, and CodeT5. We refer to~\cite{2020-R2Com} to set LSTM, and refer to~\cite{2020-Transformer-based-Approach-for-Code-Summarization} to set Transformer.} All models are implemented using the PyTorch 1.7.1 framework with Python 3.8. 
All experiments are conducted on a server equipped with one Nvidia Tesla V100 GPU with 31 GB memory, running on Centos 7.7. 
All the models in this paper are trained for the same epochs as their original paper, and we select the best model based on the lowest validation loss.

\subsection{Experimental Results}
\label{subsec:experimental_results}

\subsubsection{\textbf{RQ1:} {\toolname} vs. Baselines}
\label{subsubsec:result_of_RQ1}
\
\newline
\indent1)\textit{\;Baselines:} To answer this research question, we compare our approach {\toolname} to the following DL-based code summarization techniques.

\textbf{CODE-NN~\cite{2016-CODE-NN}} adopts a LSTM-based encoder-decoder architecture with attention mechanism. It is a classical encoder-decoder framework in NMT that encodes tokens of code snippets into embedding representations and then generates summaries in the decoder with the attention mechanism.

\textbf{DeepCom~\cite{2018-DeepCom}} also adopts a LSTM-based encoder-decoder architecture with attention mechanism. In addition, to capture the structural information, DeepCom proposes a structure-based traversal method to traverse AST sequences of the code snippet. The AST sequences are further passed to the encoder and decoder to generate summaries. 

\textbf{Hybrid-DRL~\cite{2018-Improving-Code-Summarization-via-DRL} (also shortened to RL+Hybrid2Seq in~\cite{2020-Transformer-based-Approach-for-Code-Summarization}, Hybrid2Seq in~\cite{2021-SiT})} also adopts a LSTM-based encoder-decoder architecture and is trained with reinforcement learning. It also designs additional encoder based on an AST-based LSTM to capture the structural information of the code snippet. It uses reinforcement learning to solve the exposure bias problem during decoding, which obtains better performance.

\textbf{TL-CodeSum~\cite{2018-TL-CodeSum} (also shortened to API+Code in~\cite{2021-SiT})} adopts a GRU-based encoder-decoder architecture with attention mechanism. It encodes the Application Programming Interface (API) sequence along with the code token sequence, and then generates a summary from the source code with transferred API knowledge. It introduces an API sequence summarization task, aiming to train an API sequence encoder by using an external dataset so that it can learn more abundant representations of the code snippet.  

\textbf{Dual Model~\cite{2019-Code-Generation-Summarization}} also adopts a LSTM-based encoder-decoder architecture with an attention mechanism. It treats code summarization and code generation as a dual task. It trains the two tasks jointly by a dual training framework to simultaneously improve the performance of code summarization and code generation tasks.

\textbf{Transformer-based~\cite{2020-Transformer-based-Approach-for-Code-Summarization} (also shortened to Transformer in~\cite{2021-SiT}, NCS in~\cite{2021-Neural-Code-Summarization-How-Far})} adopts a Transformer-based encoder-decoder architecture. It incorporates the copying mechanism~\cite{2017-Get-To-Point} in the Transformer to allow both generating words from vocabulary and copying from the source code.

\textbf{SiT~\cite{2021-SiT}} adopts a Transformer-based encoder-decoder architecture. It proposes structure-induced transformer to capture long-range dependencies and more global information in AST sequences of code snippets. 

\revise{
\textbf{Re2Com~\cite{2020-R2Com}} adopts an LSTM-based encoder-decoder architecture with an attention mechanism. It first uses an information retrieval technique to retrieve a similar code snippet and treat its comment as an exemplar. Then, it uses an LSTM-based seq2seq neural network that takes the given code, its AST, its similar code, and its exemplar as input, and leverages the information from the exemplar to generate summaries. 

\textbf{SCRIPT~\cite{2022-SCRIPT}} adopts a Transformer-based encoder-decoder architecture. It proposes two types of Transformer encoders to capture the structural relative positions between tokens for better learning code semantics.

\textbf{CAST~\cite{2021-CAST}} hierarchically splits an AST into a set of subtrees and devises a recursive neural network to encode the subtrees. The embeddings are then aggregated for generating the summary. They adopt Transformer as the backbone of the decoder.
}

\revise{
\textbf{CodeBERT~\cite{2020-CodeBERT}} is a representative pre-trained model for source code. CodeBERT uses the same model architecture as RoBERTa-base~\cite{2019-RoBERTa}. It is trained with the Masked Language Modeling (MLM) task and the Replaced Token Detection (RTD) task. The authors of CodeBERT fine-tune and test it on the code summarization task (also called the code documentation generation task in their paper). 

\textbf{CodeT5~\cite{2021-CodeT5}} is the state-of-the-art pre-trained model for source code. CodeT5 builds on an encoder-decoder framework with the same architecture as T5~\cite{2020-T5}. It is trained with four pre-training tasks, including Masked Span Prediction (MSP) task, Identifier Tagging (IT), Masked Identifier Prediction (MIP), and Bimodal Dual Generation (BDG). Different from CodeBERT, CodeT5 has a pre-trained decoder. The authors of CodeT5 also conduct experiments on the code summarization task.
}

\revision{It should be noted that, strictly speaking, CodeBERT and CodeT5 are two pre-trained models for source code, not code summarization techniques. Both of them can be used for multiple downstream software engineering tasks (such as code search, code clone detection, and code summarization) by fine-tuning on the corresponding downstream task datasets. CodeBERT is essentially a pre-trained encoder, which can be used to transform input code snippets into numerical representations (i.e., embeddings). CodeT5 consists of a pre-trained encoder and a pre-trained decoder. CodeT5’s pre-trained encoder does the same thing as that of CodeBERT. CodeT5’s pre-trained decoder can decode the embeddings produced by the pre-trained encoder into the expected output, such as natural language summaries in the code summarization task.} \revise{In this paper, we fine-tune the pre-trained CodeBERT and CodeT5 on the code summarization task in the three datasets, i.e., JCSD, PCSD, and CodeXGLUE.
}

\revise{In addition to the abstractive methods introduced above, we implement the following two extractive methods as baselines.

\textbf{TR-based~\cite{2010-Program-Comprehension-with-Code-Summarization}} is an extractive method based on text retrieval (TR). As described in~\cite{2010-Program-Comprehension-with-Code-Summarization, 2010-Automated-Text-Summarization-Summarizing-Code}, TR-based methods implement the generation of summaries through the following two processes:
(1) Extract the text from the code snippet and convert it into a corpus.
(2) Determine the most relevant terms for the given code snippet in the corpus and include them in the summary.
In process (2), various TR techniques can be integrated, such as Vector Space Model (VSM)~\cite{1975-Vector-Space-Model}, Latent Semantic Analysis (LSA)~\cite{1990-Latent-Semantic-Analysis, 2008-LSA-Extractive-Summarization}, Hierarchical Pachinko Allocation Model (hPAM)~\cite{2007-Mixtures-of-Hierarchical-Topics}, to generate code summaries~\cite{2013-Evaluating-Source-Code-Summarization}. 
Since the implementation code for early TR-based methods is no longer available, in this paper, we follow~\cite{2010-Program-Comprehension-with-Code-Summarization, 2010-Automated-Text-Summarization-Summarizing-Code} and reproduce a LSA-based method (LSA-based, for short). The LSA-generated summary can contain terms that do not appear in the summarized code snippet but that appear somewhere else in the corpus~\cite{2013-Evaluating-Source-Code-Summarization}.

\textbf{Extractor-based} (Ex-based, for short) is an extractive method based on our extractor. The design of the extractor is borrowed from extractive text summarization~\cite{2018-Unified-Model-for-Extractive-Abstractive-Summarization}. It should be noted that existing extractors cannot be directly used to summarize code as the extracted important texts essentially are still code. We hence try to generate summaries by directly feeding the important texts extracted by the extractor to CodeT5. In other words, the extractor-based method fine-tunes the pre-trained CodeT5 with only the extracted important statements.
}

\textbf{{\toolname}.} \deletion{In the extractor component, the encoder is built on the pre-trained encoder provided by CodeT5 and followed by a fully connected neural network as the classification layer. In the abstracter, the components ExEncoder and AbEncoder are also built on the same pre-trained encoder as the extractor, and the component Decoder is built on the pre-trained decoder provided by CodeT5.}\revision{As described in Section~\ref{sec:methodology}, our {\toolname} has two core components, i.e., an extractor and an abstracter. The extractor consists of an encoder and a classification layer. The abstracter consists of two encoders (i.e., ExEncoder and AbEncoder) and a decoder. In total, {\toolname} has three encoders and one decoder. When combining {\toolname} with CodeBERT, we build the three encoders on the pre-trained encoder provided by CodeBERT, and build the decoder on the Transformer architecture. When combining {\toolname} with CodeT5, we build the three encoders and the decoder on the pre-trained encoder and decoder provided by CodeT5, respectively.} Compared to fine-tuning \revision{CodeBERT and} CodeT5 directly, although {\toolname} needs to train an additional extractor, the training of the extractor is a one-time offline task. In addition, the abstracter of {\toolname} has two encoders that are used to transform the important statements and the entire code snippet into embeddings in parallel.

\indent2)\textit{\;Results:} 
\revision{
According to the description in baselines’ papers, except for CodeBERT and CodeT5 evaluated on the CodeXGLUE dataset, the rest of the baselines (e.g., Transformer-based, SiT, and SCRIPT) are mainly evaluated on the JCSD and PCSD datasets. Therefore, for a fair and convenient comparison, we are consistent with most baselines and compare our {\toolname} with them on the JCSD and PCSD datasets. In addition, several important baselines (including SiT and SCRIPT) require special data preprocessing, which can not be applied to some programming languages in the CodeXGLUE dataset (e.g., Go, PHP, and Ruby). Therefore, in this section, we only present experimental results on the JCSD and PCSD datasets. We compare Transformer-based, CodeBERT, and CodeT5 on the CodeXGLUE dataset and the results are discussed in Section~\ref{subsubsec:results_of_combine_with_diff_models}.
}

\begin{table*}[htbp]
  \scriptsize
  \tabcolsep=1.5pt
  \renewcommand{\arraystretch}{1.2}
  
  \caption{Performance of Our {\toolname} and Baselines. The results in rows 3--8 are reported from Transformer-based~\cite{2020-Transformer-based-Approach-for-Code-Summarization}. 
  \revise{
  The results in rows 9--10 are reported from SiT~\cite{2021-SiT}. $^{\dagger}$ refers to the method that SiT rerun. The results in row 11 is reported from CAST~\cite{2021-CAST}, which currently only supports the Java programming language. $^{\ddag}$ refers to methods we rerun.}
  \delete{Note that we only rerun SiT since it is much stronger than the other baselines.}\revision{$\mathcal{A}$: average value $\uparrow$; $\mathcal{M}$: median $\uparrow$; $\mathcal{S}$: standard deviation $\downarrow$. $\uparrow$ denotes bigger is better, and $\downarrow$ denotes smaller is better.}
  }
  \label{tab:comparison_with_baselines}
  \centering
  \begin{tabular}{|l|ccc|ccc|ccc|ccc|ccc|ccc|}
    \hline
    \multirow{3}{*}{Methods (Year)} & \multicolumn{9}{c|}{JCSD} & \multicolumn{9}{c|}{PCSD}\\
    \cline{2-19}
    & \multicolumn{3}{c|}{\deletion{BLEU-4}\revision{BLEU}} & \multicolumn{3}{c|}{METEOR} & \multicolumn{3}{c|}{ROUGE-L} & \multicolumn{3}{c|}{\deletion{BLEU-4}\revision{BLEU}} & \multicolumn{3}{c|}{METEOR} & \multicolumn{3}{c|}{ROUGE-L}\\
    \cline{2-19}
    & $\mathcal{A}$ & $\mathcal{M}$ & $\mathcal{S}$ & $\mathcal{A}$ & $\mathcal{M}$ & $\mathcal{S}$ & $\mathcal{A}$ & $\mathcal{M}$ & $\mathcal{S}$ & $\mathcal{A}$ & $\mathcal{M}$ & $\mathcal{S}$ & $\mathcal{A}$ & $\mathcal{M}$ & $\mathcal{S}$ & $\mathcal{A}$ & $\mathcal{M}$ & $\mathcal{S}$ \\ 
    \hline
    CODE-NN (2016) & 27.60 & -- & -- & 12.61 & -- & -- & 41.10 & -- & -- & 17.36 & -- & -- & 9.29 & -- & --& 37.81 & -- & --\\ 
    DeepCom (2018) & 39.75 & -- & -- & 23.06 & -- & -- & 52.67 & -- & -- & 20.78 & -- & -- & 9.98 & -- & -- & 37.35 & -- & --\\
    Hybrid-DRL (2018) & 38.22 & -- & -- & 22.75 & -- & -- & 51.91 & -- & -- & 19.28 & -- & -- & 9.75 & -- & -- & 39.34 & -- & --\\
    TL-CodeSum (2018) & 41.31 & -- & -- & 23.73 & -- & -- & 52.25 & -- & -- & 15.36 & -- & -- & 8.57 & -- & -- & 33.65 & -- & --\\
    Dual Model (2019) & 42.39 & -- & -- & 25.77 & -- & -- & 53.61 & -- & -- & 21.80 & -- & -- & 11.14 & -- & -- & 39.45 & -- & --\\
    Transformer-based (2020) & 44.58 & -- & -- & 26.43 & -- & -- & 54.76 & -- & -- & 32.52 & -- & -- & 19.77 & -- & -- & 46.73 & -- & -- \\
    Transformer-based$^{\dagger}$(2020) & 44.87 & -- & -- & 26.58 & -- & -- & 54.95 & -- & -- & 32.85 & -- & -- & 19.86 & -- & -- & 46.93 & -- & --\\
    SiT (2021) & \textbf{45.76} & -- & -- & \textbf{27.58} & -- & -- & \textbf{55.58} & -- & -- & \textbf{34.11} & -- & -- & \textbf{21.11} & -- & -- & \textbf{48.35} & -- & --\\
    CAST (2021) & 45.19 & -- & -- & 27.88 & -- & -- & 55.08 & -- & -- & -- & -- & -- & -- & -- & -- & -- & -- & --\\
    \hline
    \hline
    Re2Com$^{\ddag}$ (2020) & 35.65 & 15.23 & 0.36 & 16.26 & 12.54 & 0.35 & 44.95 & 32.05 & 0.35 & 14.68 & 13.89 & 0.09 & 6.43 & 4.13 & 0.05 & 25.16 & 20.96 & 0.12\\
    SiT$^{\ddag}$ (2021) & 45.22 & \revision{22.28} & \revision{0.41} & 27.10 & \revision{21.46} & \revision{0.46} & 55.38 & \revision{45.66} & \revision{0.37} & 33.75 & \revision{17.25} & \revision{0.32} & 21.02 & \revision{18.42} & \revision{0.41} & 48.33 & \revision{35.36} & \revision{0.31} \\
    
    SCRIPT$^{\ddag}$ (2022) & \textbf{46.41} & \revision{23.67} & \revision{0.42} & \textbf{28.47} & \revision{23.51} & \revision{0.41} & 56.57 & \revision{48.16} & \revision{0.37} & 33.52 & \revision{19.07} & \revision{0.32} & 20.80 & \revision{15.86} & \revision{0.32} & 48.09 & \revision{38.61} & \revision{0.30} \\
    
    CodeBERT$^{\ddag}$ (2020) & 43.23 & 21.96 & 0.40 & 26.13 & 21.09 & 0.31 & 54.74 & 45.57 & 0.33 & 32.65 & 18.30 & 0.32 & 20.55 & 17.50 & 0.36 & 48.45 & 39.87 & 0.32 \\
    
    CodeT5$^{\ddag}$ (2021) & 46.08 & \revision{\textbf{25.69}} & \revision{0.40} & 27.93 & \revision{\textbf{25.40}} & \revision{0.51} & \textbf{57.28} & \revision{\textbf{50.73}} & \revision{0.35} & \textbf{34.39} & \revision{\textbf{19.15}} & \revision{0.33} & \textbf{22.66} & \revision{\textbf{19.41}} & \revision{0.43} & \textbf{49.90} & \revision{\textbf{42.55}} & \revision{0.29} \\
    
    \hline
    
    TR-based$^{\ddag}$ & 5.59 & \revision{4.37} & \revision{0.04} & 5.24 & \revision{5.61} & \revision{0.06} & 7.58 & \revision{7.38} & \revision{0.07} & 6.95 & \revision{8.97} & \revision{0.06} & 7.12 & \revision{6.84} & \revision{0.06} & 8.04 & \revision{8.94} & \revision{0.08} \\
    
    Ex-based$^{\ddag}$ & \textbf{41.00} & \revision{19.90} & \revision{0.39} & \textbf{25.41} & \revision{21.44} & \revision{0.44} & \textbf{53.10} & \revision{22.24} & \revision{0.35} & \textbf{32.15} & \revision{19.07} & \revision{0.30} & \textbf{21.30} & \revision{18.42} & \revision{0.41} & \textbf{48.45} & \revision{40.76} & \revision{0.28} \\
    
    \hline
    \hline
    {\toolname} & \textbf{47.66} & \revision{\textbf{25.99}} & \revision{0.40} & \textbf{30.39} & \revision{\textbf{26.54}} & \revision{\textbf{0.52}} & \textbf{58.77} & \revision{\textbf{52.16}} & \revision{0.35} & \textbf{35.96} & \revision{\textbf{21.79}} & \revision{0.30} & \textbf{23.70} & \revision{\textbf{22.86}} & \revision{\textbf{0.45}} & \textbf{51.83} & \revision{\textbf{44.66}} & \revision{0.28} \\
    
    \hline
 \end{tabular}
\end{table*}

\begin{figure}[htbp]
\centering
    \subfigure[\deletion{BLEU-4}\revision{BLEU} Scores on JCSD]
    {
        \includegraphics[width=0.31\linewidth]{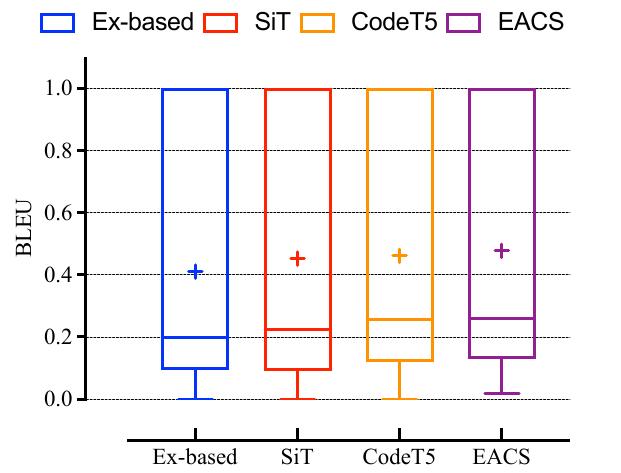}
        \label{fig:compare_BLEU_distribution_JCSD}
    }
    \subfigure[METEOR Scores on JCSD]
    {
        \includegraphics[width=0.31\linewidth]{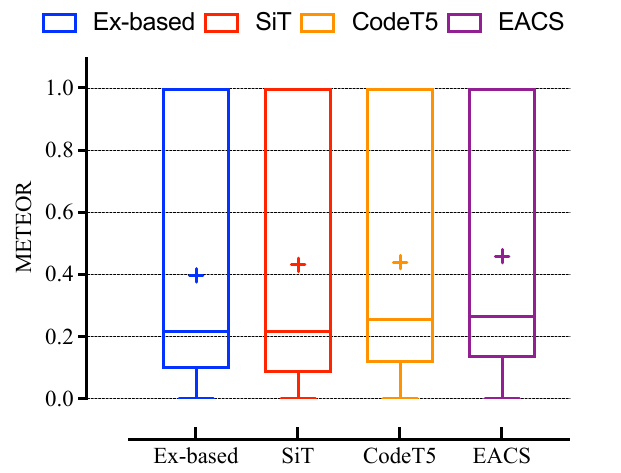}
        \label{fig:compare_METEOR_distribution_JCSD}
    }
    \subfigure[ROUGE-L Scores on JCSD]
    {
        \includegraphics[width=0.31\linewidth]{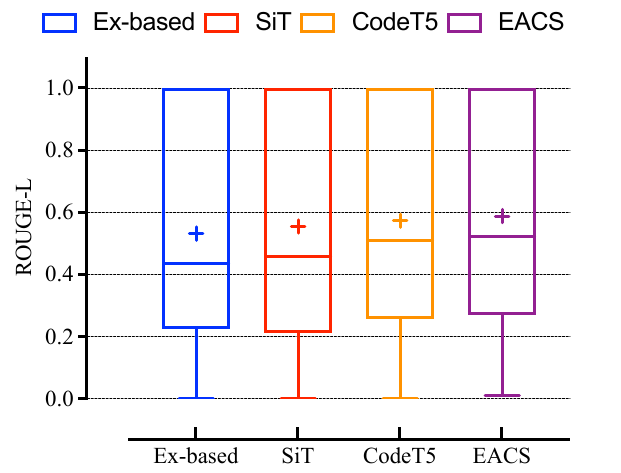}
        \label{fig:compare_ROUGE-L_distribution_JCSD}
    }
    \subfigure[\deletion{BLEU-4}\revision{BLEU} Scores on PCSD]
    {
        \includegraphics[width=0.31\linewidth]{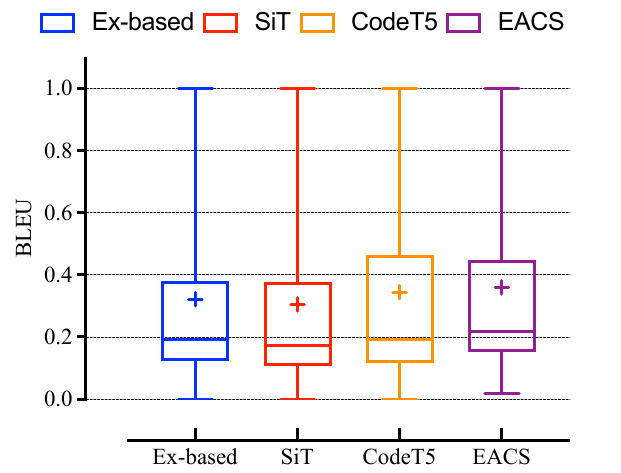}
        \label{fig:compare_BLEU_distribution_PCSD}
    }
    \subfigure[METEOR Scores on PCSD]
    {
        \includegraphics[width=0.31\linewidth]{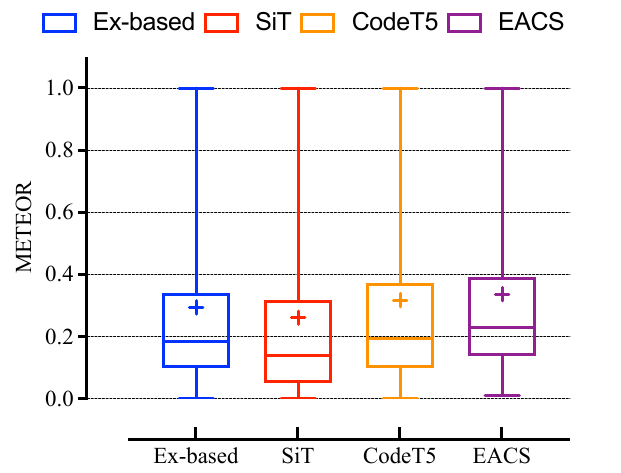}
        \label{fig:compare_METEOR_distribution_PCSD}
    }
    \subfigure[ROUGE-L Scores on PCSD]
    {
        \includegraphics[width=0.31\linewidth]{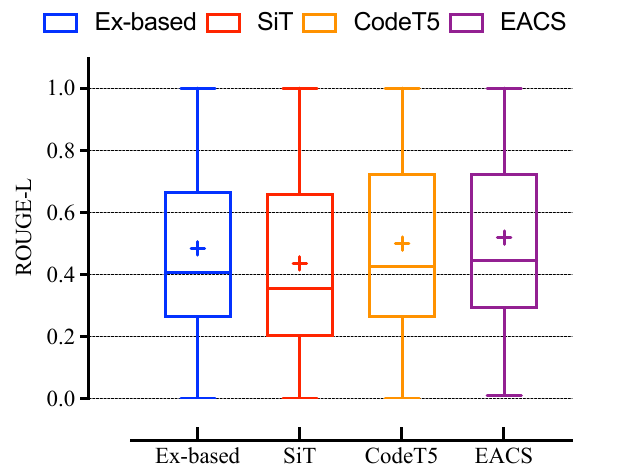}
        \label{fig:compare_ROUGE-L_distribution_PCSD}
    }
    \caption{Distributions of Metrics' Scores on the JCSD and PCSD datasets}
    \label{fig:compare_distribution}
\end{figure}

Table~\ref{tab:comparison_with_baselines} shows the performance of our {\toolname} and baselines in terms of the three evaluation metrics, i.e., \deletion{BLEU-4}\revision{BLEU}, METEOR, and ROUGE-L.
\delete{From Table~\ref{tab:comparison_with_baselines}, in all baselines, SiT performs the best on both datasets in terms of all three metrics. However, our {\toolname} is more powerful than SiT and achieves more impressive performance.} 
\revise{From rows 9--11 of Table~\ref{tab:comparison_with_baselines}, we can observe that SiT outperforms Transformer-based and earlier baselines on both datasets in terms of all three metrics. Therefore, we rerun the best baseline SiT. Rows 12--16 show the results of the five baselines we rerun. We can observe that SCRIPT slightly outperforms SiT and CodeT5 on the JCSD dataset in terms of \deletion{BLEU-4}\revision{BLEU} and METEOR, while CodeT5 performs best on the PCSD dataset, followed by SiT and SCRIPT. Overall, SiT is comparable to SCRIPT\revision{,} and CodeT5 \deletion{and}\revision{is} better than SiT and SCRIPT. It should be noted that we rerun SCRIPT on its preprocessed JCSD and PCSD datasets, which differ from the data splits provided by~\cite{2018-TL-CodeSum} and ~\cite{2018-Improving-Code-Summarization-via-DRL}. Therefore, in the rest of this paper, we mainly compare {\toolname} with SiT and CodeT5.
Rows 17--18 show the results of the two extractive methods, i.e., TR-based and Ex-based. We can observe that the TR-based extractive method is significantly worse than the extractor-based extractive method.
The last row of Table~\ref{tab:comparison_with_baselines} shows the results of our {\toolname}.} On the JCSD dataset, compared with the the-state-of-the-art (CodeT5), {\toolname} improves by 3.56\% in \deletion{BLEU-4}\revision{BLEU}, 8.81\% in METEOR, and 2.60\% in ROUGE-L. 
\revision{In addition, {\toolname} has a higher median score than CodeT5 on each metric. For example, in ROUGE-L, {\toolname} gets a median score of 52.16, while CodeT5 gets 50.73. In terms of standard deviation, {\toolname} is comparable to CodeT5 with a mean of 0.42. }
On the PCSD dataset, {\toolname} also outperforms CodeT5, improving by 4.57\% in \deletion{BLEU-4}\revision{BLEU}, 4.59\% in METEOR, and 3.87\% in ROUGE-L. \revision{{\toolname} also has a higher median score than CodeT5 on each metric. For example, in METEOR, {\toolname} gets a median score of 22.86 (19.41 for CodeT5). In terms of standard deviation, {\toolname} obtains a mean of 0.34, better than CodeT5 (0.35).} 
\revision{Based on the above observations, we can conclude that our {\toolname} consistently outperforms CodeT5 in terms of three metrics across both datasets.} 

It should be noted that the values in Table~\ref{tab:comparison_with_baselines} are the average scores of all test samples. For a more comprehensive comparison, we further \delete{statistic}\revise{analyze} the distribution of the scores of \revise{Ex-based, SiT, CodeT5, and} {\toolname} on all test samples, and the statistical results are shown in Figure~\ref{fig:compare_distribution}. In Figure~\ref{fig:compare_distribution}, `+' denotes the mean, which is the value filled in Table~\ref{tab:comparison_with_baselines}. 
\revision{
Figure~\ref{fig:compare_distribution}(a)-(c) show the box-and-whiskers plots of the BLEU, METEOR, and ROUGE-L scores obtained by the three baselines and {\toolname} on the JCSD dataset. It can be observed that on all three metrics, the median and first quartile associated with {\toolname} are better than those associated with all three baselines. Figure~\ref{fig:compare_distribution}(d)-(f) show the box-and-whiskers plots of the scores by the three baselines and {\toolname} on the PCSD dataset. From left to right, the figure shows: (1) in BLEU, the median and first quartile associated with {\toolname} are better than those associated with all three baselines; (2) in METEOR, the median, first quartile, and third quartile associated with {\toolname} are better than those associated with all three baselines; and (3) in ROUGE-L, the median and first quartile associated with {\toolname} are better than those associated with all three baselines.} 
Overall, the metric score distribution of {\toolname} is better than that of Ex-based, SiT, and CodeT5.
\deletion{To test whether there is a statistically significant difference between the two methods, we perform the paired Wilcoxon-Mann-Whitney signed-rank test at a significance level of 5\%, following previously reported guidelines for inferential statistical analysis involving randomized algorithms~\cite{2014-Hitchhiker-Guide-Statistical-Tests, 2015-Practical-Test-Selection}. In Figure~\ref{fig:compare_distribution}, `*' ($0.01 < p < 0.05$), `**' ($0.001 < p < 0.01$), `***' ($0.0001 < p < 0.001$), and `****' ($p < 0.0001$) represent the differences between the two groups are Significant, Very significant, Extremely significant, and Extremely significant, respectively. `ns' ($p \geq 0.05$) means Not significant. From the figure, we can intuitively observe that in all three metrics, {\toolname} outperforms CodeT5 on the JCSD and PCSD datasets.} 

\revision{
We further run a statistical test to assess whether there is enough empirical evidence to claim that there is a difference among the four techniques regarding the three metrics. The statistical test that must be followed depends on the properties of the data~\cite{2012-Experimentation-in-SE}. Since our data does not follow a normal distribution in general, our analysis requires the use of non-parametric test methods. In addition, to provide an answer to the question under study (``Which of the techniques gives the best performance?''), each technique should be individually compared against all other alternatives~\cite{2019-Bug-Localization-in-Models}. We need to perform a pair-wise comparison among the results of each technique, providing a $p$-value that determines whether statistically significant differences exist.

Specifically, we use the statistical tool GraphPad Prism~\cite{1995-GraphPad-Prism} for non-parametric tests. With an experimental design where each row represents matched data and the data do not follow a Gaussian distribution, Prism performs the Friedman test~\cite{1995-Friedman-Test} and Dunn's multiple comparisons test~\cite{1995-Multiple-comparisons-Tests} at a significance level of 5\% by default. 
The outcome of the Friedman test tells if there are differences among the groups, but does not tell which groups are different from other groups. In order to determine which groups are different from others, the post-hoc test can be conducted. Probably the most common post-hoc test for the Friedman test is the Dunn test~\cite{1964-Dunn-Test}. Therefore, we look at post-test results (reported by Dunn's multiple comparisons test) to see which techniques differ from which other techniques.

Table~\ref{tab:compare_baseline_in_p-value_effect_size} shows the $p$-values reported by Prism after performing the Friedman test and Dunn's multiple comparisons test among Ex-based, SiT, CodeT5, and our {\toolname}. From rows 9, 11, and 13 of Table~\ref{tab:compare_baseline_in_p-value_effect_size}, it is observed that, in all three metrics (i.e., BLEU, METEOR, and ROUGE-L), all $p$-values between {\toolname} and baselines are less than 0.0001 and smaller than the significant threshold value of 0.05. Based on this, combining the results in Table~\ref{tab:comparison_with_baselines} and Figure~\ref{fig:compare_distribution}, we can conclude that the performance of {\toolname} is significantly better than the three baselines.

Further, to assess the magnitude of the improvement of one technique (e.g., {\toolname}) relative to another (e.g., Ex-based), we calculate the effect size for each comparison group (e.g., {\toolname} vs. Ex-based). Dunn's test is a normal approximation to the exact rank sum test statistics. Therefore, according to~\cite{2012-Effect-Size-Estimates, 2014-Overview-Measures-of-Effect-Size}, we use correlation coefficients to estimate the effect size of each comparison group, which is computed as $r= z/\sqrt{n}$. $z$ is $Z$-score and $n$ is the number of observations in all groups~\cite{2012-Effect-Size-Estimates}. Rows 4, 6, 8, 10, 12, and 14 of TABLE~\ref{tab:compare_baseline_in_p-value_effect_size} show the effect size of each comparison group. From rows 10, 12, and 14, in terms of BLEU, the effect size values of the comparison groups {\toolname} vs. Ex-based, {\toolname} vs. SiT, and {\toolname} vs. CodeT5 are 0.38, 0.21, and 0.11, respectively. It means that {\toolname} has a relatively large improvement over Ex-based on BLEU, and a relatively medium improvement over SiT and CodeT5. In terms of METEOR, {\toolname} has a relatively medium improvement over Ex-based and SiT, and a relatively small improvement over CodeT5. In terms of ROUGE-L, {\toolname} has a relatively medium improvement over all three baselines. Similarly, the effect size values of the comparison groups {\toolname} vs. Ex-based, {\toolname} vs. SiT, and {\toolname} vs. CodeT5 on the PCSD dataset also show a superiority of {\toolname}. 
}

In summary, the results and observations above demonstrate that under all experimental settings, our {\toolname} consistently achieves higher performance in all three metrics, which indicates better code summarization performance.

\begin{table*}[!t]
    \footnotesize
    \tabcolsep=2.5pt
    \centering
    \renewcommand{\arraystretch}{1.1} 
    \caption{\revision{$p$-values and effect sizes for metrics used in the automatic evaluation. Effect size: $<0.1$: small effect; $0.1-<0.3$: medium effect; and $\geq 0.3$: large effect.}}
    \label{tab:compare_baseline_in_p-value_effect_size}
  
  \begin{tabular}{|l|c|ccc|ccc|}
    \hline
     \revision{\multirow{2}{*}{Comparison Group}} & \revision{\multirow{2}{*}{Result}} & \multicolumn{3}{c|}{\revision{JCSD}} & \multicolumn{3}{c|}{\revision{PCSD}}\\
     \cline{3-8}
     & & \revision{BLEU} & \revision{METEOR} & \revision{ROUGE-L} & \revision{BLEU} & \revision{METEOR} & \revision{ROUGE-L} \\
    \hline
    \revision{\multirow{2}{*}{SiT vs. Ex-based}} & \revision{$p$-value} & \revision{$<0.0001$} & \revision{$<0.0001$} & \revision{$>0.9999$} & \revision{$<0.0001$} & \revision{$<0.0001$} & \revision{$<0.0001$} \\
    & \revision{effect size} & \revision{0.17} & \revision{0.05} & \revision{0.00} & \revision{0.06} & \revision{0.18} & \revision{0.14} \\
    \hline
    \revision{\multirow{2}{*}{CodeT5 vs. Ex-based}} & \revision{$p$-value} & \revision{$<0.0001$} & \revision{$<0.0001$} & \revision{$<0.0001$} & \revision{$<0.0001$} & \revision{$<0.0001$} & \revision{$<0.0001$} \\
    & \revision{effect size} & \revision{0.27} & \revision{0.16} & \revision{0.15} & \revision{0.06} & \revision{0.06} & \revision{0.05} \\
    \hline
    \revision{\multirow{2}{*}{CodeT5 vs. SiT}} & \revision{$p$-value} & \revision{$<0.0001$} & \revision{$<0.0001$} & \revision{$<0.0001$} & \revision{$>0.9999$} & \revision{$<0.0001$} & \revision{$<0.0001$} \\
    & \revision{effect size} & \revision{0.10} & \revision{0.20} & \revision{0.15} & \revision{0.00} & \revision{0.12} & \revision{0.08} \\
    \hline
    \revision{\multirow{2}{*}{{\toolname} vs. Ex-based}} & \revision{$p$-value} & \revision{$<0.0001$} & \revision{$<0.0001$} & \revision{$<0.0001$} & \revision{$<0.0001$} & \revision{$<0.0001$} & \revision{$<0.0001$} \\
    & \revision{effect size} & \revision{0.38} & \revision{0.21} & \revision{0.26} & \revision{0.29} & \revision{0.21} & \revision{0.13} \\
    \hline
    \revision{\multirow{2}{*}{{\toolname} vs. SiT}} & \revision{$p$-value} & \revision{$<0.0001$} & \revision{$<0.0001$} & \revision{$<0.0001$} & \revision{$<0.0001$} & \revision{$<0.0001$} & \revision{$<0.0001$} \\
    & \revision{effect size} & \revision{0.21} & \revision{0.26} & \revision{0.26} & \revision{0.35} & \revision{0.38} & \revision{0.26} \\
    \hline
    \revision{\multirow{2}{*}{{\toolname} vs. CodeT5}} & \revision{$p$-value} & \revision{$<0.0001$} & \revision{$<0.0001$} & \revision{$<0.0001$} & \revision{$<0.0001$} & \revision{$<0.0001$} & \revision{$<0.0001$} \\
    & \revision{effect size} & \revision{0.11} & \revision{0.06} & \revision{0.11} & \revision{0.35} & \revision{0.27} & \revision{0.18} \\
    \hline
 \end{tabular}
\end{table*}

\subsubsection{\textbf{RQ2:} 
\deletion{Effectiveness of {\toolname} When Combing with Different Networks Architectures or Pre-trained Models}\revision{Generality of {\toolname}}}
\label{subsubsec:results_of_combine_with_diff_models}
\
\newline
The extractive-and-abstractive framework we proposed is a general code summarization framework\deletion{that does not depend on a specific deep learning network/model}. 
\deletion{In other words, {\toolname} can be combined with many advanced neural network architectures or pre-trained models\revise{ for source code}.}
\revision{We mainly verify the generality of {\toolname} from two aspects: scalability (model-agnostic) and generalizability (language-agnostic). To demonstrate that {\toolname} is model-agnostic, we verify the feasibility of combining {\toolname} with many advanced neural network architectures or code pre-trained models. To demonstrate that {\toolname} is language-agnostic, we evaluate the effectiveness of {\toolname} on a multiple-language dataset.}

In this section, we combine {\toolname} with several popular and advanced neural network architectures (including LSTM~\cite{1997-LSTM} and Transformer~\cite{2017-Transformer}) and pre-trained models (including CodeBERT~\cite{2020-CodeBERT} and CodeT5~\cite{2021-CodeT5}) \deletion{to}\revision{and} explore the potential of {\toolname} in multilingual code summarization tasks\deletion{(i.e., the CodeXGLUE dataset)}. These neural network architectures and pre-trained models are widely used in code summarization~\cite{2018-DeepCom, 2019-Code-Summarization-with-Extended-Tree-LSTM, 2020-Transformer-based-Approach-for-Code-Summarization, 2020-CodeBERT, 2021-SiT, 2021-CodeT5}. 
\revision{
Although intuitively ``{\toolname}+architecture/pre-trained model'' seems always better than architecture/pre-trained model, further experimental verification is still necessary. Only when the experimental results show that ``{\toolname}+architecture/pre-trained model'' is better than architecture/pre-trained model can we conclude that {\toolname} does make an additional contribution to improving code summarization.
}
The experimental results are shown in Table~\ref{tab:results_of_combination_with_architecture_pre-trained-model} \revise{and Table~\ref{tab:results_of_combination_with_architecture_pre-trained-model_1}}. In the two tables, rows 3 (``LSTM'') and 5 (``Transformer'') represent that we implement purely LSTM-based and Transformer-based encoder-decoder frameworks, respectively, to perform the code summarization task. Rows 4 (``{\toolname} + LSTM'') and 6 (``{\toolname} + Transformer'') represent that on the basis of the above (Rows 3 and 5), the well-trained extractor module is added. 
Rows 7 and 9 represent that we fine-tune the pre-trained models released on GitHub (i.e., CodeBERT~\cite{2021-CodeBERT-GitHub} and CodeT5~\cite{2021-CodeT5-GitHub}) on the code summarization task. Analogously, rows 8 (``{\toolname} + CodeBERT'') and 10 (``{\toolname} + CodeT5'') represent that on the basis of the above (Rows 7 and 9), the well-trained extractor module is added. 
\revision{As mentioned in Section~\ref{subsubsec:result_of_RQ1}, CodeBERT and CodeT5 can be used for many downstream software engineering tasks. For the code summarization task, CodeBERT and CodeT5 use the CodeXGLUE dataset for evaluation, so we also use this dataset and keep in line with them. In addition, the CodeXGLUE dataset contains a large number of code-comment pairs across six programming languages (including Go, Java, JavaScript, PHP, Python, and Ruby), which can also be used to evaluate the language-agnostic of {\toolname}.
}

\begin{table*}[htbp]
    \footnotesize
    \tabcolsep=1.5pt
    \renewcommand{\arraystretch}{1.1}
    \caption{Effectiveness of {\toolname} when Combined with Different Neural Network Architectures (LSTM and Transformer) and Pre-trained Models (CodeBERT and CodeT5) on \delete{the CodeSearchNet corpus (Six Programming Languages, including Go, Java, and JavaScript)}\revise{the CodeXGLUE dataset, including Go, Java, and JavaScript.}}
  \label{tab:results_of_combination_with_architecture_pre-trained-model}
  \begin{tabular}{|l|ccc|ccc|ccc|}
    \hline
    \multirow{2}{*}{Methods} &
    \multicolumn{3}{c|}{Go} & \multicolumn{3}{c|}{Java} &
    \multicolumn{3}{c|}{JavaScript} \\
    \cline{2-10}
    & \deletion{BLEU-4}\revision{BLEU} ($\mathcal{B}$) & METEOR ($\mathcal{M}$) & ROUGE-L ($\mathcal{R}$) & $\mathcal{B}$ & $\mathcal{M}$ & $\mathcal{R}$ & $\mathcal{B}$ & $\mathcal{M}$ & \multicolumn{1}{c|}{$\mathcal{R}$} \\
    \hline
    LSTM & 17.8 & 15.1 & 35.6 & 12.2 & 10.1 & 24.6 & 10.4 & 6.2 & \multicolumn{1}{c|}{17.2} \\ 
    {\toolname} + LSTM & \textbf{17.9} & \textbf{15.2} & \textbf{35.8} & \textbf{13.4} & \textbf{10.2} & \textbf{24.7} & \textbf{10.5} & \textbf{6.4} & \multicolumn{1}{c|}{\textbf{17.4}} \\
    \hline
    Transformer & 19.8 & 16.2 & 38.4 & 15.3 & 11.8 & 30.6 & 11.2 & 7.4 & \multicolumn{1}{c|}{20.5} \\
    {\toolname} + Transformer & \textbf{20.1} & \textbf{16.8} & \textbf{39.2} & \textbf{15.8} & \textbf{12.3} & \textbf{31.2} & \textbf{11.5} & \textbf{7.6} & \multicolumn{1}{c|}{\textbf{21.3}}\\
    \hline
    CodeBERT & 21.1 & 17.5 & 43.6 & 18.0 & 12.4 & 35.5 & 13.3 & 8.7 & \multicolumn{1}{c|}{24.3} \\
    {\toolname} + CodeBERT & \textbf{22.2} & \textbf{18.7} & \textbf{44.8} & \textbf{19.3} & \textbf{13.8} & \textbf{36.8} & \textbf{14.3} & \textbf{9.6} & \multicolumn{1}{c|}{\textbf{25.1}} \\
    \hline
    CodeT5 & 21.5 & 18.9 & 45.9 & 20.2 & 15.3 & 39.3 & 15.8 & 11.2 & \multicolumn{1}{c|}{28.9} \\
    {\toolname} + CodeT5 & \textbf{23.9} & \textbf{20.0} & \textbf{47.3} & \textbf{21.3} & \textbf{16.8} & \textbf{41.1} & \textbf{16.7} & \textbf{12.1} & \multicolumn{1}{c|}{\textbf{30.3}}\\
    \hline
 \end{tabular}
\end{table*}

\begin{table*}[!t]
    \footnotesize
    \tabcolsep=1.5pt
    \renewcommand{\arraystretch}{1.1}
    \caption{Effectiveness of {\toolname} when Combined with Different Neural Network Architectures (LSTM and Transformer) and Pre-trained Models (CodeBERT and CodeT5) on \delete{the CodeSearchNet corpus (Six Programming Languages, including PHP, Python, and Ruby)}\revise{the CodeXGLUE dataset, including PHP, Python, and Ruby.}}
  \label{tab:results_of_combination_with_architecture_pre-trained-model_1}
  \begin{tabular}{|l|ccc|ccc|ccc|}
    \hline
    \multirow{2}{*}{Methods} &
    \multicolumn{3}{c|}{PHP} & \multicolumn{3}{c|}{Python} & \multicolumn{3}{c|}{Ruby} \\
    \cline{2-10}
    & \deletion{BLEU-4}\revision{BLEU} ($\mathcal{B}$) & METEOR ($\mathcal{M}$) & ROUGE-L ($\mathcal{R}$) & $\mathcal{B}$ & $\mathcal{M}$ & $\mathcal{R}$ & $\mathcal{B}$ & $\mathcal{M}$ & \multicolumn{1}{c|}{$\mathcal{R}$} \\
    \hline
    LSTM & 19.5 & 12.2 & 29.8 & 13.9 & 9.1 & 23.3 & 9.4 & 5.3 & \multicolumn{1}{c|}{16.3}\\ 
    {\toolname} + LSTM &  \textbf{19.6} & \textbf{12.4} & \textbf{30.0} & \textbf{14.1} & \textbf{9.2} & \textbf{23.4} & \textbf{9.9} & \textbf{5.3} & \multicolumn{1}{c|}{\textbf{16.5}}\\
    \hline
    Transformer & 21.5 & 13.9 & 34.2 & 15.8 & 10.6 & 31.3 & 10.3 & 6.4 & \multicolumn{1}{c|}{18.3}\\
    {\toolname} + Transformer & \textbf{22.3} & \textbf{14.1} & \textbf{34.8} & \textbf{16.3} & \textbf{10.9} & \textbf{32.1} & \textbf{10.7} & \textbf{6.7} & \multicolumn{1}{c|}{\textbf{18.8}}\\
    \hline
    CodeBERT & 24.6 & 15.3 & 39.4 & 18.7 & 12.4 & 34.8 & 11.2 & 7.1 & \multicolumn{1}{c|}{20.6}\\
    {\toolname} + CodeBERT &  \textbf{25.4} & \textbf{16.1} & \textbf{40.2} & \textbf{19.4} & \textbf{13.2} & \textbf{35.8} & \textbf{12.4} & \textbf{8.2} & \multicolumn{1}{c|}{\textbf{22.6}}\\
    \hline
    CodeT5 & 25.9 & 18.2 & 43.6 & 20.0 & 15.1 & 37.8 & 14.9 & 10.8 & \multicolumn{1}{c|}{27.9}\\
    {\toolname} + CodeT5 &  \textbf{26.9} & \textbf{18.8} & \textbf{44.2} & \textbf{20.5} & \textbf{16.0} & \textbf{38.9} & \textbf{15.2} & \textbf{11.8} & \multicolumn{1}{c|}{\textbf{29.2}}\\
    \hline
 \end{tabular}
\end{table*}

From rows 3--6 of \delete{table} \revise{both tables}, we can observe that the performance of the combination of {\toolname} and architecture is significantly better than that of only architecture-based methods in terms of all three metrics. Among them, the combination of {\toolname} and Transformer (i.e., ``{\toolname} + Transformer'') performs the best. 
In addition, from rows 7--10, we can observe that the performance of the combination of {\toolname} and pre-trained models is significantly better than that of purely pre-trained model-based methods in terms of all three metrics. Among them, the ``{\toolname} + CodeT5'' performs the best. 
We can also observe that the combinations of {\toolname} with pre-trained models significantly outperform the combinations with neural network architectures. This is mainly because pre-trained models for code representation are usually trained on larger-scale datasets, so they have stronger code representation capabilities and can represent code semantics more accurately. Based on the above results and observations, we can make a conclusion that our {\toolname} is a general code summarization framework and can be easily combined with advanced neural network architectures or pre-trained models. We have reasons to believe that {\toolname} can be combined with more advanced neural network architectures or models to exert more powerful performance in the future.

\subsubsection{\textbf{RQ3:} Influence of Fusion Ways of Extractor and Abstracter on {\toolname}} 
\label{subsubsec:effect_of_fusion_fashion}
\
\newline
In this section, we further explore the influence of fusion ways between the extractor and abstracter on the performance of {\toolname}. In practice, as mentioned earlier, we try to fuse the outputs of the ExEncoder and AbEncoder by the following two concatenated ways: $[\bm{e}^{Ex};\bm{e}^{Ab}]$ and $[\bm{e}^{Ab};\bm{e}^{Ex}]$. The experimental results are shown in Table~\ref{tab:impact_of_fusion_fashion}. In the table, rows 3 and 4 ($[\bm{e}^{Ex};\bm{e}^{Ab}]$ and $[\bm{e}^{Ab};\bm{e}^{Ex}]$) mean that we use the way of concatenation to fuse the embeddings generated by ExEncoder/AbEncoder and AbEncoder/ExEncoder.

\begin{table}[htbp]
    \small
    \renewcommand{\arraystretch}{1.1}
  \caption{Influence of Fusion Ways on {\toolname}}
  \label{tab:impact_of_fusion_fashion}
  \centering
  \begin{tabular}{|c|ccc|ccc|}
    \hline
    \multirow{2}{*}{Fusion Ways} & \multicolumn{3}{c|}{JCSD} & \multicolumn{3}{c|}{PCSD}\\
    \cline{2-7}
    & \deletion{BLEU-4}\revision{BLEU} & METEOR & ROUGE-L & \deletion{BLEU-4}\revision{BLEU} & METEOR & ROUGE-L \\
    \hline
    $[\bm{e}^{Ex};\bm{e}^{Ab}]$ & 47.27 & 30.08 & 58.14 & 35.56 & 23.53 & 51.69\\
    $[\bm{e}^{Ab};\bm{e}^{Ex}]$ & \textbf{47.66} & \textbf{30.39} & \textbf{58.77} & \textbf{35.96} & \textbf{23.70} & \textbf{51.83}\\
    \hline
    \hline
    $p$-value & \revision{$<0.0001$} & \revision{0.0052} & \revision{0.0001} & \revision{$<0.0001$} & \revision{$<0.0001$} & \revision{$<0.0001$}\\
    \revision{effect size} & \revision{0.21} & \revision{0.11} & \revision{0.19} & \revision{0.11} & \revision{0.06} & \revision{0.06}\\
     \hline
 \end{tabular}
\end{table}

From Table~\ref{tab:impact_of_fusion_fashion}, we can observe that, in general, the fusion way $[\bm{e}^{Ab};\bm{e}^{Ex}]$ is better than $[\bm{e}^{Ex};\bm{e}^{Ab}]$. \delete{This is in line with human intuition. When summering a code snippet, human should first browse the whole code snippet and then write a summary according to the key content.}
\revise{
For each metric, we perform the paired Wilcoxon-Mann-Whitney signed-rank test on all scores of both fusion ways at a significance level of 5\%. The test results are presented in the penultimate row of Table~\ref{tab:impact_of_fusion_fashion}. \deletion{In the ``$p$-value'' row, the symbol ``*'' has the same meaning as that in Figure~\ref{fig:compare_distribution}. For example, ``****'' in the \deletion{BLEU-4}\revision{BLEU} column of the $p$-value row indicates that there is an extremely significant difference between $[\bm{e}^{Ab};\bm{e}^{Ex}]$ and $[\bm{e}^{Ex};\bm{e}^{Ab}]$ in terms of the \deletion{BLEU-4}\revision{BLEU} score  (i.e., $p$-value < 0.0001).} \revision{It is observed that all $p$-values are smaller than the significant threshold value of 0.05. The last row shows the effect size values of the comparison group $[\bm{e}^{Ab};\bm{e}^{Ex}]$ vs. $[\bm{e}^{Ex};\bm{e}^{Ab}]$. It is observed that on the JCSD dataset, $[\bm{e}^{Ab};\bm{e}^{Ex}]$ has a relatively medium improvement over $[\bm{e}^{Ex};\bm{e}^{Ab}]$ in terms of all three metrics. On the PCSD dataset, $[\bm{e}^{Ab};\bm{e}^{Ex}]$ has a relatively medium improvement over $[\bm{e}^{Ex};\bm{e}^{Ab}]$ in terms of BLEU, while the improvement in METEOR and ROUGE-L is relatively small. It also indicates that $[\bm{e}^{Ab};\bm{e}^{Ex}]$ has a significantly larger effect on the performance of {\toolname} than $[\bm{e}^{Ex};\bm{e}^{Ab}]$.}

During model training, the word sequences $w^{AbEx}$ and $w^{ExAb}$ predicted based on $[\bm{e}^{Ab};\bm{e}^{Ex}]$ and $[\bm{e}^{Ex};\bm{e}^{Ab}]$ would be compared with the same reference summary to calculate the loss and update the model parameters. 
We find that, under the same training epoch, the loss value computed based on $w^{AbEx}$ is less than that based on $w^{ExAb}$. We conjecture that, compared with $w^{ExAb}$, the order of words in $w^{AbEx}$ is closer to that in the reference summary. 
In other words, compared with $[\bm{e}^{Ex};\bm{e}^{Ab}]$, the fusion way of $[\bm{e}^{Ab};\bm{e}^{Ex}]$ makes the model easily learn features of code snippets. 
}

\subsubsection{\textbf{RQ4:} Robustness of {\toolname}}
\label{subsubsec:robustness_of_our_approach}
\
\newline

\begin{figure}[htbp]
\centering
    \subfigure[Code Snippets in JCSD]
    {
        \includegraphics[width=0.23\linewidth]{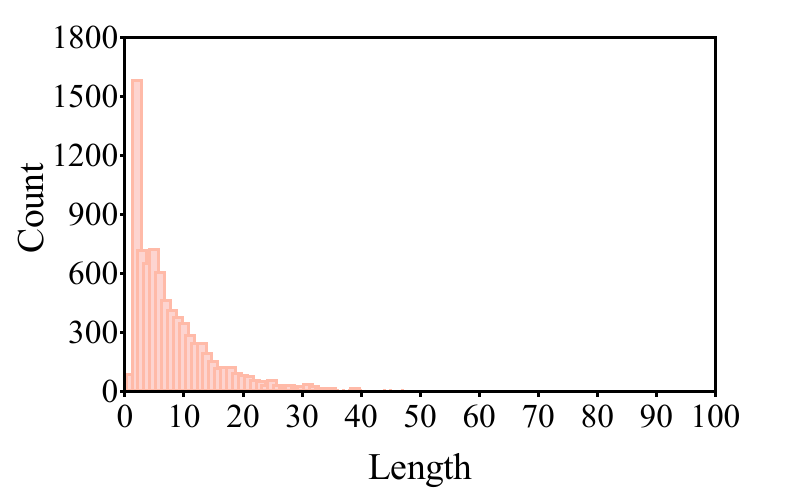}
        \label{fig:distribution_of_code_length_JCSD}
    }
    \subfigure[Comments in JCSD]
    {
        \includegraphics[width=0.23\linewidth]{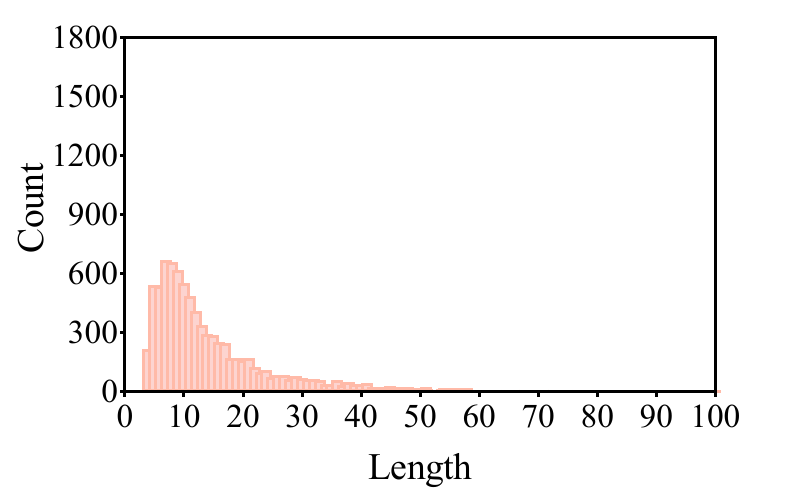}
        \label{fig:distribution_of_comment_length_JCSD}
    }
    \subfigure[Code Snippets in PCSD]
    {
        \includegraphics[width=0.23\linewidth]{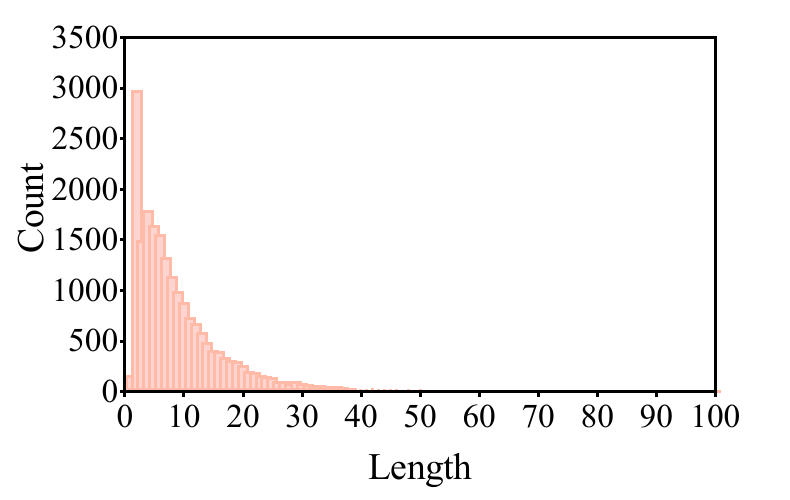}
        \label{distribution_of_code_length_PCSD}
    }
    \subfigure[Comments in PCSD]
    {
        \includegraphics[width=0.23\linewidth]{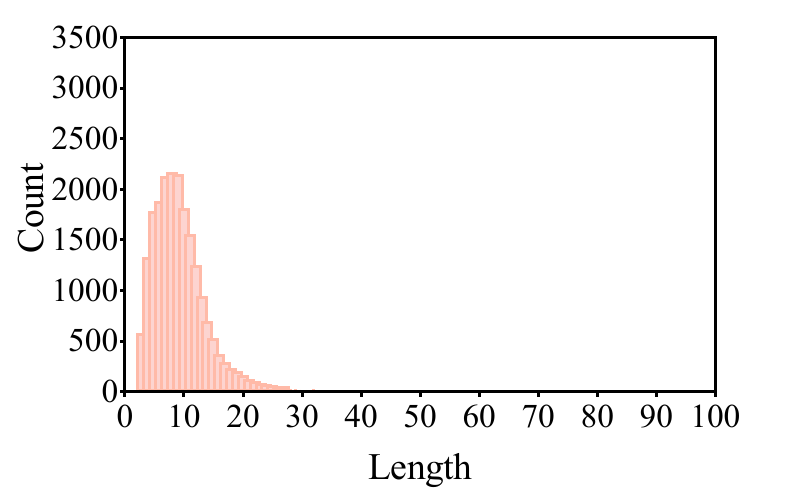}
        \label{fig:distribution_of_comment_length_PCSD}
    }
    \caption{Length distribution of code snippets and comments in test sets}
    \label{fig:distribution_of_length}
\end{figure}

\begin{figure}[htbp]
\centering
    \subfigure[Code Snippets in JCSD]
    {
        \includegraphics[width=0.23\linewidth]{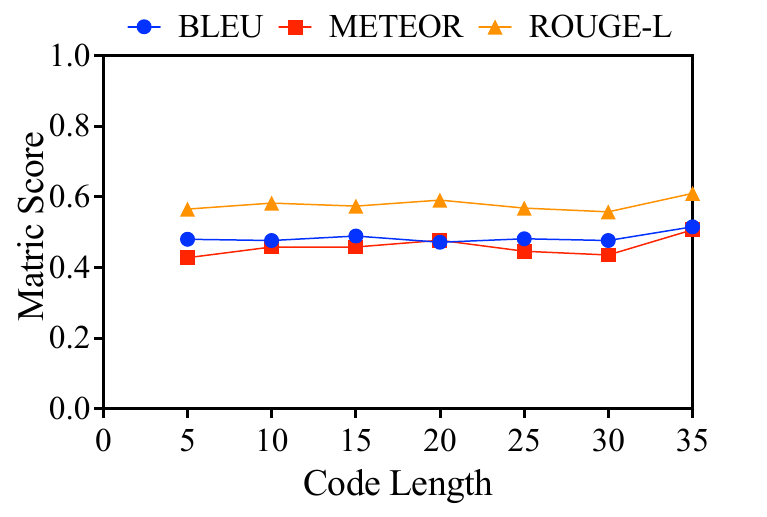}
        \label{fig:robustness_on_vary_code_length}
    }
    \subfigure[Comments in JCSD]
    {
        \includegraphics[width=0.23\linewidth]{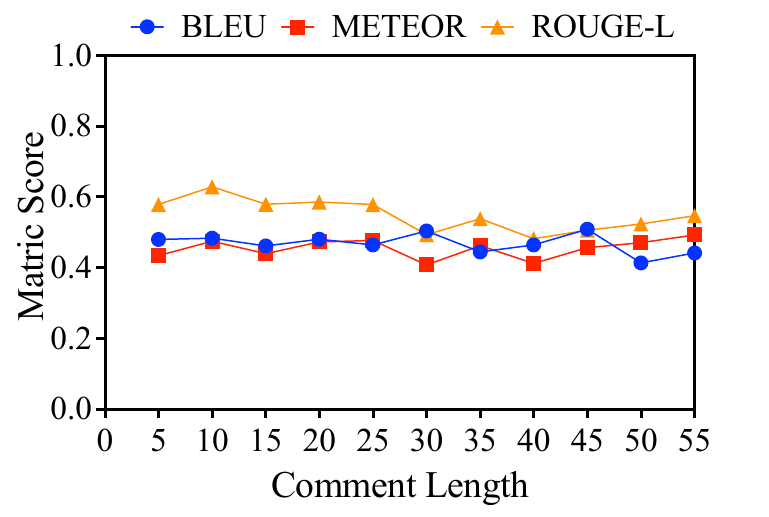}
        \label{fig:robustness_on_comment_length}
    }
    \subfigure[Code Snippets in PCSD]
    {
        \includegraphics[width=0.23\linewidth]{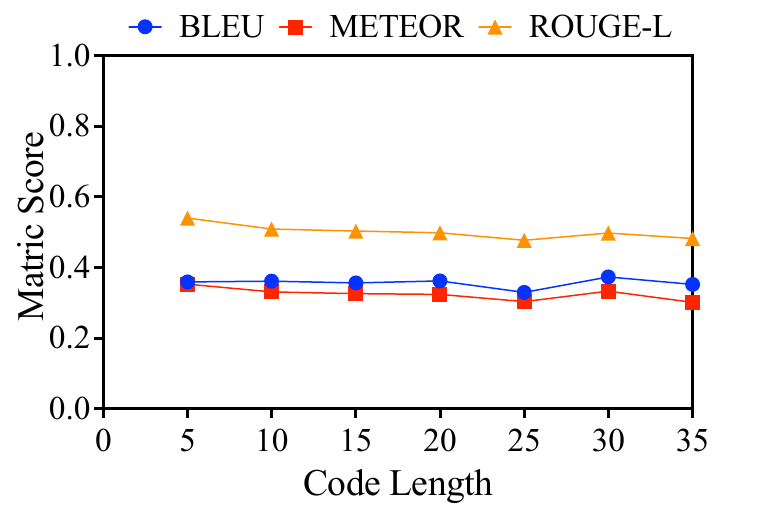}
        \label{fig:robustness_on_vary_code_length}
    }
    \subfigure[Comments in PCSD]
    {
        \includegraphics[width=0.23\linewidth]{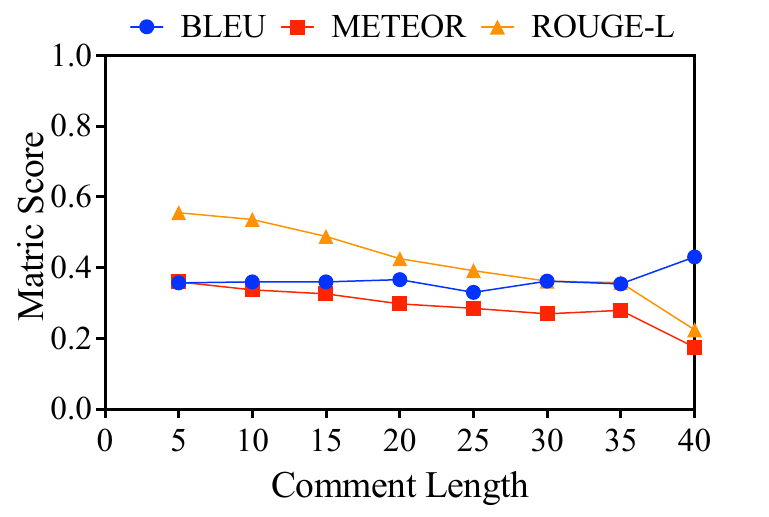}
        \label{fig:robustness_on_comment_length}
    }
    \caption{Effect of code snippet and comment length on the robustness of {\toolname}}
    \label{fig:robustness_of_EACS}
\end{figure}

To study the robustness of {\toolname}, we analyze two parameters (i.e., code length and comment length) that may have an effect on the embedding representations of code snippets and comments.
Figure~\ref{fig:distribution_of_length} shows the length distributions of code snippets and comments on the test sets of the JCSD and PCSD datasets. For a code snippet, its length refers to the number of lines of the code snippet. For a comment, its length refers to the number of words in the comment. 
From Figure~\ref{fig:distribution_of_length}(a) and (c), it can be observed that the lengths of most code snippets are less than 20. \revise{This was also observed in the quote in~\cite{2009-Clean-Code, 2022-TranCS} ``Functions should hardly ever be 20 lines long''.} From Figure~\ref{fig:distribution_of_length}(b) and (d), it is noticed that almost all comments are less than 20 in length. This also confirms the challenge of capturing the correlation between the long code snippet with its corresponding short comment (summary).

\begin{figure}[htbp]
\centering
    \subfigure[Robustness of SiT]
    {
        \includegraphics[width=0.44\linewidth]{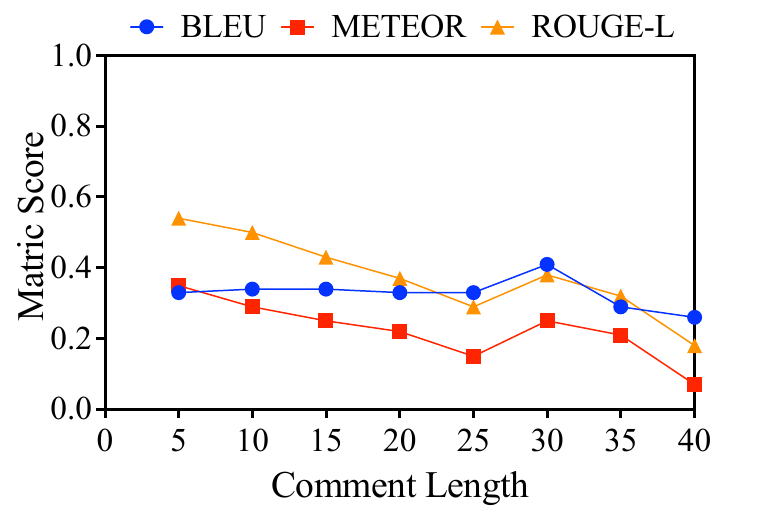}
        \label{fig:Robustness_on_vary_comment_length_PCSD_test_set_SiT}
    }
    \subfigure[Robustness of CodeT5]
    {
        \includegraphics[width=0.44\linewidth]{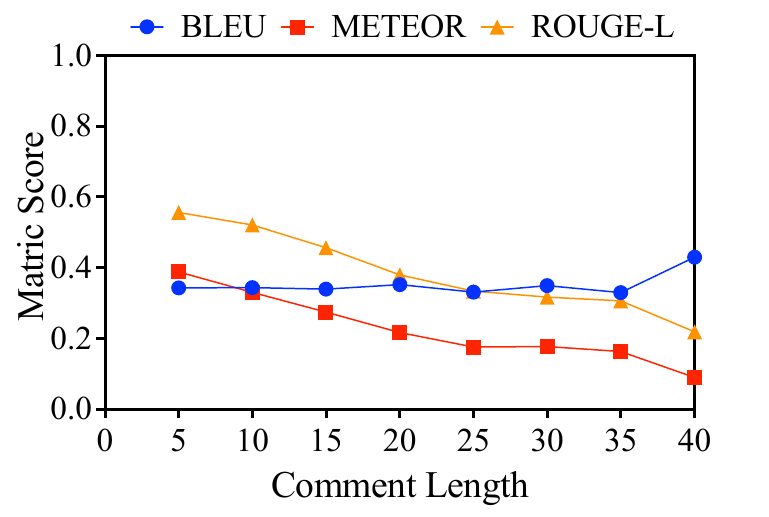}
        \label{fig:CodeT5_robustness_on_vary_comment_length_PCSD}
    }
    \caption{Effect of comment length on the robustness of SiT and CodeT5}
    \label{fig:robustness_of_baselines}
\end{figure}

Figure~\ref{fig:robustness_of_EACS} shows the performance of {\toolname} based on different evaluation metrics with varying parameters. 
\delete{From Figure~\ref{fig:robustness_of_EACS}, we can observe that in most cases, {\toolname} maintains a stable performance even though the code snippet length or comment length increases, which can be attributed to the extractive-and-abstractive framework we proposed.} 
\revise{From Figure~\ref{fig:robustness_of_EACS}(a) and (b), it can be observed that on the JCSD test set, {\toolname} maintains stable performance even though the code snippet length or comment length increases, which can be attributed to the extractive-and-abstractive framework we proposed.}
\delete{When the length of the comment exceeds 25 (a common range shown in~\ref{fig:distribution_of_comment_length_test_set}), the METEOR and ROUGE-L scores of {\toolname} decreases as the length increases. It means that when the length of the comments exceeds the common range, as the expected length of the generated comment continues to increase, it will be more difficult to generate.\fcr{generate what? summary? it?}} 
\revise{On the PCSD test set, from Figure~\ref{fig:robustness_of_EACS}(c) and (d), it can be observed that {\toolname} also maintains stable performance when the code snippet length increases, however, the performance of {\toolname} degrades significantly in terms of METEOR and ROUGE-L as the comment length increases. We further analyze the performance of SiT and CodeT5 on varying the PCSD comments, and the results are shown in Figure~\ref{fig:robustness_of_baselines}. From this figure, we can observe the same phenomenon as {\toolname}. It means that as the expected length of the generated summary continues to increase, it will be more challenging to generate.}
Overall, the results verify the robustness of our {\toolname}.

\subsubsection{\revise{\textbf{RQ5:} Human Evaluation}}
\label{subsubsec:human_evaluation}
\
\newline
Many works~\cite{2016-CODE-NN, 2020-Hybrid-DeepCom, 2020-R2Com, 2020-Rencos, 2021-CAST, 2021-EditSum, 2021-Reassessing-Metrics-for-Code-Summarization, 2022-Evaluation-Neural-Code-Summarization} show that the automatic evaluation metrics (BLEU, METEOR, and ROUGE-L) mainly calculate the textual similarity rather than the semantic similarity between the reference summaries and generated summaries. Hence, we conduct a human evaluation by following the previous works~\cite{2021-SiT, 2020-Hybrid-DeepCom, 2020-R2Com, 2020-Rencos, 2021-CAST} to evaluate the summaries generated by the baselines Ex-based, SiT, CodeT5, and our {\toolname}. Specially, we invite 10 volunteers with more than 3 years of software development experience and excellent English ability to carry out the evaluation. Each volunteer is asked to assign scores from 0 to 4 (the higher, the better) to generated summaries from four aspects: 
\textit{similarity} (similarity of the generated summaries and the reference summaries), \textit{naturalness} (grammaticality and fluency), \textit{informativeness} (the amount of content carried over from the input code snippets to the generated summaries, ignoring fluency) and \textit{relevance} (the degree to which the generated summaries are relevant with the input code snippets). We randomly select 100 code snippets, including 50 from the JCSD dataset and 50 from the PCSD dataset, the corresponding summaries generated by Ex-based, SiT, CodeT5, and our {\toolname}, and the reference summaries (i.e., ground-truth), respectively. We divide the 100 samples into two groups, and each of them includes 50 samples, of which 25 belong to the JCSD dataset and 25 belong to the PCSD dataset. To reduce the workload of volunteers and ensure the fairness of experimental results, each volunteer randomly evaluates only one group of samples. Each summary is evaluated by five volunteers, and the final score is the average of them.

\begin{table*}[htbp]
    \small
    \centering
    \renewcommand{\arraystretch}{1.2}
    \caption{Results of human evaluation. The values in parentheses represent standard deviations.}
    \label{tab:human_evaluation}
    
    \begin{tabular}{|c|l|l|l|l|l|}
    \hline
    Dataset & Metrics & Ex-based & SiT & CodeT5 & {\toolname}\\
    \hline
    \multirow{4}{*}{JCSD} & Similarity & 2.44 (0.48) & 2.60 (0.48) & 2.77 (0.40) &\textbf{2.90 (0.37)}\\
    & Naturalness & 3.36 (0.56) & 3.51 (0.47) & 3.52 (0.44) & \textbf{3.57 (0.43)} \\
    & Informativeness & 2.65 (0.52) & 2.92 (0.56) & 3.00 (0.51) & \textbf{3.17 (0.52)} \\
    & Relevance & 2.70 (0.59) & 2.95 (0.53) & 3.02 (0.51) & \textbf{3.23 (0.50)} \\
    \hline
    \multicolumn{2}{|c|}{Average} & 2.79 & 3.00 & 3.08 & \textbf{3.22} \\
    \hline
    \hline
    \multirow{4}{*}{PCSD} & Similarity & 2.09 (0.47) & 2.38 (0.49) & 2.47 (0.56) & \textbf{2.66 (0.41)} \\
    & Naturalness & 3.10 (0.65) & 3.28 (0.60) & 3.37 (0.54) & \textbf{3.43 (0.54)} \\
    & Informativeness & 2.22 (0.59) & 2.66 (0.61) & 2.70 (0.65) & \textbf{2.86 (0.59)} \\
    & Relevance & 2.14 (0.55) & 2.53 (0.63) & 2.58 (0.66) & \textbf{2.74 (0.52)} \\
    \hline
    \multicolumn{2}{|c|}{Average} & 2.39 & 2.71 & 2.78 & \textbf{2.92} \\
    \hline
 \end{tabular}
\end{table*}

The results of the human evaluation are shown in Table~\ref{tab:human_evaluation}. The standard deviations of all methods (the values in all parentheses) are small, which indicates that their scores by humans are about the same degree of concentration~\cite{2021-EditSum}. From Table~\ref{tab:human_evaluation}, it can be observed that overall our {\toolname} consistently outperforms Ex-based, SiT, and CodeT5 in all four aspects. On the JCSD dataset, compared with Ex-based, SiT, and CodeT5, {\toolname} improves on average by 15.41\%, 7.33\%, and 4.55\% in four aspects, respectively, while on the PCSD dataset, {\toolname} improves on average by 22.18\%, 7.75\%, and 5.04\%, respectively. 
\revision{
Although from the average values in rows 6 and 11 of Table~\ref{tab:human_evaluation}, {\toolname} is slightly better than the best baseline CodeT5, the improvement of {\toolname} to CodeT5 is greater than that of CodeT5 to SiT. Specifically, {\toolname} improves CodeT5 by 4.55\% and 5.04\% on the JCSD and PCSD datasets, while CodeT5 improves SiT by 2.67\% and 2.58\%, respectively. In other words, {\toolname} improves CodeT5 nearly twice as much as CodeT5 improves SiT.
}

\begin{figure}[htbp]
\centering
    \subfigure[Similarity Scores on JCSD]
    {
        \includegraphics[width=0.4\linewidth]{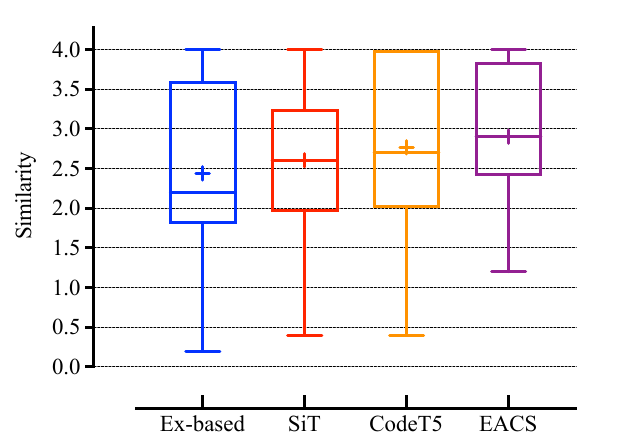}
        \label{fig:compare_similarity_on_PCSD}
    }
    \subfigure[Naturalness Scores on JCSD]
    {
        \includegraphics[width=0.4\linewidth]{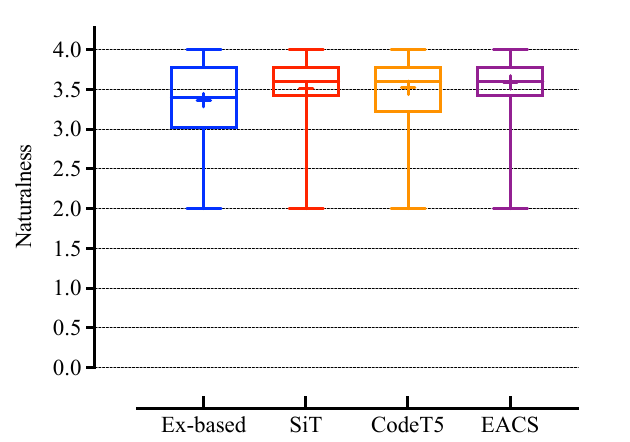}
        \label{fig:compare_naturalness_on_JCSD}
    }
    \subfigure[Informativeness Scores on JCSD]
    {
        \includegraphics[width=0.4\linewidth]{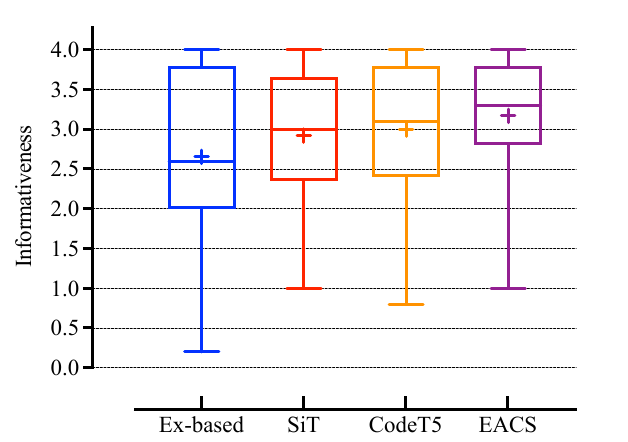}
        \label{fig:compare_informativeness_on_JCSD}
    }
    \subfigure[Relevance Scores on JCSD]
    {
        \includegraphics[width=0.4\linewidth]{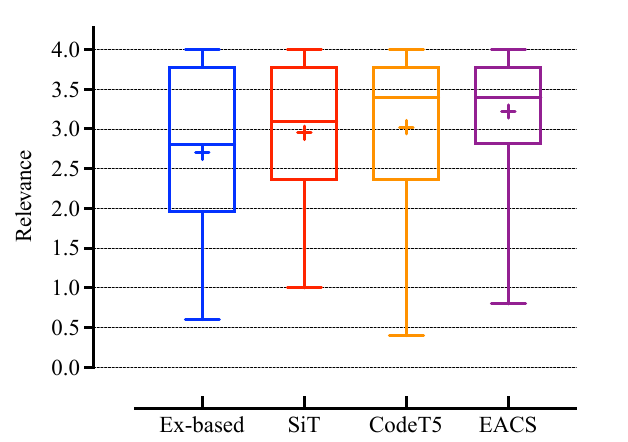}
        \label{fig:compare_relevance_on_JCSD}
    }
    \subfigure[Similarity Scores on PCSD]
    {
        \includegraphics[width=0.4\linewidth]{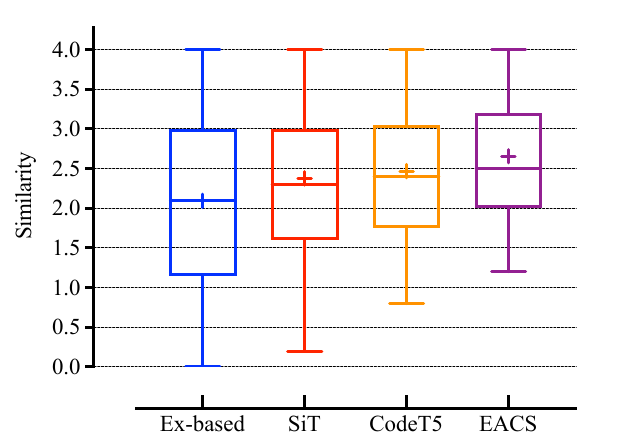}
        \label{fig:compare_similarity_on_PCSD}
    }
    \subfigure[Naturalness Scores on PCSD]
    {
        \includegraphics[width=0.4\linewidth]{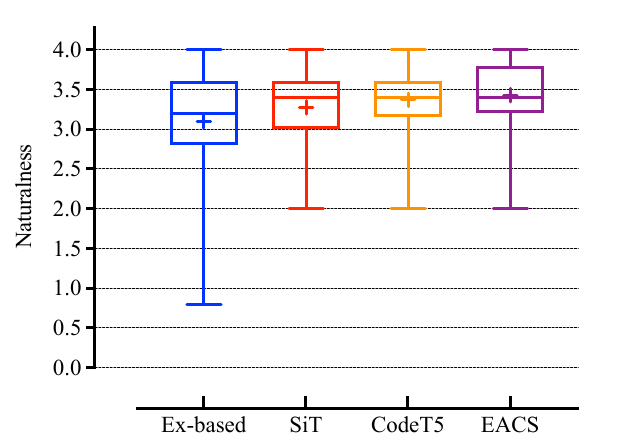}
        \label{fig:compare_naturalness_on_PCSD}
    }
    \subfigure[Informativeness Scores on PCSD]
    {
        \includegraphics[width=0.4\linewidth]{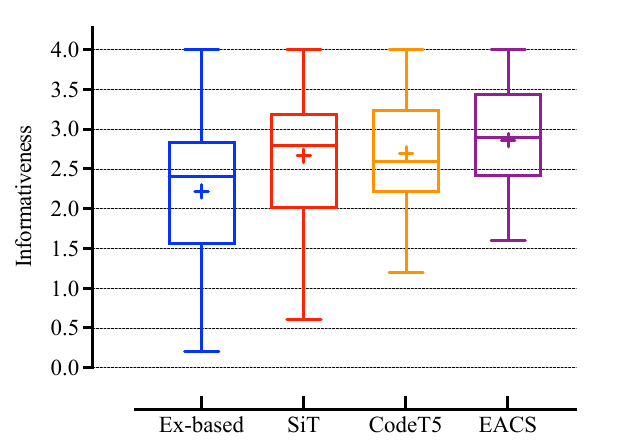}
        \label{fig:compare_informativeness_on_PCSD}
    }
    \subfigure[Relevance Scores on PCSD]
    {
        \includegraphics[width=0.4\linewidth]{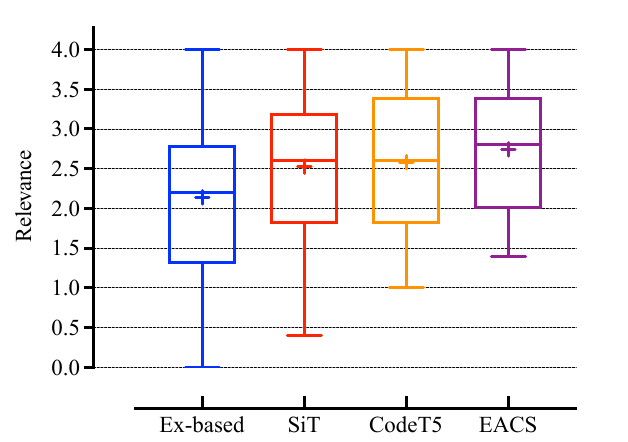}
        \label{fig:compare_relevance_on_PCSD}
    }
    \caption{Distributions of scores of human evaluation on the JCSD and PCSD datasets}
    \label{fig:compare_human_score_distribution}
\end{figure}

\deletion{Note that the values in Table~\ref{tab:human_evaluation} are the average scores of all 50 samples. }For a more comprehensive comparison, we further analyze the distribution of the scores of Ex-based, SiT, CodeT5, and {\toolname} on all samples, and the statistical results are shown in Figure~\ref{fig:compare_human_score_distribution}. In Figure~\ref{fig:compare_human_score_distribution}, `+' denotes the mean, which is the value filled in Table~\ref{tab:human_evaluation}. \revision{Figure~\ref{fig:compare_human_score_distribution}(a)-(d) show the box-and-whiskers plots of the similarity, naturalness, informativeness, and relevance scores obtained by the three baselines and {\toolname} on the JCSD dataset. It can be observed that on all four metrics, the first quartile associated with {\toolname} is better than those associated with all three baselines. Figure~\ref{fig:compare_distribution}(e)-(h) show the box-and-whiskers plots of the scores by the three baselines and {\toolname} on the PCSD dataset. In similarity, the median, first quartile, and third quartile associated with {\toolname} are better than those associated with all three baselines. In 
naturalness, the third quartile associated with {\toolname} is better than those associated with all three baselines. In informativeness, the median, first quartile, and third quartile associated with {\toolname} are better than those associated with all three baselines. In relevance, the median and first quartile associated with {\toolname} are better than those associated with all three baselines.}
Overall, the metric score distribution of {\toolname} is better than that of Ex-based, SiT, and CodeT5. 
\revision{In particular, the first quartile associated with {\toolname} is better than those associated with the best baseline CodeT5 on all four metrics. This means that {\toolname} generates fewer low-scoring summaries compared to CodeT5.}

\begin{table*}[!t]
    \scriptsize
    \tabcolsep=1.5pt
    \centering
    \renewcommand{\arraystretch}{1.1} 
    \caption{\revision{$p$-values and effect sizes for metrics used in the human evaluation. Effect size: $<0.1$: small effect; $0.1-<0.3$: medium effect; and $\geq 0.3$: large effect.}}
    \label{tab:compare_human_evaluation_in_p-value_effect_size}
  
  \begin{tabular}{|l|c|cccc|cccc|}
    \hline
     \revision{\multirow{2}{*}{Comparison Group}} & \revision{\multirow{2}{*}{Result}} & \multicolumn{4}{c|}{\revision{JCSD}} & \multicolumn{4}{c|}{\revision{PCSD}}\\
     \cline{3-10}
     & & \revision{Similarity} & \revision{Naturalness} & \revision{Informativeness}& \revision{Relevance} & \revision{Similarity} & \revision{Naturalness} & \revision{Informativeness} & \revision{Relevance} \\
    \hline
    
    \revision{\multirow{2}{*}{SiT vs. Ex-based}} & \revision{$p$-value} & \revision{$>0.9999$} & \revision{0.5302} & \revision{0.9794} & \revision{0.728} & \revision{$>0.9999$} & \revision{$>0.9999$} & \revision{0.1208} & \revision{0.6229} \\
    & \revision{effect size} & \revision{0.14} & \revision{0.24} & \revision{0.20} & \revision{0.22} & \revision{0.15} & \revision{0.09} & \revision{0.33} & \revision{0.23} \\
    \hline
    
    \revision{\multirow{2}{*}{CodeT5 vs. Ex-based}} & \revision{$p$-value} & \revision{0.0791} & \revision{0.1089} & \revision{0.1339} & \revision{0.2641} & \revision{$>0.9999$} & \revision{0.1339} & \revision{\textbf{0.0358}} & \revision{0.2406} \\
    & \revision{effect size} & \revision{0.35} & \revision{0.33} & \revision{0.32} & \revision{0.28} & \revision{0.18} & \revision{0.32} & \revision{0.39} & \revision{0.29} \\
    \hline
    
    \revision{\multirow{2}{*}{CodeT5 vs. SiT}} & \revision{$p$-value} & \revision{0.7856} & \revision{$>0.9999$} & \revision{$>0.9999$} & \revision{$>0.9999$} & \revision{$>0.9999$} & \revision{0.575} & \revision{$>0.9999$} & \revision{$>0.9999$} \\
    & \revision{effect size} & \revision{0.21} & \revision{0.09} & \revision{0.13} & \revision{0.07} & \revision{0.03} & \revision{0.24} & \revision{0.06} & \revision{0.06} \\
    \hline
    
    \revision{\multirow{2}{*}{{\toolname} vs. Ex-based}} & \revision{$p$-value} & \revision{\textbf{0.0002}} & \revision{\textbf{0.0078}} & \revision{\textbf{0.0005}} & \revision{\textbf{0.0003}} & \revision{\textbf{0.0014}} & \revision{\textbf{0.0172}} & \revision{\textbf{0.0003}} & \revision{\textbf{0.0019}} \\
    & \revision{effect size} & \revision{0.59} & \revision{0.45} & \revision{0.55} & \revision{0.57} & \revision{0.52} & \revision{0.42} & \revision{0.58} & \revision{0.51} \\
    \hline
    
    \revision{\multirow{2}{*}{{\toolname} vs. SiT}} & \revision{$p$-value} & \revision{\textbf{0.009}} & \revision{0.7856} & \revision{0.0709} & \revision{0.0791} & \revision{0.0568} & \revision{0.1089} & \revision{0.4882} & \revision{0.2895} \\
    & \revision{effect size} & \revision{0.45} & \revision{0.21} & \revision{0.36} & \revision{0.35} & \revision{0.37} & \revision{0.33} & \revision{0.25} & \revision{0.28} \\
    \hline
    
    \revision{\multirow{2}{*}{{\toolname} vs. CodeT5}} & \revision{$p$-value} & \revision{0.575} & \revision{$>0.9999$} & \revision{0.6229} & \revision{0.2641} & \revision{0.098} & \revision{$>0.9999$} & \revision{$>0.9999$} & \revision{0.728}\\
    & \revision{effect size} & \revision{0.24} & \revision{0.21} & \revision{0.23} & \revision{0.28} & \revision{0.34} & \revision{0.10} & \revision{0.19} & \revision{0.22} \\
    \hline
 \end{tabular}
\end{table*}

\revision{We further run a statistical test to assess whether there is enough empirical evidence to claim that there is a difference among the four techniques regarding the four metrics. Table~\ref{tab:compare_human_evaluation_in_p-value_effect_size} shows the $p$-values reported by Prism after performing the Friedman test and Dunn's multiple comparisons test among Ex-based, SiT, CodeT5, and our {\toolname}. From this table, we can observe that the $p$-values for most comparison groups are larger than the significance threshold of 0.05. Only the scores of {\toolname} and Ex-based on all four indicators are significantly different, as shown in row 9 of Table~\ref{tab:compare_human_evaluation_in_p-value_effect_size}. The main reason is that the experimental samples in human evaluation are relatively small. In other words, more samples are needed to detect statistically significant differences. This is corroborated by the automated evaluation (as the $p$-values in Table~\ref{tab:compare_baseline_in_p-value_effect_size}). 

To assess the magnitude of the improvement of one technique relative to another, we calculate the effect size for each comparison group (e.g., {\toolname} vs. Ex-based). Rows 4, 6, 8, 10, 12, and 14 of TABLE~\ref{tab:compare_human_evaluation_in_p-value_effect_size} show the effect size of each comparison group. For example, from row 10, we can observe that all effect size values of the comparison group {\toolname} vs. Ex-based are larger than 0.3, which means that {\toolname} has a relatively large improvement over Ex-based on all four metrics. 
}

\begin{table*}[!t]
    \footnotesize
    \tabcolsep=2pt
    \centering
    \renewcommand{\arraystretch}{1.1} 
    \caption{\revision{$p$-values and effect sizes for metrics used in the human evaluation. In this table, $p$-values are reported by the paired Wilcoxon-Mann-Whitney signed-rank test~\cite{1963-Wilcoxon} at a significance level of 5\%. Effect size: $<0.1$: small effect; $0.1-<0.3$: medium effect; and $\geq 0.3$: large effect.}}
    \label{tab:compare_human_evaluation_in_p-value_effect_size_with_CodeT5}
  
  \begin{tabular}{|c|cccc|cccc|}
    \hline
     \revision{\multirow{2}{*}{Result}} & \multicolumn{4}{c|}{\revision{JCSD}} & \multicolumn{4}{c|}{\revision{PCSD}}\\
     \cline{2-9}
     & \revision{Similarity} & \revision{Naturalness} & \revision{Informativeness}& \revision{Relevance} & \revision{Similarity} & \revision{Naturalness} & \revision{Informativeness} & \revision{Relevance} \\
    \hline
    
    \revision{$p$-value} & \revision{0.031} & \revision{0.070} & \revision{0.032} & \revision{0.018} & \revision{0.002} & \revision{0.105} & \revision{0.024} & \revision{0.012} \\
    
    \hline
    
    \revision{effect size} & \revision{0.24} & \revision{0.21} & \revision{0.23} & \revision{0.28} & \revision{0.34} & \revision{0.10} & \revision{0.19} & \revision{0.22} \\
    \hline
 \end{tabular}
\end{table*}

\revision{The $p$-values shown in Table~\ref{tab:compare_human_evaluation_in_p-value_effect_size} are reported by Dunn's multiple comparisons test with correction. If we compare {\toolname} and CodeT5 alone, we can get the statistical test results as shown in Table~\ref{tab:compare_human_evaluation_in_p-value_effect_size_with_CodeT5}. From row 2 of Table~\ref{tab:compare_human_evaluation_in_p-value_effect_size_with_CodeT5}, it is observed that except for naturalness, the p-values are less than the significance threshold of 0.05. Combining the effect sizes shown in row 3 of Table~\ref{tab:compare_human_evaluation_in_p-value_effect_size_with_CodeT5}, we can conclude that overall {\toolname} is superior to CodeT5, especially with significantly medium improvements in similarity, informativeness, and relevance of generated summaries.
}

\begin{table*}[htbp]
    \footnotesize
    \renewcommand{\arraystretch}{1.2}
  \caption{\revision{Number statistic. Values in bold in each column are the maximum values. Columns $\geq 1$, $\geq 2$, and $\geq 3$ show the number of samples that get a metric score larger than 1, 2, and 3, respectively.}}
  \label{tab:human_evaluation_number_statistic}
  \centering
  \begin{tabular}{|c|l|ccc|ccc|ccc|ccc|}
    \hline
    \multirow{2}{*}{\revision{Dataset}} & \multirow{2}{*}{\revision{Methods}} & \multicolumn{3}{c|}{\revision{Similarity}} & \multicolumn{3}{c|}{\revision{Naturalness}} & \multicolumn{3}{c|}{\revision{Informativeness}} & \multicolumn{3}{c|}{\revision{Relevance}} \\
    \cline{3-14}
    & & \revision{$\geq 1$} & \revision{$\geq 2$} & \revision{$\geq 3$} & \revision{$\geq 1$} & \revision{$\geq 2$} & \revision{$\geq 3$} & \revision{$\geq 1$} & \revision{$\geq 2$} & \revision{$\geq 3$} & \revision{$\geq 1$} & \revision{$\geq 2$} & \revision{$\geq 3$} \\
    
    \hline
    \multirow{4}{*}{\revision{JCSD}} & \revision{Ex-based} & \revision{45} & \revision{36} & \revision{17} & \revision{\textbf{50}} & \revision{\textbf{50}} & \revision{41} & \revision{47} & \revision{39} & \revision{19} & \revision{48} & \revision{39} & \revision{24} \\
    
    & \revision{SiT} & \revision{48} & \revision{38} & \revision{20} & \revision{\textbf{50}} & \revision{\textbf{50}} & \revision{46} & \revision{\textbf{50}} & \revision{44} & \revision{26} & \revision{\textbf{50}} & \revision{43} & \revision{27} \\
    
    & \revision{CodeT5} & \revision{48} & \revision{40} & \revision{23} & \revision{\textbf{50}} & \revision{\textbf{50}} & \revision{44} & \revision{49} & \revision{43} & \revision{28} & \revision{48} & \revision{44} & \revision{31} \\

    \cline{2-14}
    
    & \revision{{\toolname}} & \revision{\textbf{50}} & \revision{\textbf{42}} & \revision{\textbf{25}} & \revision{\textbf{50}} & \revision{\textbf{50}} & \revision{\textbf{47}} & \revision{\textbf{50}} & \revision{\textbf{47}} & \revision{\textbf{34}} & \revision{\textbf{50}} & \revision{\textbf{47}} & \revision{\textbf{35}} \\

    \hline
    \hline
    \multirow{4}{*}{\revision{PCSD}} & \revision{Ex-based} & \revision{41} & \revision{28} & \revision{14} & \revision{\textbf{50}} & \revision{48} & \revision{37} & \revision{45} & \revision{33} & \revision{13} & \revision{42} & \revision{32} & \revision{11} \\
    
    & \revision{SiT} & \revision{48} & \revision{30} & \revision{17} & \revision{\textbf{50}} & \revision{\textbf{50}} & \revision{40} & \revision{48} & \revision{42} & \revision{19} & \revision{48} & \revision{35} & \revision{15} \\
    
    & \revision{CodeT5} & \revision{48} & \revision{32} & \revision{15} & \revision{\textbf{50}} & \revision{\textbf{50}} & \revision{44} & \revision{\textbf{50}} & \revision{41} & \revision{19} & \revision{\textbf{50}} & \revision{37} & \revision{14} \\

    \cline{2-14}
    
    & \revision{{\toolname}} & \revision{\textbf{50}} & \revision{\textbf{42}} & \revision{\textbf{20}} & \revision{\textbf{50}} & \revision{\textbf{50}} & \revision{\textbf{47}} & \revision{\textbf{50}} & \revision{\textbf{47}} & \revision{\textbf{25}} & \revision{\textbf{50}} & \revision{\textbf{40}} & \revision{\textbf{18}} \\

    \hline
 \end{tabular}
\end{table*}

\revision{We count the number of samples for different score bands. Table~\ref{tab:human_evaluation_number_statistic} shows the statistical results. From the table, it is observed that, {\toolname} overall outperforms all three baselines on all four metrics, especially for scores $\geq 2$ and $\geq 3$. For example, among the randomly selected 100 code snippets, in similarity, our {\toolname} can generate comments with scores $\geq 3$ for 45 code snippets, while the best baseline CodeT5 can only achieve 38. In naturalness, {\toolname} can generate comments with scores $\geq 3$ for 94 code snippets, outperforming CodeT5 (88). In informativeness, {\toolname} can generate comments with scores $\geq 3$ for 59 code snippets and better than CodeT5 (47). In relevance, our {\toolname} can generate comments with scores $\geq 3$ for 53 code snippets and outperforms CodeT5 (45). Referring to~\cite{2020-Rencos}, a summary with a score $\geq 3$ can be regarded as a good summary. Based on this, we can conclude that {\toolname} can generate good summaries for more code snippets than CodeT5.
}

\subsubsection{\revision{\textbf{RQ6:} Effect of Similarity Metrics}}
\label{subsubsec:influence_of_similarity_metrics}
\
\newline
\revision{In ground-truth important statement generation, we follow existing extractors~\cite{2018-Fast-Abstractive-Summarization, 2017-SummaRuNNer, 2019-Text-Summarization-Pretrained-Encoders, 2018-Unified-Model-for-Extractive-Abstractive-Summarization} in NLP and use the ROUGE-L score to measure the informativity of each statement. In existing research (including code summarization and text summarization), ROUGE-L is only one of many commonly used similarity metrics. Other similar metrics, such as BLEU and METEOR, can also be used as a replacement for ROUGE-L in the extractor. In addition, some recent studies~\cite{2021-Reassessing-Metrics-for-Code-Summarization, 2022-Semantic-Metrics-for-Evaluating-Code-Summarization} have also found that some metrics (e.g., chrF~\cite{2015-chrF} and Sentence-BERT~\cite{2019-Sentence-BERT}) are more suitable for code summarization tasks than the above three metrics. chrFz works solely on character n-grams rather than word n-grams~\cite{2015-chrF}. It can be seen as a character n-gram F-score. Haque et. al~\cite{2022-Semantic-Metrics-for-Evaluating-Code-Summarization} experimentally found that the Sentence-BERT~\cite{2019-Sentence-BERT} produced vectorized representations of the summaries that have the highest correlation to perceived similarity by the human experts. These metrics may also be good choices. Therefore, we conduct an experiment to explore the effect of different similarity metrics on the performance of {\toolname}. And experimental results are shown in Table~\ref{tab:impact_of_similarity_metric}.
}

\begin{table}[htbp]
    \footnotesize
    \centering
    \tabcolsep=1pt
    \renewcommand{\arraystretch}{1.1}
    \caption{\revision{Effect of similarity metrics used in the extractor on {\toolname}. SBERT: Sentence-BERT. $\mathcal{A}$: average value $\uparrow$; $\mathcal{M}$: median $\uparrow$; $\mathcal{S}$: standard deviation $\downarrow$.}}
  \label{tab:impact_of_similarity_metric}
  \centering
  \begin{tabular}{|c|ccc|ccc|ccc|ccc|ccc|ccc|}
    \hline
    \multirow{3}{*}{\revision{Metric}} & \multicolumn{9}{c|}{\revision{JCSD}} & \multicolumn{9}{c|}{\revision{PCSD}} \\
    \cline{2-19}
    & \multicolumn{3}{c|}{\revision{BLEU}} & \multicolumn{3}{c|}{\revision{METEOR}} & \multicolumn{3}{c|}{\revision{ROUGE-L}} & \multicolumn{3}{c|}{\revision{BLEU}} & \multicolumn{3}{c|}{\revision{METEOR}} & \multicolumn{3}{c|}{\revision{ROUGE-L}} \\
    \cline{2-19}
    & \revision{$\mathcal{A}$} & \revision{$\mathcal{M}$} & \revision{$\mathcal{S}$} & \revision{$\mathcal{A}$} & \revision{$\mathcal{M}$} & \revision{$\mathcal{S}$} & \revision{$\mathcal{A}$} & \revision{$\mathcal{M}$} & \revision{$\mathcal{S}$} & \revision{$\mathcal{A}$} & \revision{$\mathcal{M}$} & \revision{$\mathcal{S}$} & \revision{$\mathcal{A}$} & \revision{$\mathcal{M}$} & \revision{$\mathcal{S}$} & \revision{$\mathcal{A}$} & \revision{$\mathcal{M}$} & \revision{$\mathcal{S}$} \\
    \hline
    
    \revision{BLEU} & \revision{46.86} & \revision{25.60} & \revision{0.41} & \revision{29.32} & \revision{\textbf{26.80}} & \revision{0.40} & \revision{58.60} & \revision{\textbf{52.70}} & \revision{\textbf{0.35}} & \revision{35.33} & \revision{20.30} & \revision{0.32} & \revision{23.49} & \revision{20.59} & \revision{0.32} & \revision{51.67} & \revision{44.44} & \revision{0.29} \\
    
    \revision{METEOR} & \revision{46.74} & \revision{25.27} & \revision{0.41} & \revision{29.21} & \revision{26.67} & \revision{0.40} & \revision{58.34} & \revision{52.14} & \revision{\textbf{0.35}} & \revision{35.31} & \revision{20.16} & \revision{0.32} & \revision{23.52} & \revision{20.50} & \revision{0.32} & \revision{51.68} & \revision{44.31} & \revision{0.29} \\
    
    \revision{ROUGE-L} & \revision{\textbf{47.66}} & \revision{\textbf{25.99}} & \revision{\textbf{0.40}} & \revision{\textbf{30.39}} & \revision{26.54} & \revision{\textbf{0.39}} & \revision{\textbf{58.77}} & \revision{52.16} & \revision{\textbf{0.35}} & \revision{\textbf{35.96}} & \revision{\textbf{21.79}} & \revision{\textbf{0.30}} & \revision{\textbf{23.70}} & \revision{\textbf{22.86}} & \revision{\textbf{0.30}} & \revision{\textbf{51.83}} & \revision{\textbf{44.66}} & \revision{\textbf{0.28}} \\

    \revision{chrF} & \revision{46.73} & \revision{25.41} & \revision{0.41} & \revision{29.23} & \revision{26.64} & \revision{\textbf{0.39}} & \revision{58.45} & \revision{52.36} & \revision{\textbf{0.35}} & \revision{35.30} & \revision{20.30} & \revision{0.32} & \revision{23.53} & \revision{20.63} & \revision{0.32} & \revision{51.72} & \revision{44.44} & \revision{0.29} \\

    \revision{SBERT} & \revision{46.80} & \revision{25.56} & \revision{\textbf{0.40}} & \revision{29.30} & \revision{26.74} & \revision{\textbf{0.39}} & \revision{58.53} & \revision{52.66} & \revision{\textbf{0.35}} & \revision{35.31} & \revision{20.31} & \revision{0.31} & \revision{23.50} & \revision{20.61} & \revision{0.31} & \revision{51.70} & \revision{44.32} & \revision{\textbf{0.28}} \\
    
    \hline
 \end{tabular}
\end{table}

\revision{In Table~\ref{tab:impact_of_similarity_metric}, the first column shows the similarity metrics used in the extractor. From this table, it is observed that compared with BLEU, METEOR, chrF, and Sentence-BERT (SBERT), the similarity metric ROUGE-L overall contributes more to the score of the final generated summaries (e.g., with the highest average values and the lowest standard deviations). In other words, if we use the similarity metric ROUGE-L to guide the selection of ground-truth important statements in the extractor, the final summaries generated by the abstracter will achieve higher BLEU, METEOR, and ROUGE-L scores.
Although some differences between programming languages and natural language, we get the same observation as researchers in NLP, that ROUGE-L is suitable for generating ground-truth important statements/sentences and thereby helps to improve code/text summarization.
}

\subsection{Case Study}
\label{subsec:case_study}
In this section, we provide case studies to understand the generated summaries of {\toolname} compared with Ex-based, SiT, and CodeT5 to demonstrate the usefulness of our {\toolname}. \revision{Specifically, we will present several cases, including two successful cases, two moderate cases, and two weak cases. We apply Ex-based, SiT, CodeT5, and our {\toolname} to generate summaries for comparison. In all examples, as in Section~\ref{sec:motivating_example}, we treat comments of code snippets as reference summaries.}

\subsubsection{Case Study on Java Code Summarization}
\begin{figure}[htbp]
  \centering
  \includegraphics[width=0.8\linewidth]{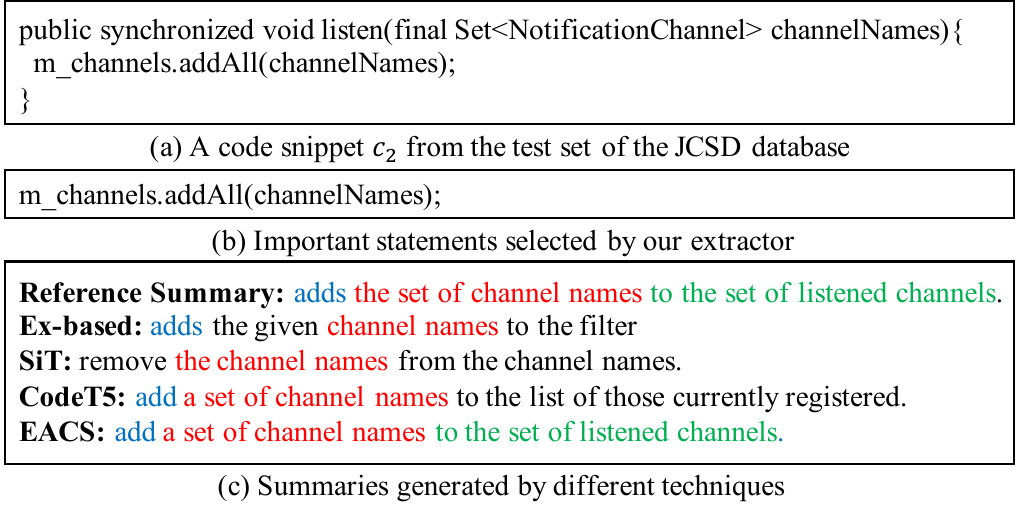}
  \caption{Successful Java Case}
  \label{fig:successful_Java_case}
\end{figure}

\
\newline
In this section, \deletion{we take the Java code snippet $c_2$ shown in Figure~\ref{fig:successful_Java_case}(a) as an example and apply \revise{Ex-based, }SiT\revise{, CodeT5,} and our {\toolname} to generate summaries for comparison. It is a real-world example from \delete{the test set of the JCSD dataset} \revise{the JCSD test set}.}\revision{we present three Java cases, including a successful case, a moderate case, and a weak case. They are real-world examples from the JCSD test set.}

\revision{\textbf{Successful Case.} Figure~\ref{fig:successful_Java_case} shows a successful Java case.} \deletion{As in Section~\ref{sec:motivating_example}, we consider the comment of $c_2$ as the reference summary, as shown in the first line of Figure~\ref{fig:successful_Java_case}(c).}\revision{The first line of Figure~\ref{fig:successful_Java_case}(c) shows the reference summary.} The summaries generated by \revise{Ex-based,} SiT\revise{, CodeT5,} and {\toolname} for $c_2$ are shown in \delete{the second and third lines, respectively}\revise{lines 2--5}. From the figure, we can observe that compared with the reference summary, the summary generated by SiT only covers the main semantics of the second part, i.e., ``the channel names'' (Red font). What is even worse is that the first core word ``remove'' generated by SiT and that of ``adds'' in the reference summary have completely opposite semantics, which makes the semantics of the generated summary and the reference summary completely opposite.
\revise{Compared with SiT, the summaries generated by Ex-based and CodeT5 cover the core word ``add'' and the main semantics of the second part (``channel names'' ).
In summary, in all three baselines, CodeT5 performs the best, followed by Ex-based and SiT.
Compared to CodeT5, the summary generated by {\toolname} covers all three parts of the reference summary. Technically, CodeT5 can be considered as an abstracter. The abstracter of {\toolname} has one additional component, i.e., ExEncoder, which is responsible for embedding important statements extracted by the extractor. Therefore, we can attribute the successful generation of the third part of the reference summary (i.e., ``to the set of listened channels'') to the extractive-and-abstractive framework we proposed. Intuitively, the text ``the set of listened channels'' in the third part is the summary of the code identifier ``m\_channels'' contained in the important statement in Figure~\ref{fig:successful_Java_case}(b). Based on the above, we can conclude that our {\toolname} is also a very competitive technique for the Java code summarization task.
}
\delete{Compared with the summary generated by SiT, the summary generated by {\toolname} covers all three parts of the reference summary. In addition, {\toolname} successfully generates the core word ``add'', which we attribute to the extractive-and-abstractive framework we proposed. This is demonstrated by the important statement shown in Figure~\ref{fig:successful_Java_case}(b) selected by the extractor module in our framework. Only the important statement contains the core word ``add''.} 

\begin{figure}[htbp]
  \centering
  \includegraphics[width=0.8\linewidth]{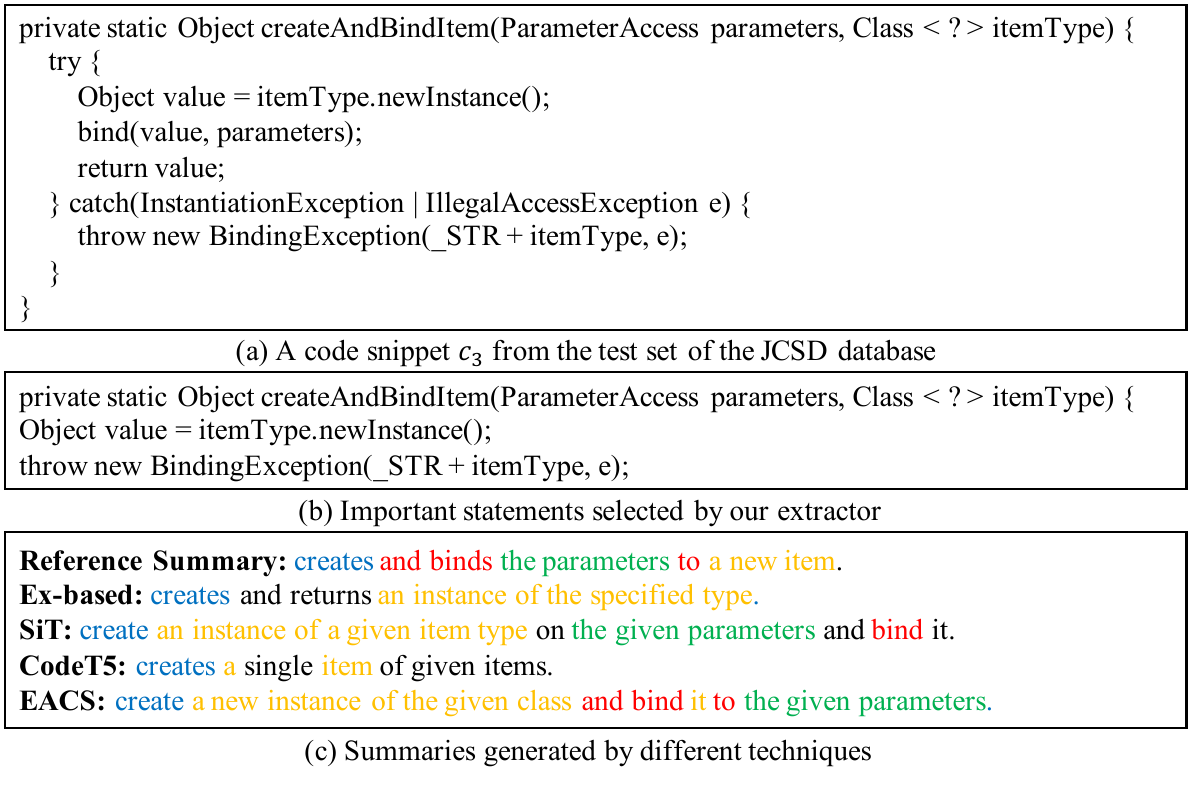}
  \caption{Moderate Java Case}
  \label{fig:moderate_Java_case}
\end{figure}

\revision{\textbf{Moderate Case.} Figure~\ref{fig:moderate_Java_case} shows a moderate Java case. In this example, according to the specific semantics of the code snippet $c_3$, the reference summary in the first line of Figure~\ref{fig:moderate_Java_case}(c) can be split into two clauses: ``creates a new item'' and ``binds the parameters to the new item''. From lines 2-5 of Figure~\ref{fig:moderate_Java_case}(c), it is observed that 1) all four techniques cover the core semantic of the first clause, i.e., ``creates a new item'' (highlighted in Blue and Orange font); 2) SiT and {\toolname} cover the most factual details (e.g., ``bind'' and ``parameters'') contained in the reference summary and outperform Ex-based and CodeT5. However, for the second clause ``binds the parameters to the new item'', the summary generated by SiT is incorrect. Similarly, {\toolname} erroneously summarizes ``bind it (i.e., the new item) to the (given) parameters''. Of course, it is undeniable that our extractor still performs well in extracting important statements. From this case, we can find that it is still challenging for the current code summarization models (including our {\toolname}) to understand the fine-grained positional relationships among code elements, especially when such relationships are not explicitly presented in the code (e.g., ``bind(value, parameters)'') or in a form that is prone to misleading the model (e.g., ``createAndBindItem'').
}

\begin{figure}[htbp]
  \centering
  \includegraphics[width=0.8\linewidth]{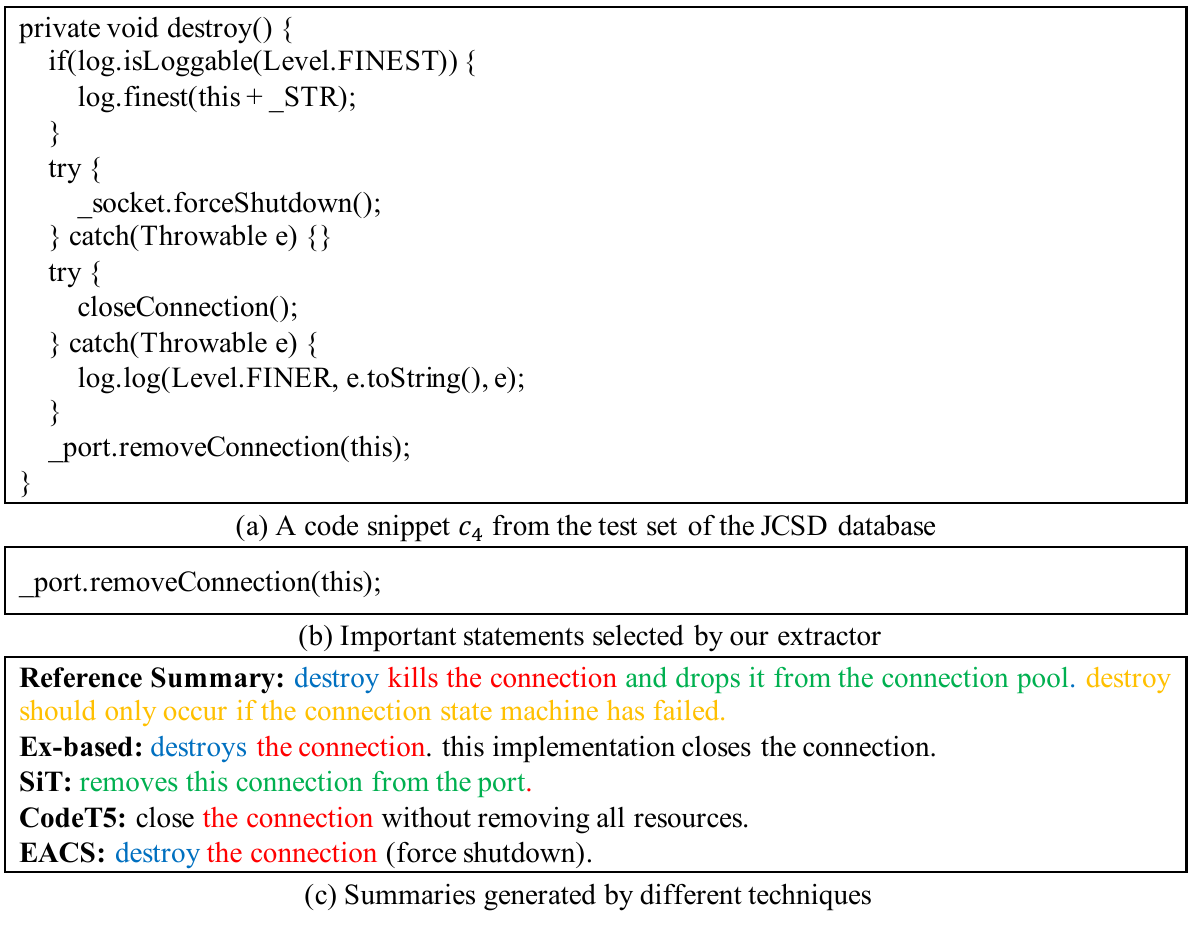}
  \caption{Weak Java Case}
  \label{fig:weak_Java_case}
\end{figure}

\revision{\textbf{Weak Case.} Figure~\ref{fig:weak_Java_case} shows a weak Java case. In this example, the reference summary in the first line of Figure~\ref{fig:weak_Java_case}(c) contains two sentences. The first sentence introduces the functionality of the code snippet $c_4$. The second sentence describes the condition that triggers such a functionality (highlighted in Orange font). From lines 2-5 of Figure~\ref{fig:weak_Java_case}(c), it is observed that 1) Ex-based, CodeT5, and our {\toolname} only cover the first half of the first sentence, while SiT only covers the second half; 2) all four techniques fail to cover the second sentence. As shown in Figure~\ref{fig:weak_Java_case}(b), our extractor successfully extracts the important statement, which provides key information for the abstracter to summarize ``drops it from the connection pool''. Unfortunately, the abstracter still fails to summarize the second half of the first sentence. This means that our abstracter needs to be further optimized and improved. In addition, although the second sentence does not describe the core functionality of $c_4$, it provides very valuable information for developers to understand and use $c_4$. From this case, we can find that the summaries generated by current code summarization models (including our {\toolname}) are still not as good as professional human developers, and there is still much room for improvement.
}

\subsubsection{Case Study on Python Code Summarization}
\
\newline
\revision{In this section, we present three Python cases, including a successful case, a moderate case, and a weak case. They are real-world examples from the PCSD test set.}

\begin{figure}[htbp]
  \centering
  \includegraphics[width=0.8\linewidth]{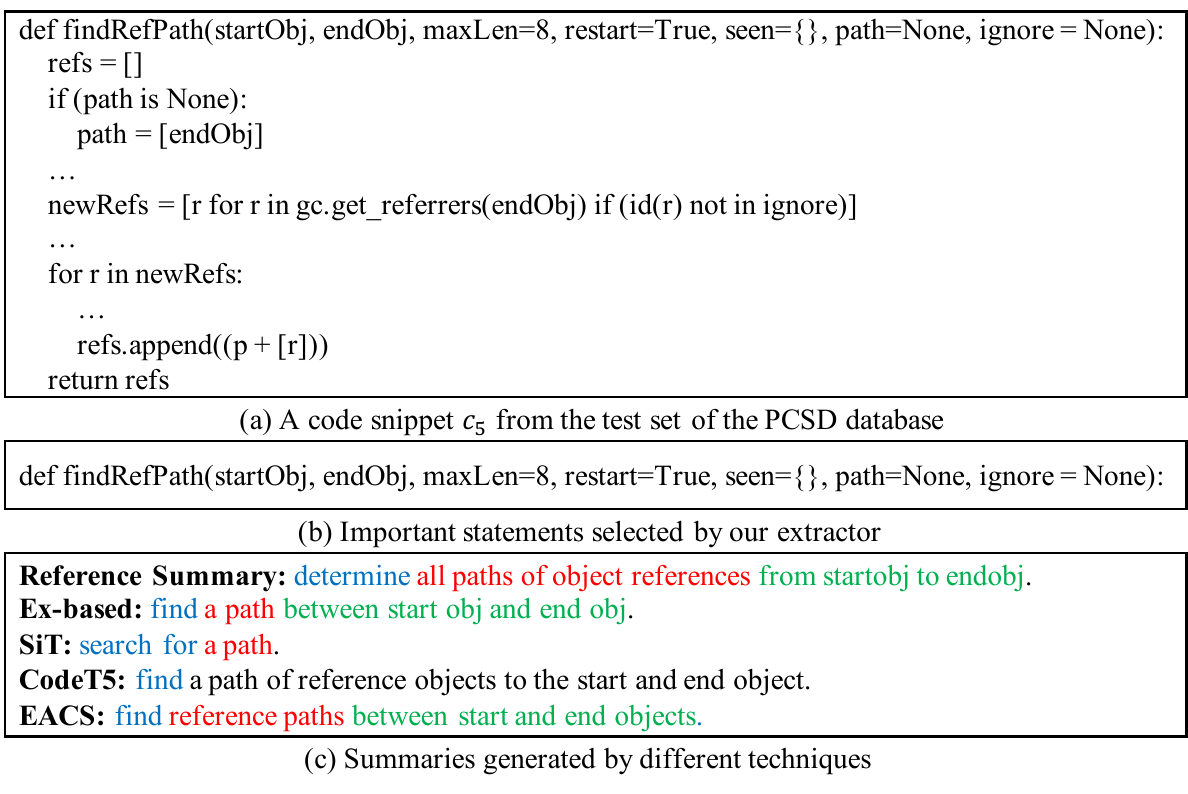}
  \caption{Successful Python Case}
  \label{fig:successful_Python_case}
\end{figure}

\revision{\textbf{Successful Case.} Figure~\ref{fig:successful_Python_case} shows a successful Java case.} \deletion{Figure~\ref{fig:successful_Python_case}(a) shows the Python code snippet $c_3$ from \delete{the test set of the PCSD dataset}\revise{the PCSD test set}. $c_3$}\revision{$c_5$ is a long piece that consists of 51 lines of code.} Figure~\ref{fig:successful_Python_case}(b) presents the important statements selected by our extractor from $c_3$. \delete{Figure~\ref{fig:successful_Python_case}(c) contains three summaries, the first from the comment of \deletion{$c_3$}\revision{$c_5$}, and the second and last generated by SiT and our EACS, respectively.}
\revise{Figure~\ref{fig:successful_Python_case}(c) contains five summaries, the first from the comment of \deletion{$c_3$}\revision{$c_5$}, the second to five summaries are generated by Ex-based, SiT, CodeT5, and our {\toolname}, respectively.}
From the figure, we can observe that compared with the reference summary, 1) the summary generated by SiT misses many factual details, such as ``from startobj to endobj'';
\delete{2) although it looks a little different literally, the summary generated by our {\toolname} is able to cover all of the three parts and are semantically equivalent.}
\revise{2) although the summary generated by CodeT5 is literally similar to the reference summary, it has completely different semantics; 3) the summaries generated by Ex-based and {\toolname} cover the main semantics of the reference summary although they look different literally. In addition, compared with Ex-based, {\toolname} can correctly cover more complete semantics. For example, the summary text ``reference paths'' generated by {\toolname} is semantically equivalent to the text ``paths of object references'' in the reference summary, although the expressions are different. In addition, the expression ``find reference paths'' is aligned with the function name ``findRefPath''. It should be noticed that, as shown in Figure~\ref{fig:successful_Python_case}(b), our extractor successfully extracts the important statement in which the factual detail ``reference'' appears. Therefore, we can attribute the successful generation of the text ``reference paths'' to the extractive-and-abstractive framework we proposed. In summary, our {\toolname} significantly outperforms the other three baselines in terms of the semantic completeness of the generated summaries. Based on the above, we can conclude that our {\toolname} is also a very competitive technique for the Python code summarization task.}

\begin{figure}[htbp]
  \centering
  \includegraphics[width=0.8\linewidth]{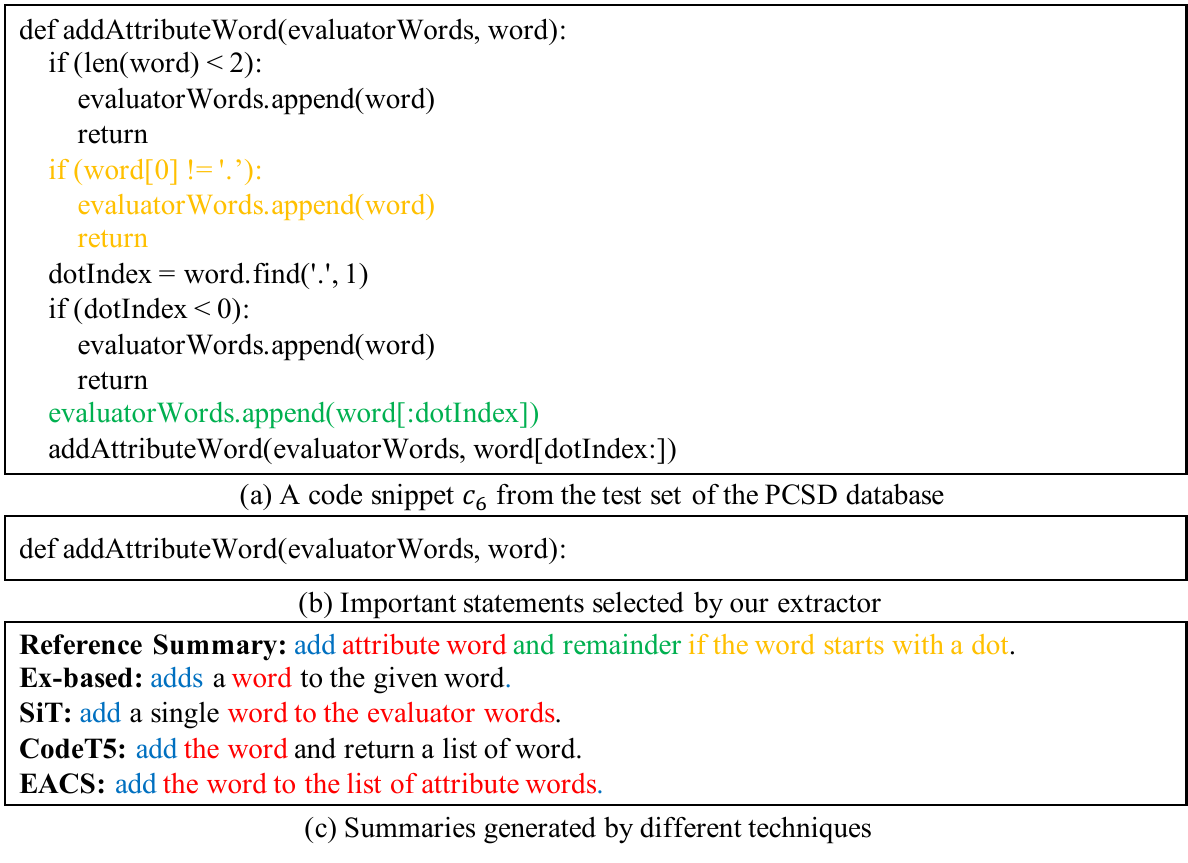}
  \caption{Moderate Python Case}
  \label{fig:moderate_Python_case}
\end{figure}

\revision{\textbf{Moderate Case.} Figure~\ref{fig:moderate_Python_case} shows a moderate Python case. Similarly, we can divide the reference summary of $c_6$ shown in the first line of Figure~\ref{fig:moderate_Python_case}(c) into four parts: ``add'', ``attribute word'', ``and remainder'', and ``if the word starts with a dot''. Lines 2-5 show the summaries generated by four techniques. From Figure~\ref{fig:moderate_Python_case}(a), we can know that ``evaluatorWords'' is a list used to store the attribute words. From a semantic point of view, only SiT and {\toolname} correctly cover the first and second parts of the reference summary. The third part of the reference summary (i.e., ``and remainder'') is a summary of the ``word[:dotIndex]'' shown in the penultimate line of Figure~\ref{fig:moderate_Python_case}(a). This summary is highly abstract, reflecting the summarization ability and word habits of professional human developers. Obviously, current code summarization models (including our {\toolname}) do not yet have such capabilities. The fourth part of the reference summary, i.e., ``if the word starts with a dot'', is a summary of lines 5-7 of $c_6$ (highlighted in Orange font). This summary is also very abstract. ``the word starts with a dot'' and ``word[0]!='.''' are opposite conditions. The developer uses ``if (false) $\cdots$ return $\cdots$'' to achieve the same functionality as ``if (true) $\cdots$ else $\cdots$'', thereby achieving the goal of simplifying code. All four techniques fail to generate such abstract summaries. This case shows that current code summarization models (including our {\toolname}) are not yet able to make a high-level abstract summarization of the code, and there is still much room for improvement. In addition, as shown in Figure~\ref{fig:moderate_Python_case}(b), our extractor also fails to extract those important statements related to highly abstract summaries. We will further explore how to capture the alignment between highly abstract summaries and code statements in future work.
}

\begin{figure}[htbp]
  \centering
  \includegraphics[width=0.8\linewidth]{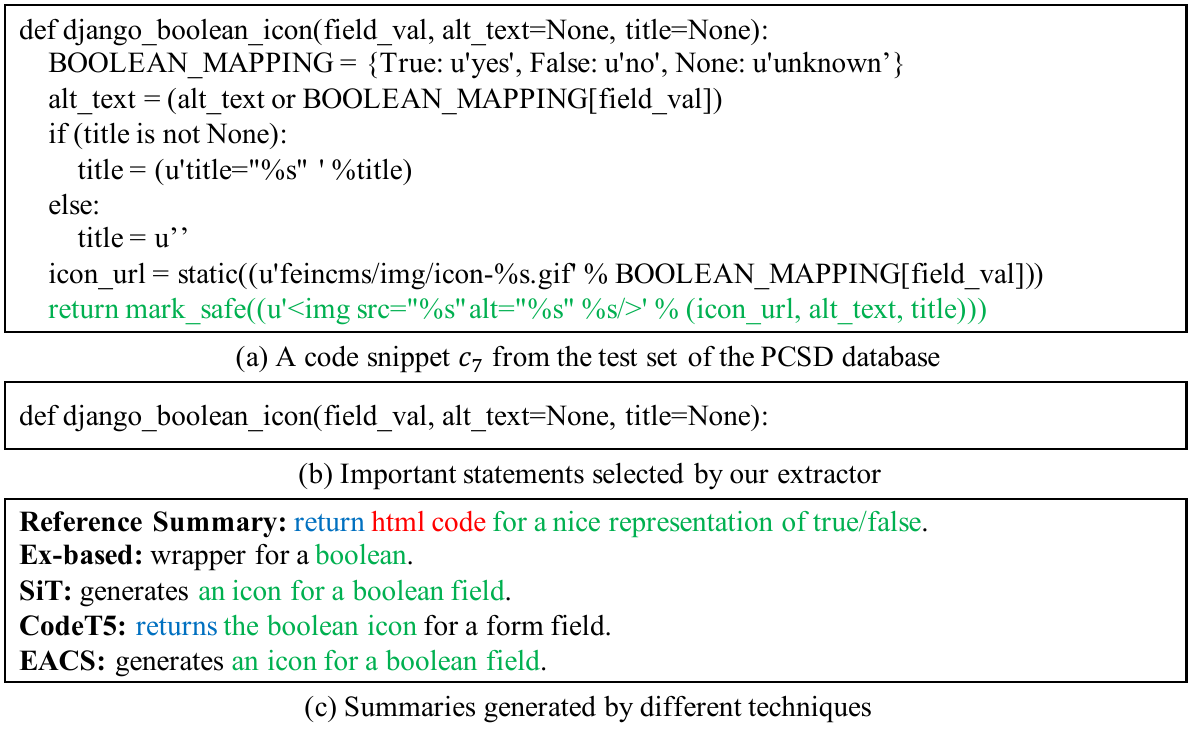}
  \caption{Weak Python Case}
  \label{fig:weak_Python_case}
\end{figure}

\revision{\textbf{Weak Case.} Figure~\ref{fig:weak_Python_case} shows a weak Python case. The reference summary of $c_7$ shown in the first line of Figure~\ref{fig:weak_Python_case}(c) can be divided into three parts: ``return'', ``html code'', and ``for a nice representation of true/false''. Combining the code snippet of $c_7$ shown in Figure~\ref{fig:weak_Python_case}(a), we can know that the ``a nice representation of true/false'' indeed refers to a boolean icon. Therefore, from a semantic perspective, the summaries generated by SiT, CodeT5, and {\toolname} correctly cover the third part of the reference summary. Of course, literally, the four generated summaries are far from the reference summary. Most of the words used in the generated summaries come from the code itself. Similar to what is observed in the previous case, these models are not yet able to make a high-level abstract summarization of the code like professional human developers. It is worth noting that all four techniques failed to cover the second part. Figure~\ref{fig:weak_Python_case}(a), we can observe that the last statement is closely related to the keyword ``html''. Although ``<img />'' is one of the commonly used tags in HTML, the four code summarization models do not seem to learn this association. Our extractor also fails to extract this important statement. We check the vocabulary and find that the tokenizer used in data preprocessing would break HTML tags. For example, the ``<img src'' in the last statement of Figure~\ref{fig:weak_Python_case}(a) is split into ``< '', ``imgs'', and ``rc'' by the CodeT5 tokenizer. Obviously, this not only breaks the tag mode, but also wrongly splits the ``img src'' into ``imgs'' and ``rc''. This may be the main reason why the four code summarization models and our extractor do not work well on this example. How to strengthen the model to learn special domain knowledge (e.g., HTML tags) may be a future research point, which is worthy of further exploration.
}

\section{Threats to Validity}
\label{sec:threats_to_validity}
There are three main threats to the validity of our approach
as follows:

\textbf{Dataset.} The first threat to validity is the evaluation dataset. The JCSD and PCSD datasets may contain duplications (as mentioned in~\cite{2020-Rencos}) in the training and test sets to make the model overfit easily.
To compare fairly with previous baselines, we follow~\cite{2018-TL-CodeSum, 2021-SiT} to directly use the same data as them and do not perform any additional processing on the two datasets. 
To alleviate this threat, we also conduct experiments on a large dataset CodeXGLUE~\cite{2021-CodeXGLUE}, which is very carefully de-duplicated and includes six programming languages, and thus the performance numbers we report here can be expected to be fairly good.
We will use the preprocessing script provided by~\cite{2020-Rencos} to deduplicate the dataset and conduct experiments in future work.

\textbf{Evaluation Method.} The second threat to validity is the evaluation method. In this paper, we employ three automatic evaluation metrics (i.e., \deletion{BLEU-4}\revision{BLEU}, METEOR, and ROUGE-L) to evaluate the quality of the generated summaries. Although these metrics are widely used in existing code summarization works, they still have limitations. The three metrics can calculate the textual difference between generated summaries and references. However, in some cases, the model may produce valid summaries that do not align with the ground truth, i.e., the metrics cannot truly reflect the semantic similarity~\cite{2022-SCRIPT, 2022-M2TS}. To alleviate this threat, we also perform a human evaluation from four aspects (i.e., similarity, naturalness, informativeness, and relevance). Furthermore, we use the average results of multiple evaluators to alleviate the subjectivity of human evaluation. 

\textbf{Baselines.} The third threat to validity is about baselines. 
Since the implementation code for early TR-based methods is no longer available, in this paper, to alleviate this threat, we try our best to re-implement a TR-based method with the guide of the early works~\cite{2010-Program-Comprehension-with-Code-Summarization, 2010-Automated-Text-Summarization-Summarizing-Code}. 
Since this paper adopts the very recent datasets (i.e., JCSD and PCSD), the early TR-based methods did not conduct experimental evaluations on these datasets before. To proceed with our study, we perform LSA on our evaluation dataset. We process these data through our understanding of papers~\cite{2010-Program-Comprehension-with-Code-Summarization, 2010-Automated-Text-Summarization-Summarizing-Code} and choose recommended settings to minimize experimental bias. We make the implementation code public for subsequent researchers to check and use.

\section{Related Work}
\label{sec:related_work}
Code summarization has always been one of the hottest research topics in software engineering. As mentioned earlier, we can categorize existing work related to code summarization into extractive methods, abstractive methods, and others.

\textbf{Extractive methods.} Most early (prior to 2016) code summarization techniques~\cite{2010-Program-Comprehension-with-Code-Summarization, 2010-Towards-Generating-Summary-Java-Methods, 2010-Automated-Text-Summarization-Summarizing-Code, 2013-Automatic-Generation-Summaries-for-Java-Classes} are extractive methods. Such methods work by extracting a subset of the statements and keywords from the code, and then including information from those statements and keywords in summary. For example, Sonia Haiduc et al.~\cite{2010-Program-Comprehension-with-Code-Summarization} first propose an extractive method to generate extractive summaries for code snippets automatically. Extractive summaries are obtained from the contents of a document by selecting the most important information in that document. \textit{Extractive methods} use text retrieval (TR) techniques (e.g., Vector Space Model~\cite{1975-Vector-Space-Model}, Latent Semantic Indexing~\cite{1990-Latent-Semantic-Analysis}, and Hierarchical PAM~\cite{2007-Mixtures-of-Hierarchical-Topics}) to determine the most important $n$ terms for each code snippet. Considering that the quality of the summaries generated by extractive methods depends heavily on the process of extracting the subset, Paige Rodeghero et al.~\cite{2014-Code-Summarization-Eye-tracking-Study} present an eye-tracking study of programmers and propose a tool for selecting keywords based on the findings of the eye-tracking study. The extractive methods rely on high-quality identifier names and method signatures from the source code. These techniques may fail to generate accurate comments if the source code contains poorly named identifiers or method names~\cite{2015-CloCom}. 

\textbf{Abstractive methods.} Nowadays, DL-based (abstractive) code summarization techniques have been proposed one after another. By combining Seq2Seq models trained on large-scale code-comment datasets, abstractive methods can generate words that do not appear in the given code snippet to overcome the limitations of extractive methods. For example, Srinivasan Iyer et al.~\cite{2016-CODE-NN} present the first completely abstractive method for generating short, high-level summaries of code snippets. Their experimental results demonstrate that abstractive methods significantly outperform extractive methods in the naturalness of generated summaries. In other words, the summaries generated by abstractive methods are human-written-like. Code representation plays a key role in abstractive methods. To produce semantic-preserving code embedding representations, multiple aspects of the code snippet have been explored, including tokens~\cite{2016-CODE-NN, 2018-Structured-Neural-Summarization, 2018-TL-CodeSum, 2018-Improving-Code-Summarization-via-DRL, 2019-Ast-attendgru, 2019-Convolutional-Neural-Network-Code-Summarization, 2020-Hybrid-DeepCom, 2020-RL-Guided-Code-Summarization, 2020-Rencos, 2020-R2Com, 2020-Improved-Code-Summarization-via-GNN, 2021-BASTS, 2021-CoTexT, 2020-Transformer-based-Approach-for-Code-Summarization}, abstract syntactic trees (ASTs)~\cite{2018-Code2seq, 2018-DeepCom, 2018-Improving-Code-Summarization-via-DRL, 2019-Code-Summarization-with-Extended-Tree-LSTM, 2019-Ast-attendgru, 2020-Hybrid-DeepCom, 2020-Rencos, 2020-RL-Guided-Code-Summarization, 2020-R2Com, 2020-Improved-Code-Summarization-via-GNN, 2021-Code-Summarization-for-Smart-Contracts, 2021-BASTS}, control flows~\cite{2020-RL-Guided-Code-Summarization}, code property graph~\cite{2020-FusionGNN}. In addition, existing abstractive methods have tried various neural network architectures, such as LSTM~\cite{2016-CODE-NN, 2018-DeepCom, 2020-RL-Guided-Code-Summarization}, Bidirectional-LSTM~\cite{2021-EditSum, 2020-Rencos, 2020-R2Com}, GRU~\cite{2020-Hybrid-DeepCom, 2019-Ast-attendgru}, Transformer~\cite{2020-Transformer-based-Approach-for-Code-Summarization, 2021-BASTS} and GNN~\cite{2020-Improved-Code-Summarization-via-GNN, 2020-FusionGNN}. Although deep learning-based abstractive methods show great potential for generating human-written-like summaries, we find that the generated summaries often miss important factual details.

\textbf{Others.} Paul W. McBurney~\cite{2016-Code-Summarization-of-Context} takes the context of code snippets into account when generating summaries. The context includes the dependencies of the method and any other methods which rely on the output of the method~\cite{2006-Context-on-Program-Slicing}. The works~\cite{2021-Project-Level-Encoding-Code-Summarization, 2021-Generate-Comment-from-Class-Hierarchies, 2021-CoCoSum, 2021-API2Com} also use the information beyond the code snippet itself, e.g., project or class context~\cite{2021-Project-Level-Encoding-Code-Summarization, 2021-Generate-Comment-from-Class-Hierarchies, 2021-CoCoSum}, application programming interface documentations/knowledge (API Docs)~\cite{2021-API2Com}. These techniques cannot be tested on existing commonly used datasets (e.g., JCSD, PCSD, and CodeSearchNet Corpus) because methods (code snippets) in these datasets are context-missing. There are techniques using summaries of similar code snippets as the summary of the current snippet or as a starting point to generate a new summary. For example, Edmund Wong et al.~\cite{2015-CloCom} apply code clone detection techniques to discover similar code snippets and use the comments from some code snippets to describe the other similar code snippets. The works~\cite{2021-EditSum, 2020-Rencos, 2020-R2Com} retrieve similar code snippets and then use the information contained in the similar code snippets or their summaries to generate a summary of the current code snippet. They rely on whether similar code snippets can be retrieved and how similar they are. All the works mentioned above emphasize using external knowledge sources (e.g., contexts, API Docs, and similar code snippets) to improve the quality of the generated comments. We will explore the effect of combining our {\toolname} with the above external knowledge in future work.

\textbf{Our method.} {\toolname} is a general code summarization framework that adopts extractive and abstract schemes at the same time. Different from existing extractive methods that use TR techniques to extract important statements/words, the extractive module of {\toolname} applies a deep learning-based classifier to predict the importance of each statement. Unlike existing abstractive methods, our abstracter receives and processes the entire code snippet and important statements as input in parallel.

\section{Conclusion}
\label{sec:conclusion}
In this paper, we propose an extractive-and-abstractive framework named {\toolname} for source code summarization. {\toolname} consists of two modules, extractor and abstracter. The extractor has the ability to extract important statements from the code snippet. The important statements contain important factual details that should be included in the final generated summary. The abstracter takes in the important statements extracted by the extractor and the entire code snippet and is able to generate human-written-like summary in natural language. We conduct comprehensive experiments on three databases to evaluate the performance of {\toolname}. And the experimental results demonstrate that our {\toolname} is an effective code summarization technique and significantly outperforms the state-of-the-art. Extensive human evaluations demonstrate that the summaries generated by {\toolname} have higher naturalness and informativeness and are more relevant to given code snippets.

\revision{
As mentioned earlier, the extractive-and-abstractive framework we proposed is highly scalable that does not depend on a specific deep learning network/model. Based on the experimental results of \textbf{RQ2} shown in Section~\ref{subsubsec:results_of_combine_with_diff_models}, we can speculate that replacing the pre-trained model in the framework with more advanced large language models (LLMs), such as ChatGPT~\cite{2022-ChatGPT} (if available), has the opportunity to further improve code summarization performance. 
We plan to combine {\toolname} with more advanced LLMs, to exert more powerful performance in the future.
}

\section*{Acknowledgment}
The authors would like to thank the anonymous reviewers for their insightful comments. This work is supported partially by National Natural Science Foundation of China (61932012, 62141215, 62372228) and the Program B for Outstanding PhD Candidate of Nanjing University (202201B054).

\bibliographystyle{ACM-Reference-Format}
\bibliography{reference}

\end{document}